\documentclass[12pt]{article}

\usepackage{amsmath}
\usepackage{amssymb}
\usepackage{latexsym}
\usepackage{graphicx}
\usepackage{epsfig}

\usepackage{tabularx} % extra features for tabular environment
\usepackage{amsmath}  % improve math presentation
\usepackage{graphicx} % takes care of graphic including machinery
\usepackage{cite} % takes care of citations
\usepackage[final]{hyperref} % adds hyper links inside the generated pdf file
\hypersetup{
	colorlinks=true,       % false: boxed links; true: colored links
	linkcolor=blue,        % color of internal links
	citecolor=blue,        % color of links to bibliography
	filecolor=magenta,     % color of file links
	urlcolor=blue         
	}

\def\beq{\begin{equation}}
\def\eeq{\end{equation}}
\def\bea{\begin{eqnarray}}
\def\eea{\end{eqnarray}}

\def\a{\alpha}

\def\m{\mu}
\def\n{\nu}

\def\f{\phi}
\def\L{\Lambda}

\def\nn{\nonumber}
\def\2{\;\;}
\def\4{\;\;\;\;}
\begin{document}

\begin{titlepage}

\vspace*{1cm}
\begin{center}
{\bf \Large Large and Ultra-compact Gauss-Bonnet Black Holes\\[3mm] 
with a Self-interacting Scalar Field} \\[4mm]

\bigskip \bigskip \medskip

{\bf A. Bakopoulos}$^{\,(a)}$\,\footnote{Email: a.bakop@uoi.gr},
{\bf P. Kanti}$^{\,(a)}$\,\footnote{Email: pkanti@uoi.gr} and
{\bf N. Pappas}$^{\,(a,b)}$\,\footnote{Email: npappas@uoi.gr}

\bigskip
$^{(a)}${\it Division of Theoretical Physics, Department of Physics,\\
University of Ioannina, Ioannina GR-45110, Greece}

\medskip
$^{(b)}${\it Nuclear and Particle Physics Section, Physics Department,\\ National and
Kapodistrian University of Athens, Athens GR-15771, Greece}

\bigskip \medskip
{\bf Abstract}
\end{center}
We consider the Einstein-scalar-Gauss-Bonnet theory, and study the case where a
negative cosmological constant is replaced by a more realistic, negative scalar-field
potential. We study different forms of the coupling function between the scalar
field and the Gauss-Bonnet term as well as of the scalar potential. In all cases, we obtain
asymptotically-flat, regular black-hole solutions with a non-trivial scalar field which
naturally dies out at large distances. For a quadratic negative potential, two distinct
subgroups of solutions emerge: the first comprises light black holes with a large horizon
radius, and the second includes massive, ultra-compact black holes. The most
ultra-compact solutions, having approximately the 1/20 of the horizon radius of the
Schwarzschild solution with the same mass, emerge for the exponential and linear
coupling functions. For other polynomial forms of the scalar potential, the subgroup of ultra-compact
solutions disappears, and the black holes obtained may have a horizon radius larger or
smaller than the Schwarzschild solution depending on the particular value of their mass.

\end{titlepage}

\setcounter{page}{1}
%%%%%%%%%%%%%%%%%%%%%%%%%%%%%%%%%%%%%%%%%%%%%%%%%%%%%%%%%%%%%%%%%%%%%%

\section{Introduction}

In our quest for a fundamental theory of gravity, the generalised gravitational theories,
containing extra fields or curvature terms compared to the traditional General Relativity
(GR) \cite{Stelle,General}, provide a valuable framework of study. These theories are considered
to be only effective theories of gravity which reduce to GR at the weak-field limit but
predict modifications of it at regimes of strong curvature. The successful detection of
gravitational waves during the recent years \cite{LIGO, VIRGO} has provided an additional
tool for the study and verification, or disproof, of these theories, and thus has refueled
the interest in their predictions. In particular, since modifications are expected only at
regimes of strong gravity, predictions for the existence of novel black-hole solutions
or compact objects have attracted a wide interest in the community.

The study of generalised gravitational theories was initiated only a few years after the
the classification of black-hole solutions in GR was completed, and included extensions
of GR \cite{Lovelock} and its connection to other fields \cite{NH-scalar}. Soon, solutions
of black holes with additional characteristics, or ``hair'', compared to those of GR were
found \cite{YM, Skyrmions, Conformal}. The development of the superstring effective theory
\cite{Zwiebach, Gross, Metsaev} gave an additional boost to the study of these theories,
and led the way towards the discovery and study of a number of novel black-hole
solutions \cite{Gibbons, Callan, Campbell, Mignemi, Kanti1995, DBH, Torii, KT, Guo,
Kleihaus, Pani, Herdeiro, Ayzenberg} (see \cite{Win-review, Charmousis-rev,
Herdeiro-review, Blazquez} for a number of reviews).
Most of these solutions were found in the context of generalised theories
built around a scalar field with non-minimal couplings to gravity,  and evaded
the so-called ``no scalar-hair'' theorem \cite{Bekenstein}. Extending GR via the
addition of a scalar field and gravitational terms has also been the core of the
Horndeski \cite{Horndeski} and Galileon \cite{Galileon} theories. In the context of
these theories, the no-hair theorems were re-formulated \cite{SF, HN} but were
again evaded, and additional black-hole solutions were constructed 
\cite{SZ, Babichev, Benkel, Yunes2011}. 

The class of Einstein-scalar-Gauss-Bonnet (EsGB) theories, which include, apart
from the Einstein term, a scalar field and the quadratic Gauss-Bonnet (GB) term,
is a particularly simple but extremely rich family of generalised gravitational theories.
It is characterised by the form of the coupling function between the scalar field and
the GB term, which is not a priori fixed. Setting this coupling function to be of an
exponential form leads to the dilatonic theory, in the context of which the dilatonic
black holes \cite{DBH}, the first counter-example of the scalar no-hair theorem
\cite{Bekenstein}, were found. For a linear coupling function, the shift-symmetric
Galileon black holes \cite{Benkel} were also derived. These solutions have
the characteristic feature of scalar hair: a regular, non-trivial scalar field which
is associated with the black hole, a feature forbidden by GR. The group of theories
which may lead to such solutions was significantly expanded a few years ago, when
it was demonstrated \cite{ABK}, both analytically and numerically, that the EsGB theories
support black-hole solutions with a regular, non-trivial scalar hair for every form of
the coupling function. In addition to this {\it natural scalarisation} of the solutions,
{\it spontaneous scalarisation} of the corresponding GR solution was also shown to
take place \cite{Doneva, Silva}. 

A large number of additional works has appeared over the years, which studied novel
black holes or compact objects in these, or similar, types of theories as well as their
properties \cite{Bardoux}-\cite{Carson}.
%\cite{Charmousis, Correa, Doneva-NS, Motohashi, Radu, Doneva-Papa, Butler, Danila,
%Stetsko, Ayzen, Dolan, Kunz, Bhatta, Tatter, Mukherjee, Chakra, Berti, Brihaye, Prabhu,
%Myung, Don-Kunz, Benkel2018, Iorio, Ovalle, Barack, Gao, Lee, Witek2018, Moto, Cunha}.
Apart from asymptotically-flat black-hole solutions, solutions with an asymptotic (Anti)-de
Sitter behaviour have also been investigated \cite{Martinez-deSitter}-\cite{Tang}. 
%\cite{Martinez, Radu-Win, Anabalon, Hosler, Kolyvaris, Ohta, Saenz, Caldarelli, Gonzalez,
%Gaete, Giribet, BenAchour, Hartmann, BAK, Radu-scal-dS}. 
In addition, the Einstein-scalar-Gauss-Bonnet theory has been shown to lead to a large
number of families of wormhole solutions that require no exotic matter \cite{KKK1, ABKKK}.
Last, but not least, it supports compact, particle-like solutions with regular spacetimes,
non-trivial scalar hair and a number of interesting observable features \cite{Hartmann,
Herdeiro-Oliveira, Afonso, Canate1, Canate2, KKK2, Radu-part}. 

%%%%%%%%%%%%%%%%%%%%%%%%%%%%%%%%%%%%%%%%%%%%%%%%%%%%%%%%%%%%%%%%%%%%%%%

In a previous work \cite{BAK}, we studied the EsGB theory in the presence of a 
negative cosmological constant $\Lambda$, and demonstrated that black-hole solutions
with an asymptotically Anti-de Sitter behaviour and a scalar hair arise as easily as their
asymptotically-flat counterparts. In the present work, we will address the case where 
this constant energy density is replaced by a non-trivial potential for the scalar field. 
Our objective is to provide a more realistic scenario where the cosmological constant,
usually introduced in an ad hoc way in the theory, is now replaced by a field potential. 
We will first pose the question of whether black holes, naturally scalarised, arise in the
context of such a theory, and whether only specific forms of the scalar-field potential 
may support them. If novel black-hole solutions do emerge, we would like to investigate
which features of the previously found solutions are still preserved and which are modified.
To this end, we will consider a variety of polynomial forms for the scalar-field potential,
and combine them with various forms of the coupling function between the scalar
field and the GB term. Given that black-hole solutions arise in abundance in the
case of a negative cosmological constant \cite{BAK}, here, we will consider only
negative-definite forms of the scalar potential. Scalarised solutions in the presence
of a positive scalar potential of quadratic and quartic form have been studied in
\cite{Doneva-massive} and \cite{Macedo_potential}, respectively. In the present context,
it will also be of great interest to see whether the combined effect of the GB term with
a scalar potential -- of a form not physically preferred -- will manage to support
regular black-hole solutions. 

Our analysis will demonstrate that this is indeed the case, and that in fact regular
black holes with a scalar hair emerge for any combination of choices for the coupling
function and form of negative scalar potential we have used. We will present families of
robust solutions describing spacetimes with a regular horizon at one end and an
asymptotically-flat limit at the other. The scalar field, its effective potential, the GB
curvature-invariant term and all components of the energy-momentum tensor will
be finite reducing to zero at large distances. The solutions obtained will be larger or
smaller, compared to the Schwarzschild solution with the same mass, 
depending on the exact form of the scalar potential, the value of the black-hole 
mass and the branch they belong to. In the case of a negative, quadratic potential,
and for all forms of coupling functions, we obtain solutions which belong to two
subgroups: the first comprises light GB black holes with horizon radius and
entropy larger than the ones of the corresponding GR solution, and the second includes
the more massive black holes with an increasingly smaller horizon radius which terminates
to a class of massive, ultra-compact black holes.

The outline of this work is as follows: in Section 2, we present our theoretical framework
and in Section 3, we consider the form of the asymptotic solutions near the 
sought-for black-hole horizon and asymptotic infinity. In Section 4, we study their
thermodynamical properties, and in Section 5, we present our numerical results for 
he black-hole solutions found and their properties. We finish with our conclusions and
discussion of our results in Section 6.

%%%%%%%%%%%%%%%%%%%%%%%%%%%%%%%%%%%%%%%%%%%%%%%%%%%%%%%%%%%%%%%%%%5

\section{The Theoretical Framework}

In this work, we will study a general class of higher-curvature gravitational theories,
which includes the Einstein-Hilbert term, given by the Ricci scalar curvature $R$, 
a scalar field $\phi$, and the quadratic Gauss-Bonnet term defined as 
%%%%%%%%%%%%
\begin{equation}\label{GB-def}
R^2_{GB}=R_{\mu\nu\rho\sigma}R^{\mu\nu\rho\sigma}-4R_{\mu\nu}R^{\mu\nu}+R^2\,,
\end{equation}
%%%%%%%%%%%%
in terms of the Riemann tensor $R_{\mu\nu\rho\sigma}$, the Ricci tensor $R_{\mu\nu}$
and the Ricci scalar $R$. Therefore, the action functional of the theory has the form
%%%%%%%%%%%%%%%%%%%%
\begin{equation}
S=\frac{1}{16\pi}\int{d^4x \sqrt{-g}\left[R-\frac{1}{2}\,\partial_{\mu}\phi\partial^{\mu}\phi+f(\phi)R^2_{GB}
- 2\L V(\phi)\right]}.
\label{action}
\end{equation}
%%%%%%%%%%%%%% 
A general coupling function $f(\phi)$ provides a coupling of the scalar field to the
GB term since the latter is a total derivative in four dimensions. We have also included
in the theory a self-interacting potential $V(\phi)$ for the scalar field; this upgrades
the usually assumed cosmological constant $\Lambda$ to a dynamical potential, 
with $\Lambda$ assuming now the role of a coupling constant.
In the context of this work, we will consider only the case with $\Lambda <0$.
Therefore, by setting $V(\phi)=1$, we recover the Einstein-scalar-GB theory 
in the presence of a negative cosmological constant \cite{BAK}.

The variation of the aforementioned action with respect to the metric tensor $g_{\mu\nu}$
and the scalar field $\phi$ leads to the gravitational field equations and the equation for
the scalar field, respectively. These have the form:
%%%%%%%%%%%%%%%%
\begin{equation}
G_{\mu\nu}=T_{\mu\nu}\,, \label{field-eqs}
\end{equation}
\begin{equation}
\nabla^2 \phi+\dot{f}(\phi)R^2_{GB}-2\Lambda \dot V(\phi)=0\,. \label{phi-eq_0}
\end{equation}
%%%%%%%%%%%%%%
In the above, $G_{\mu\nu}$ is the Einstein tensor and $T_{\mu\nu}$ is the energy-momentum tensor,
given by the expression
%%%%%%%%%%%%%%%%%%%
\begin{equation}\label{Tmn}
T_{\mu\nu}=-\frac{1}{4}g_{\mu\nu}\partial_{\rho}\phi\partial^{\rho}\phi+\frac{1}{2}\partial_{\mu}\phi\partial_{\nu}\phi-\frac{1}{2}\left(g_{\rho\mu}g_{\lambda\nu}+g_{\lambda\mu}g_{\rho\nu}\right)
\eta^{\kappa\lambda\alpha\beta}\tilde{R}^{\rho\gamma}_{\quad\alpha\beta}
\nabla_{\gamma}\partial_{\kappa}f(\phi)- \L V(\phi)\,g_{\m\n}\,.
\end{equation}
%%%%%%%%%%%%%
We also note that the dot over the coupling function and potential denotes their derivatives with
respect to the scalar field (i.e. $\dot V =dV/d\phi$). We have also used the definitions
%%%%%%%%%%%%%%%%
\begin{equation}\label{tildeR}
\tilde{R}^{\rho\gamma}_{\quad\alpha\beta}=\eta^{\rho\gamma\sigma\tau}
R_{\sigma\tau\alpha\beta}=\frac{\epsilon^{\rho\gamma\sigma\tau}}{\sqrt{-g}}\,
R_{\sigma\tau\alpha\beta}\,,
\end{equation}
%%%%%%%%%%%%%%%%%%%%%%%%
and, for simplicity, we have employed units in which $G=c=1$.

We are interested in deriving regular, static, spherically-symmetric black-hole solutions with a non-trivial
scalar field. The line-element of space-time will accordingly take the form
%%%%%%%%%%%%%%%%
\begin{equation}\label{metric}
{ds}^2=-e^{A(r)}{dt}^2+e^{B(r)}{dr}^2+r^2({d\theta}^2+\sin^2\theta\,{d\varphi}^2)\,.
\end{equation}
%%%%%%%%%%%%%%%
We will assume that the scalar field has the same symmetries as the metric tensor, and
therefore $\phi=\phi(r)$. Both the coupling function $f(\phi)$ and the scalar potential
$V(\phi)$ will be assumed at the moment to have a general form, and particular choices
will be made in the forthcoming sections.  

Employing the line-element (\ref{metric}), we may easily derive the non-vanishing components
of the Einstein tensor; these are
%%%%%%%%%%%%%
\begin{align}
 G^t_{\;\, t} &= \frac{e^{-B}}{r^2}(1-e^B-rB'),\label{Gtt}\\
 G^r_{\;\, r} &= \frac{e^{-B}}{r^2}(1-e^B+rA'),\label{Grr}\\
 G^\theta_{\;\,\theta} &=G^\phi_{\;\,\phi}=
 \frac{e^{-B}}{4r}\left[r{A'}^2-2B'+A'(2-rB')+2rA''\right].\label{Gthth}
\end{align}
%%%%%%%%%%%%%%%
Throughout our analysis, the prime denotes differentiation with respect to the radial
coordinate $r$. Next, using Eq. (\ref{Tmn}), we may find the components of
the energy-momentum tensor $T^\mu_{\;\; \nu}$
%%%%%%%%%%%%%%%%%%%%
\begin{align}
T^t_{\;\,t}=&-\frac{e^{-2B}}{4r^2}\left[\phi'^2\left(r^2e^B+16\ddot{f}(e^B-1)\right)-8\dot{f}\left(B'\phi'(e^B-3)-2\phi''(e^B-1)\right)\right]-\L V, \label{Ttt}\\[-3mm]
T^r_{\;\,r}=&\frac{e^{-B}\phi'}{4}\left[\phi'-\frac{8e^{-B}\left(e^B-3\right)\dot{f}A'}{r^2}\right] -\L V, \label{Trr}\\[0mm]
T^{\theta}_{\;\,\theta}=&T^{\varphi}_{\;\,\varphi}=-\frac{e^{-2B}}{4 r}\left[\phi'^2\left(re^B-8\ddot{f}A'\right)-4\dot{f}\left(A'^2\phi'+2\phi'A''+A'(2\phi''-3B'\phi')\right)\right]
-\L V. \label{Tthth}
\end{align}
%%%%%%%%%%%%%%%%%%
The explicit form of Einstein's field equations may then be derived by 
matching the corresponding components of $G^\mu_{\;\, \nu}$ and $T^\mu_{\;\, \nu}$,
We may also derive the explicit form of the scalar-field equation (\ref{phi-eq_0}) which reads
%%%%%%%%%%%%%%%%%%%
\begin{equation}
2r\phi''+(4+rA'-rB')\,\phi'+\frac{4\dot{f}e^{-B}}{r}\left[(e^B-3)A'B'-(e^B-1)(2A''+A'^2)\right]
-4r e^B \Lambda \dot V=0\,. \label{phi-eq}
\end{equation}
%%%%%%%%%%%%%%%%%%%%%

In order to find a complete solution describing a regular black hole with scalar hair, we
need to determine three unknown functions, namely $A(r)$, $B(r)$ and $\phi(r)$. However,
the metric function $B(r)$ is in fact a dependent quantity whose form may easily be determined
once the solutions for $A(r)$ and $\phi(r)$ are found. Indeed, the $(rr)$-component of field
equations takes the form of a second-order polynomial with respect to $e^B$, i.e. $\alpha e^{2B}+\beta e^{B}+\gamma=0$, which then leads to the solution 
%%%%%%%%%%%%%%%
\begin{equation}\label{Bfunction}
e^B=\frac{-\beta\pm\sqrt{\beta^2-4\a\gamma}}{2\a},
\end{equation}
where
\beq
\a= 1-\L V r^2, \qquad
\beta=\frac{r^2{\phi'}^2}{4}-(2\dot{f}\phi'+r) A'-1,\label{abc}\qquad
\gamma=6\dot{f}\phi'A'.%\label{gamma}
\eeq
%%%%%%%%%%%%%%%%%
We may then eliminate the quantity $B'$ from all equations since the above solution,
when differentiated, leads to the expression
%%%%%%%%%%%%%%%%
\begin{equation}\label{B'}
B'=-\frac{\gamma'+\beta' e^{B}+\a' e^{2B}}{2\a e^{2B}+\beta e^{B}}.
\end{equation}
%%%%%%%%%%%%%
The remaining field equations then lead to a system of only two independent, ordinary
differential equations of second order for the functions $A(r)$ and $\phi(r)$:
%%%%%%%%%%%%%%%%%%%%%
\begin{align}
A''=&\frac{P}{S}\,,  \label{A-sys} \\
\phi''=&\frac{Q}{S}\,. \label{phi-sys}
\end{align} 
The expressions for the quantities $P$, $Q$ and $S$ in terms of $(r, \phi', A', \dot f, \dot V, \ddot f)$
are given in Appendix A, for the interested reader, as they are quite complicated. 

\section{Asymptotic Solutions} 

%\subsection{Asymptotic Solution at Black-Hole Horizon}

Before turning to the numerical integration of the system (\ref{A-sys})-(\ref{phi-sys}),
we will attempt to derive the analytical form of the solutions close to and far away
from the black-hole horizon. In fact, the asymptotic solution in the inner region will
be explicitly constructed by demanding the existence of a regular, black-hole horizon.
To this end, we assume that, as $r \rightarrow r_h$, the metric function $e^{A(r)}$
should vanish (and $e^{B(r)}$ should diverge) whereas the scalar field must remain
finite. As was explicitly shown in previous constructions \cite{ABK, BAK}, this amounts
to working in the limit $A'(r) \rightarrow \infty$ while keeping $\phi'(r)$ and $\phi''(r)$
finite as the black-hole horizon is approached. Working in these limits, 
Eq. (\ref{Bfunction}) yields\footnote{Note, that only the (+)-sign
in the expression for $e^B$ in Eq. (\ref{Bfunction}) leads to the desired black-hole
behaviour.}
\begin{align}\label{expb-limit}
e^B=\frac{2 \dot f \phi '+ r}{\left(1-r^2 \Lambda  
   V\right)}\,A'  + \frac{2 \dot f \phi ' \left(r^2 \phi '^2-12 \Lambda  r^2
   V+8\right)+r \left(r^2 \phi '^2-4\right)}{4
   \left(r^2\Lambda   V-1\right) \left(2 \dot f \phi '+r\right)} + \mathcal{O}\left(\frac{1}{A'}\right).
\end{align}
Employing the above expansion into the system (\ref{A-sys})-(\ref{phi-sys}), we obtain
\begin{align}
A''=&\frac{W_1}{W_3}\,A'^2+\mathcal{O}\left(A'\right),\label{expa}\\[3mm]
\f''=&\frac{W_2}{W_3}\,(2 \dot f \phi '+r) A' + \mathcal{O}(1),\label{expf}
\end{align}
where
\begin{align}
W_1=&+ 24 \Lambda ^2 r^4 V^2
   \dot f^2+4 r^3 \dot f (\Lambda  r \dot V-\phi ')+16
   \Lambda  r^2 \dot f^3 \phi '^2 \dot V +4
   \dot f^2 (4 \Lambda  r^3 \phi ' \dot V-r^2
   \phi '^2+6) \nn\\[0mm]
   &+   \Lambda  V \Bigl[4 r^5 \dot f \phi
   '+4 r^2 \dot f^2 \left(r^2 \phi
   '^2-16\right)-64 r \dot f^3 \phi '-64
   \dot f^4 \phi
   '^2+r^6\Bigr] -r^4, 
\end{align}
\begin{align}\label{w2}
W_2=&-32 \Lambda  V \dot f^3 \phi '^2+8 \Lambda  r \dot f ^2
   \phi ' (2 \Lambda  r^2 V^2+r \phi ' \dot V-6 V)+r^3 \Bigl[\phi ' (\Lambda  r^2
   V-1)+2 \Lambda  r \dot V\Bigr]\nn\\[0mm]
  & -2 \dot f
   \Bigl[2 \Lambda ^2 r^4 V^2-4 \Lambda  r^3 \phi ' \dot V+r^2
   \phi '^2-\Lambda  r^2 V \,(r^2 \phi'^2+4)+6\Bigr],
\end{align}
and
\begin{align}
W_3=(1-\Lambda  r^2 V)
   \Bigl[2 r^3 \dot f \phi '+16 \dot f^2 (2 \Lambda  r^2
   V-3)-32 \Lambda  r V \dot f^3 \phi '+r^4\Bigr].\label{w3}
\end{align}

From Eq. (\ref{expf}), we readily deduce that, if $\phi''$ is to remain finite near the
black-hole horizon as $A' \rightarrow \infty$, the coefficient of the latter, i.e.
$W_2\,(2 \dot f \phi '+r)$, should vanish. But, from Eq. (\ref{expb-limit}), we conclude
that the combination $(2 \dot f \phi '+r)$ must be non-vanishing for
the metric function $e^B$ to have the correct behaviour near the black-hole horizon.
Therefore, we demand that
\begin{equation}
W_2|_{r=r_h}=0.
\end{equation}
This constraint has the form of a second-order polynomial equation in terms of $\phi'$ which may
be solved to yield: 
\begin{align}\label{solphi'}
\f'_h=-\frac{r_h^3(1-\Lambda  V r_h^2) +16 \Lambda V r_h \dot f^2(3-\Lambda V r_h^2)
-8 \Lambda  \dot V r_h^3 \dot f \pm (1-\Lambda  V r_h^2)\sqrt{C}}
{4 \dot f \Bigl[r_h^2\,(1-4 \Lambda  \dot V \dot f)-\Lambda  V (r_h^4-16
   \dot f^2)\Bigr]},
\end{align}
where all quantities have been evaluated at $r=r_h$. The quantity $C$ is given by the expression
\begin{equation}\label{C-def}
C=384 \Lambda  r_h^2 \dot f^3 \dot V+32 r_h^2 \dot f^2 \left(2
   \Lambda V r_h^2 -3\right)+256 \Lambda  V \dot f^4
   \left(\Lambda V r_h^2-6\right)+r_h^6\,,
\end{equation}
and should always be positive-definite. This additional constraint imposes a bound on
one of the free parameters of the theory: for specific choices of $f(\phi)$ and $V(\phi)$
and fixed coupling parameter $\Lambda$, the aforementioned constraint imposes a bound,
or bounds, on the value of the black-hole horizon $r_h$. 

For $V(\phi)=1$, the expression
of $C$ may take the form of a second-order polynomial for $\dot f^2_h$, which then 
leads to two branches of black-hole solutions (or four, if negative values of $\dot f_h$ are
also allowed) describing solutions having either a minimum or a maximum horizon radius
\cite{Brihaye, BAK}. In reality, only solutions with a minimum horizon radius are found
while the second branch is plagued by instabilities. For $V(\phi) \neq 1$, the
expression for $C$ may take instead the form of a third-order polynomial for $r_h^2$;
this polynomial may have from one up to three real roots, therefore branches of solutions
with either a minimum mass, a maximum mass or both may emerge. Unfortunately, the
complexity of this polynomial combined with the arbitrariness in the values of the different
parameters do not allow us to analytically draw definite conclusions about the number
of positive, real roots and thus about the existence of upper or lower limits on the horizon
radius of the black-hole solutions. However, we may make the following observation: 
for $\Lambda V<0$, as will be the case in the present analysis, Eq. (\ref{expb-limit})
dictates that the combination $(2 \dot f \phi '+r)$ must be necessarily positive near
the horizon, therefore, we may write that $r_h > -2 \dot f_h \phi'_h$. If the right-hand-side
of this inequality is negative, it yields the trivial bound $r_h>0$; if, however, it is
positive, then it signals the existence of a minimum horizon radius for our black-hole
solutions. As we will see in Section 5, the right-hand-side of this inequality is indeed
positive for all the solutions found, therefore, a lower limit on the horizon radius of
the ensuing black holes always exists. 

Coming back to the constraint (\ref{solphi'}) and employing it into Eqs. (\ref{expa})-(\ref{expf}),
we readily obtain 
\begin{align}
A''=&-A'^2+\mathcal{O}\left(A'\right),\label{expa2}\\[3mm]
\f''=& \mathcal{O}(1).\label{expf2}
\end{align}
The first of the above equations may be integrated to give 
\beq
A'=\frac{1}{r-r_h}\,,\label{ahorp}
\eeq
%%%
which indeed exhibits the diverging behaviour near the black-hole horizon assumed at the
beginning of the analysis. Integrating once more leads to the solution for the metric function
$A(r)$ as the black-hole horizon is approached. Combining this
with Eq. (\ref{expb-limit}), we also obtain the near-horizon behaviour of the metric
function $B$. Putting everything together, we may therefore write the analytic form of
the solution of the field equations near $r_h$ as 
%%%%%%%%%%%%%
\bea
&&e^{A}=a_1 (r-r_h) + ... \,, \label{A-rh}\\[1.5mm]
&&e^{-B}=b_1 (r-r_h) + ... \,, \label{B-rh}\\[1mm]
&&\phi =\phi_h + \phi_h'(r-r_h)+ \phi_h'' (r-r_h)^2+ ... \,, \label{phi-rh}
\eea
%%%%%%%%%%%%%
where $a_1, b_1, \phi_h, \phi_h'$, and $\phi_h''$ are integration constants. The above
expressions describe a
regular black-hole horizon with a non-trivial, finite scalar field. Apparently, the presence
of a general form of a potential for the scalar field does not affect the existence per se 
of a near-horizon asymptotic solution, however, it is expected to modify the properties
of the ensuing solutions. 

%%%%%%%%%%%%%%%%%%%%%%%%%%%%%%%%%%%%%%%%%%%%%%%%%%%%%%%%%%%%%%%%%

We now turn to the form of the solution of the field equations at large distances from
the black-hole horizon. In this regime, the assumed form of the potential for
the scalar field is of paramount importance. Indeed, for $\Lambda=0$, and thus
a vanishing potential, we expect to recover an asymptotically-flat spacetime and a 
scalar field of the form
%%%%%%%%%%%%%%
\beq
\phi(r) \simeq \phi_\infty + \frac{D}{r} + ... \,,
\label{phi-flat}
\eeq
%%%%%%%%%%%%
for all forms of the coupling function $f(\phi)$, with $D$ being the scalar charge,
as found in \cite{ABK}. For $\Lambda<0$ and $V(\phi)=1$, black-hole solutions 
with an asymptotically Anti-de Sitter behaviour are expected to emerge as in
\cite{Brihaye, BAK}. These solutions do not possess a scalar charge since the
asymptotic behaviour of the scalar field is given by the expression
%%%%%%%%%%%%%%
\beq
\phi(r) \simeq \phi_\infty + d_1 \,\ln r + \frac{d_2}{r^2} + ... \,,
\label{phi-AdS}
\eeq
%%%%%%%%%%%%
independently of the form of the coupling function $f(\phi)$. For $\Lambda>0$, $V(\phi)=1$
and $f(\phi)=f_0-\alpha \phi^2$, black-hole solutions with a smooth scalar field both at
the black-hole and cosmological horizons of a form similar to the one given in Eq. (\ref{phi-rh})
were derived in \cite{BHR}; however, the scalar field diverges beyond the cosmological horizon
thus deviating from the expected asymptotically de Sitter behaviour. Finally, for $\Lambda=1$
and $V(\phi)=m^2 \phi^2$, the case of a massive scalar field was studied for different
forms of the coupling function $f(\phi)$ \cite{Doneva-massive}: in all cases, the spacetime
approached an asymptotically-flat solution with the scalar field exhibiting a universal profile
of a Yukawa-type form, namely
%%%%%%%%%%%%%
\beq
\phi(r) \simeq \frac{e^{-m r}}{r} + ... \,.
\label{phi-massive}
\eeq
%%%%%%%%%%%%
As a result, one cannot derive generic forms for either the spacetime or the scalar field, 
at the limit of large distances from the black-hole horizon, when the scalar potential $V(\phi)$
is kept arbitrary. To this end, we will postpone this analysis for the next sections
where specific forms of $V(\phi)$ will be chosen. 

%%%%%%%%%%%%%%%%%%%%%%%%%%%%%%%%%%%%%%%%%%%%%%%%%%%%%%

\section{Thermodynamical Analysis}

The form of the asymptotic solutions for the metric functions and scalar field near
the black-hole horizon, derived in the previous section, allows us also to calculate
the thermodynamical properties of the assumed black-hole solutions. Thus,
the temperature of the sought-for black hole may be derived from its surface gravity
$\kappa_h$ through the following definition \cite{York, GK}
%%%%%%%%%%%%%%%
\beq 
T=\frac{\kappa_h}{2\pi}=\frac{1}{4\pi}\,\left(\frac{1}{\sqrt{|g_{tt} g_{rr}|}}\,
\left|\frac{dg_{tt}}{dr}\right|\right)_{r_h}=\frac{\sqrt{a_1 b_1}}{4\pi}\,.
\label{Temp-def}
\eeq
%%%%%%%%%%%%%%%%
The above formula holds for spherically-symmetric black holes emerging even in theories
which contain higher-derivative terms such as the GB term. After employing the near-horizon
asymptotic forms (\ref{A-rh})-(\ref{B-rh}) of the metric functions, the temperature of the
black hole is expressed solely in terms of the near-horizon coefficients $a_1$ and $b_1$;
however, the exact profile and horizon values of these coefficients do depend on the exact
content of the theory. 

Next, we may calculate the entropy of the black hole. One could do this by employing the
Euclidean approach in which the entropy is given by the relation \cite{GH}
%%%%%%%%%%%%%%%%%%%
\beq
S=\beta\left[\frac{\partial (\beta F)}{\partial \beta} -F\right],
\label{entropy-def}
\eeq
%%%%%%%%%%%%%%%%%%%%
where $F=I_E/\beta$ is the Helmholtz free-energy of the system given in terms of the
Euclidean version of the action $I_E$, and $\beta=1/(k_B T)$. However, as was discussed
in \cite{HP, Dutta, BAK}, this formula needs to be appropriately modified when the spacetime
asymptotic solution deviates from the asymptotically-flat limit. Equivalently, as was demonstrated
in \cite{Dutta}, one may employ an alternative method developed in \cite{Wald, Iyer} in which the
entropy of the black hole is identified with the Noether charge on the horizon under
diffeomorphisms. In this case, we may write 
%%%%%%%%%%%
\begin{equation}
S=-2\pi \oint{d^2x \sqrt{h_{(2)}}\left(\frac{\partial \mathcal{L}}{\partial R_{abcd}}\right)_\mathcal{H}\hat{\epsilon}_{ab}\,\hat{\epsilon}_{cd}}\,,
\label{entropy_AdS}
\end{equation}
%%%%%%%%%%%%%%
where $\mathcal{L}$ is the Lagrangian of the theory, $\hat{\epsilon}_{ab}$ the binormal to
the horizon surface $\mathcal{H}$, and $h_{(2)}$ the 2-dimensional projected metric on 
$\mathcal{H}$. 

Therefore, in order to calculate the entropy of the black holes emerging in the context of
the theory (\ref{action}), one needs to calculate the derivative of the Lagrangian with respect
to the Riemann tensor. In \cite{BAK}, we have presented a straightforward, pedagogical way
to do this for the theory where $V(\phi)=1$, and we have explicitly calculated the entropy of
the asymptotically Anti-de Sitter black-hole solutions to be
%%%%%%%%%%%%%%%
\beq
S=\frac{A_h}{4} +4 \pi f(\phi_h)\,.
\label{entropy}
\eeq
%%%%%%%%%%%%%%
The above result gives the entropy of a black-hole solution emerging in the context of the
theory with a general coupling function $f(\phi)$ between the scalar field and the GB term.
Its exact expression remains the same independently of whether a positive or negative 
cosmological constant $\Lambda$ is present in the theory, or whether a scalar-field potential
$V(\phi)$ is included in the theory since all these additive terms do not bear any dependence on
the Riemann tensor. As in the case of the temperature, their presence will only implicitly modify
the entropy of the solution through the values of the metric functions and scalar field at the
vicinity of the horizon. 

%%%%%%%%%%%%%%%%%%%%%%%%%%%%%%%%%%%%%%%%%%%%%%%%%%%%%%%%%%%%%%%%%%%%%%%%%%%

\section{Numerical Solutions}

We will now present the numerical results from the integration of the system of
field equations (\ref{A-sys})-(\ref{phi-sys}). The integration starts from the vicinity of the
black-hole horizon, i.e. at $r\approx r_h+\mathcal{O}(10^{-5})$ (for simplicity, we set
$r_h=1$) and proceeds outwards until an asymptotic solution is reached. The near-horizon
solutions (\ref{A-rh}) and (\ref{phi-rh}) for the metric function $A$ and scalar field $\phi$
are used as boundary conditions. The quantity $\phi'_h$, which is an input parameter of
the problem, is uniquely determined through Eq. (\ref{solphi'}) once the coupling function 
$f(\phi)$ and the scalar potential $V(\phi)$ are chosen. We could write the
coupling function as $f(\phi) =\alpha \tilde f(\phi)$, where $\alpha$ is a coupling constant
with units of (length)$^2$ and $\tilde f(\phi)$ is dimensionless. However, $\phi_h$ and
$\alpha$ are not independent since they both appear in the expression of $f(\phi)$ and
determine the strength of the scalar-GB coupling; thus, we may in fact fix $\phi_h$
and vary only $\alpha$. The values of $\phi_h$ and $\Lambda$ are also correlated
as they both appear in the expression of $C$ given in Eq. (\ref{C-def}) -- the first
one implicitly, through the functions $\dot f, V$ and $\dot V$. Since $C$ must always be
positive, when the value of the first is chosen, an allowed range of values exists for the
second.

Our numerical code has successfully reproduced the families of asymptotically-flat
scalarised black-hole solutions derived in \cite{ABK} as well as the asymptotically
AdS solutions found in \cite{BAK}. In the first part of the present analysis, we have
chosen a specific form for the coupling function, namely the exponential (dilatonic)
form $f(\phi)=\alpha e^{\phi}$, and supplemented it with different forms of the
scalar-field potential $V(\phi)$. Due to the interesting behaviour found, we paid
particular attention to the case of the quadratic potential, $V(\phi)=\phi^2$ with
$\Lambda<0$. The corresponding results are presented in the next two subsections.
Next, we considered a variety of choices for the coupling function $f(\phi)$ combined
always with a quadratic form for $V(\phi)$. This case is presented in the last subsection. 

%%%%%%%%%%%%%%%%%%%%%%%%%%%%%%%%%%%%%%%%%%%%%%%%%%%%%%%%%%%%%%%% 

\subsection{Exponential Coupling Function with Different Potentials}

%%%%%%%%%%%%%%%%%%%%%
\begin{figure}[t]
\begin{center}
\mbox{\hspace*{-0.1cm} \includegraphics[width = 0.50 \textwidth] {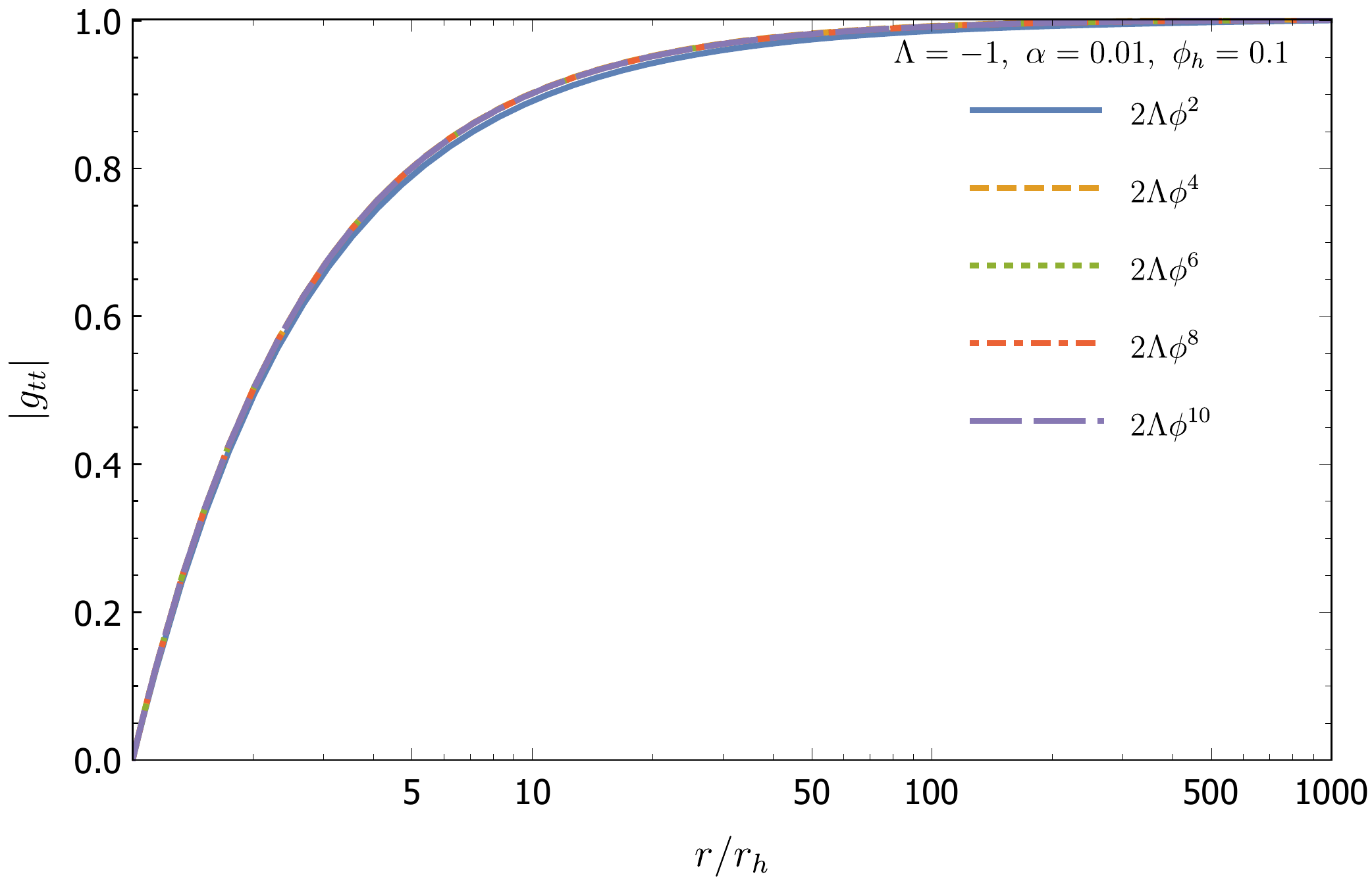}}
\hspace*{-0.2cm} {\includegraphics[width = 0.485 \textwidth]
{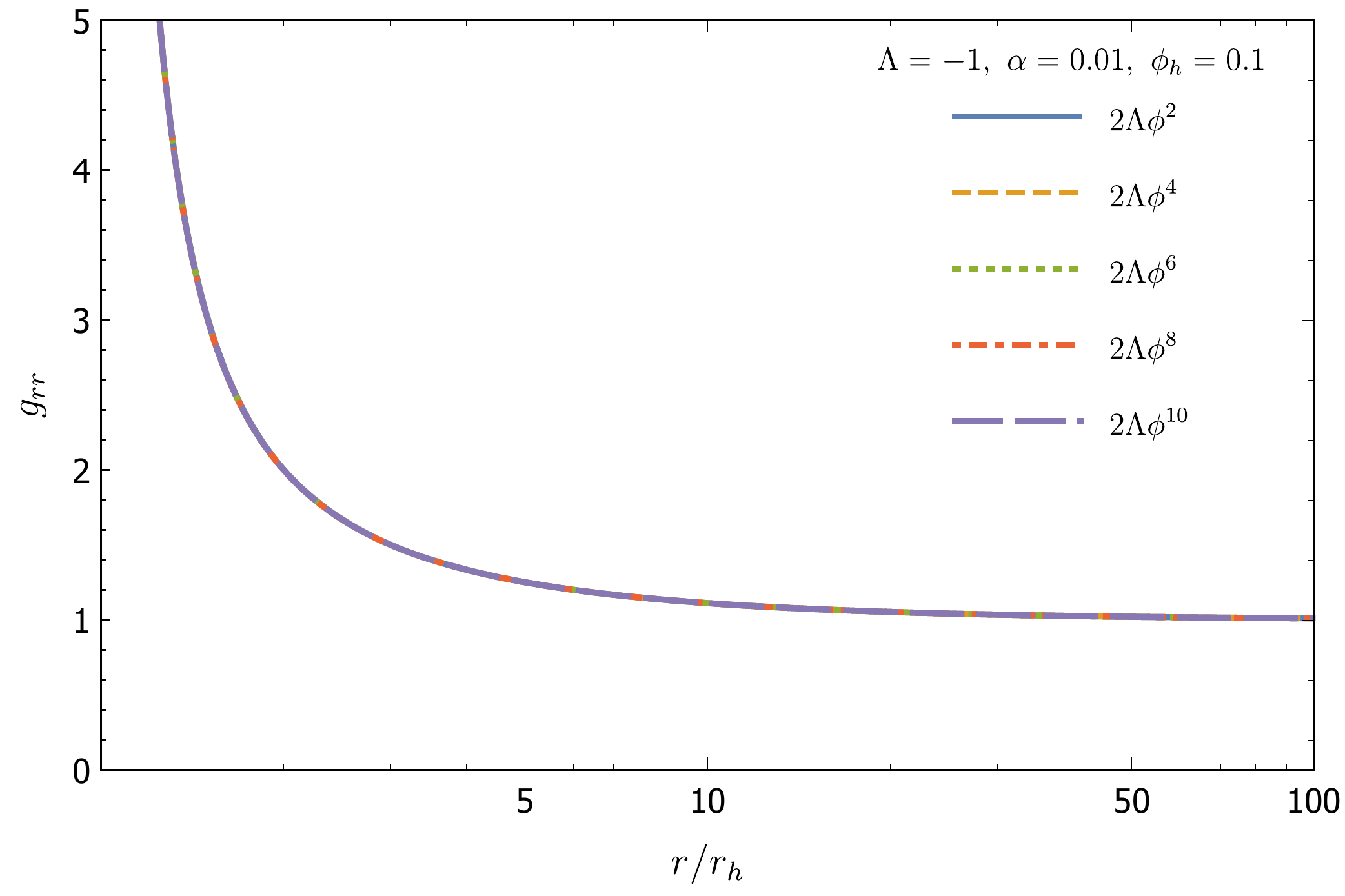}}
    \caption{The metric components $|g_{tt}|$ (left plot) and $g_{rr}$ (right plot) in terms of
the radial coordinate $r$, for $f(\phi)=\alpha e^{\phi}$ and a variety of potentials $V(\phi)$
(with $\Lambda=-1$).}
   \label{Metric_exp}
  \end{center}
\end{figure}
%%%%%%%%%%%%%%%%%

We will first discuss the case of an exponential coupling function, $f(\phi)=\alpha e^{\phi}$. The
solutions for the metric functions $e^{A(r)}$ and $e^{B(r)}$ are depicted in the left and right
plot of Fig. \ref{Metric_exp}, respectively, for a variety of forms of the scalar-field potential
$V(\phi)$. We observe the anticipated behaviour near the black-hole horizon, with $e^{A(r)}$
vanishing and $e^{B(r)}$ diverging. At large distances, the gravitational background approaches
the Minkowski spacetime. Therefore, the asymptotic forms of the metric functions read
%%%%%%%%%%%%%%%%
\beq
A(r) \simeq -\frac{2M}{r} + {\cal O}\left(\frac{1}{r^2}\right)\,, \qquad
B(r) \simeq \frac{2M}{r} + {\cal O}\left(\frac{1}{r^2}\right)\,,
\label{metric_asym}
\eeq
%%%%%%%%%%%%%%%
from which we may easily obtain the mass $M$ of the black hole. The solutions presented
correspond to the particular values $\Lambda=-1$ (in units of $r_h^{-2}$), $\alpha=0.01$
(in units of $r_h^{2}$) and $\phi_h=0.1$, however, the observed behaviour is typical for a
large range of values we have used. We note that the behaviour of the metric functions both
at small and large distances from the black-hole horizon is in fact independent of the form
of the scalar-field potential $V(\phi)$, as Fig. \ref{Metric_exp} clearly depicts \footnote{Note,
that we present only even functions of the scalar field $\phi$ for the potential $V(\phi)$. It is
only for these choices that we obtain asymptotically-flat black-hole solutions. For odd functions
of $\phi$, the asymptotic behaviour of the obtained solutions resembles the one for a de Sitter
spacetime. As these solutions seem to comprise a physically different family of black-hole solutions,
we postpone their detailed study for a subsequent work.}. We would like to stress that the 
asymptotically-flat behaviour of the $g_{rr}$ metric component at radial infinity follows
naturally by solving the field equations from the horizon outwards and without imposing any
condition at large distances. The $g_{tt}$ component also approaches naturally a finite,
constant value, and the boundary condition $g_{tt} \rightarrow -1$, as $r \rightarrow \infty$,
is imposed in order to determine the parameter $a_1$ which otherwise remains arbitrary.

\begin{figure}[t]
\begin{center}
\mbox{\hspace*{-0.1cm} \includegraphics[width = 0.47 \textwidth] {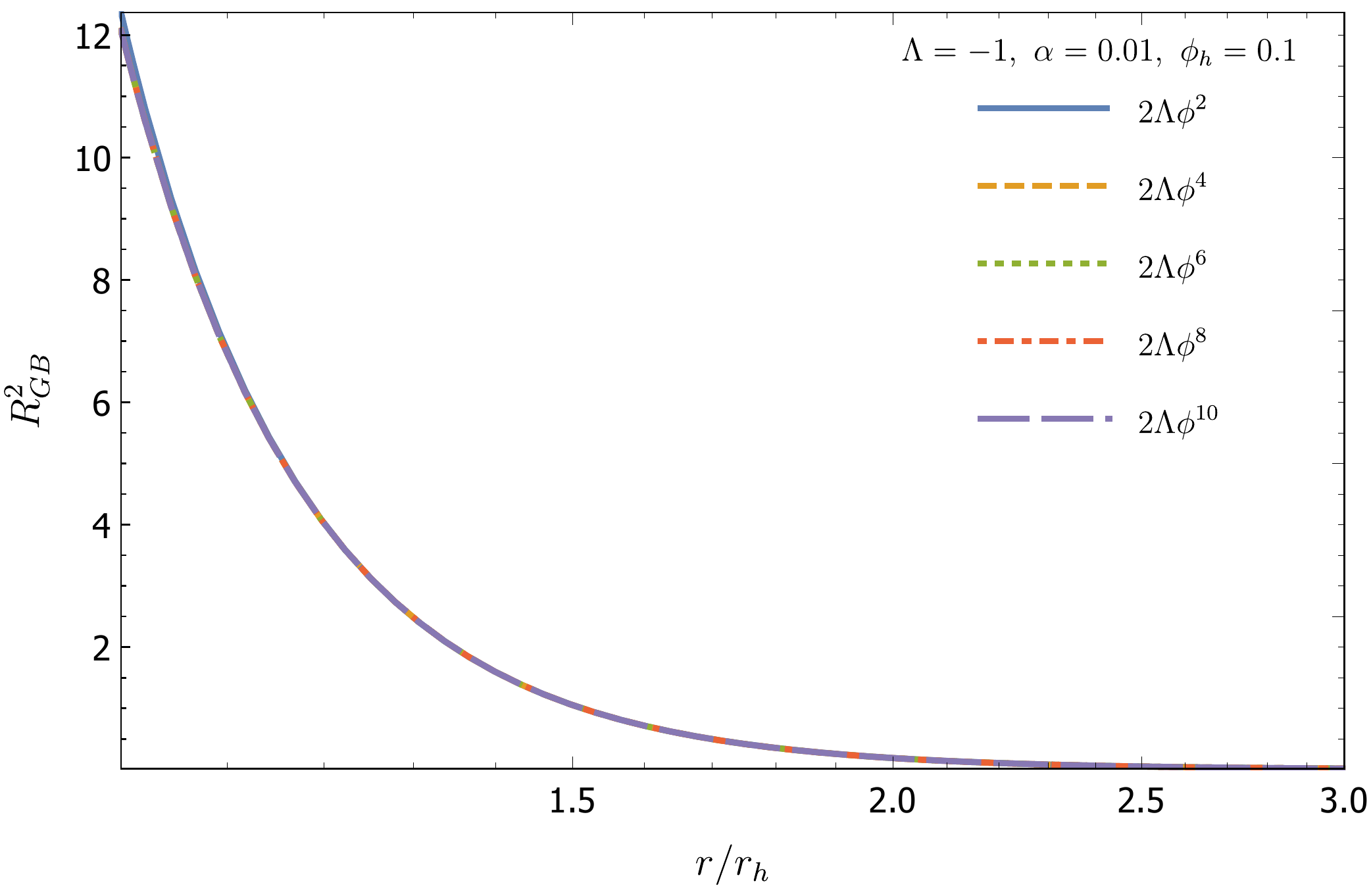}}
\hspace*{-0.4cm} {\includegraphics[width = 0.50 \textwidth]
{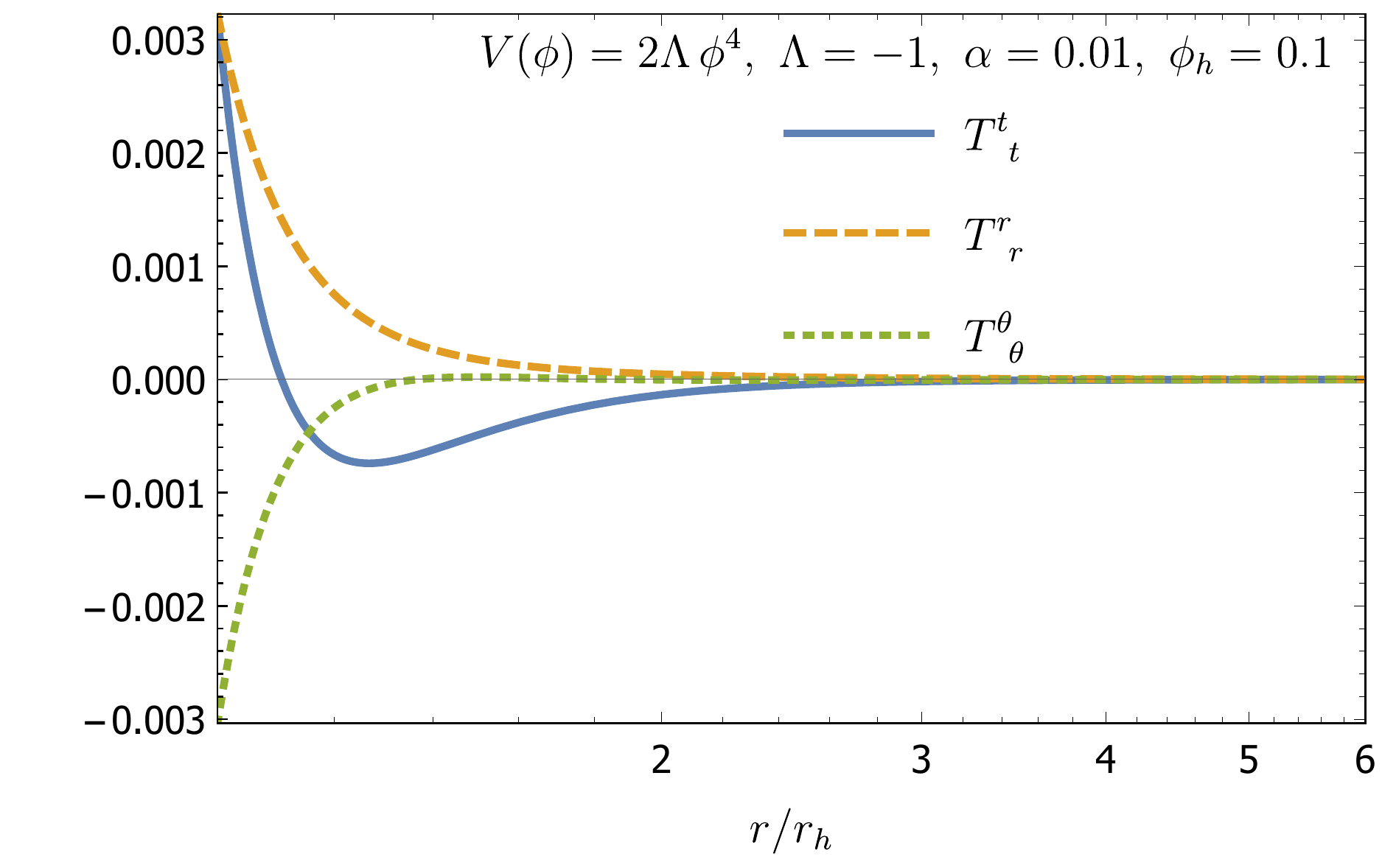}}
    \caption{The Gauss-Bonnet term $R^2_{GB}$ for a variety of potentials $V(\phi)$ (left plot) \
and the energy-momentum tensor $T^\mu_{\;\,\nu}$ (right plot) for the indicative case of the quartic
potential, $V(\phi)=\phi^4$, in terms of the radial coordinate $r$, for $f(\phi)=\alpha e^{\phi}$ and
$\Lambda=-1$.}
   \label{GB_Tmn}
  \end{center}
\end{figure}

The spacetime remains everywhere regular as this may be seen from the profile of the
curvature-invariant GB term depicted in the left plot of Fig. \ref{GB_Tmn}. As expected, the GB
term assumes its maximum value near the black-hole horizon, where the curvature is
maximum, and quickly decreases eventually vanishing far way from the horizon. Once again,
the exact form of the scalar-field potential $V(\phi)$ does not affect the profile of the
gravitational GB term. The regularity of the spacetime is also reflected in the finiteness of
all components of the energy-momentum tensor presented on the right plot of Fig. \ref{GB_Tmn}.
Here, the indicative case of the quartic potential, $V(\phi)=\phi^4$, is presented, however,
the observed profile remains unaffected as the form of $V(\phi)$ varies. We may clearly see
the characteristic behaviour caused by the presence of the GB in the theory: the radial
pressure $p=T^r_{\;\,r}$ is positive near the horizon while the energy density $\rho =-T^t_{\;\,t}$
is negative -- both these features cause the evasion of the scalar no-hair theorems
\cite{DBH, ABK} and the emergence of regular black-hole solutions with scalar hair. 
The depicted profiles of $T^\mu_{\;\,\nu}$ also reveal the asymptotic flatness of spacetime
as all components assume a zero value away from the black-hole horizon.

%%%%%%%%%%%%%%%%%%%%%
%%%%%%%%%%%%%%%%%

%%%%%%%%%%%%%%%%%%%%%
\begin{figure}[b!]
\begin{center}
\mbox{\hspace*{-0.1cm} \includegraphics[width = 0.49 \textwidth] {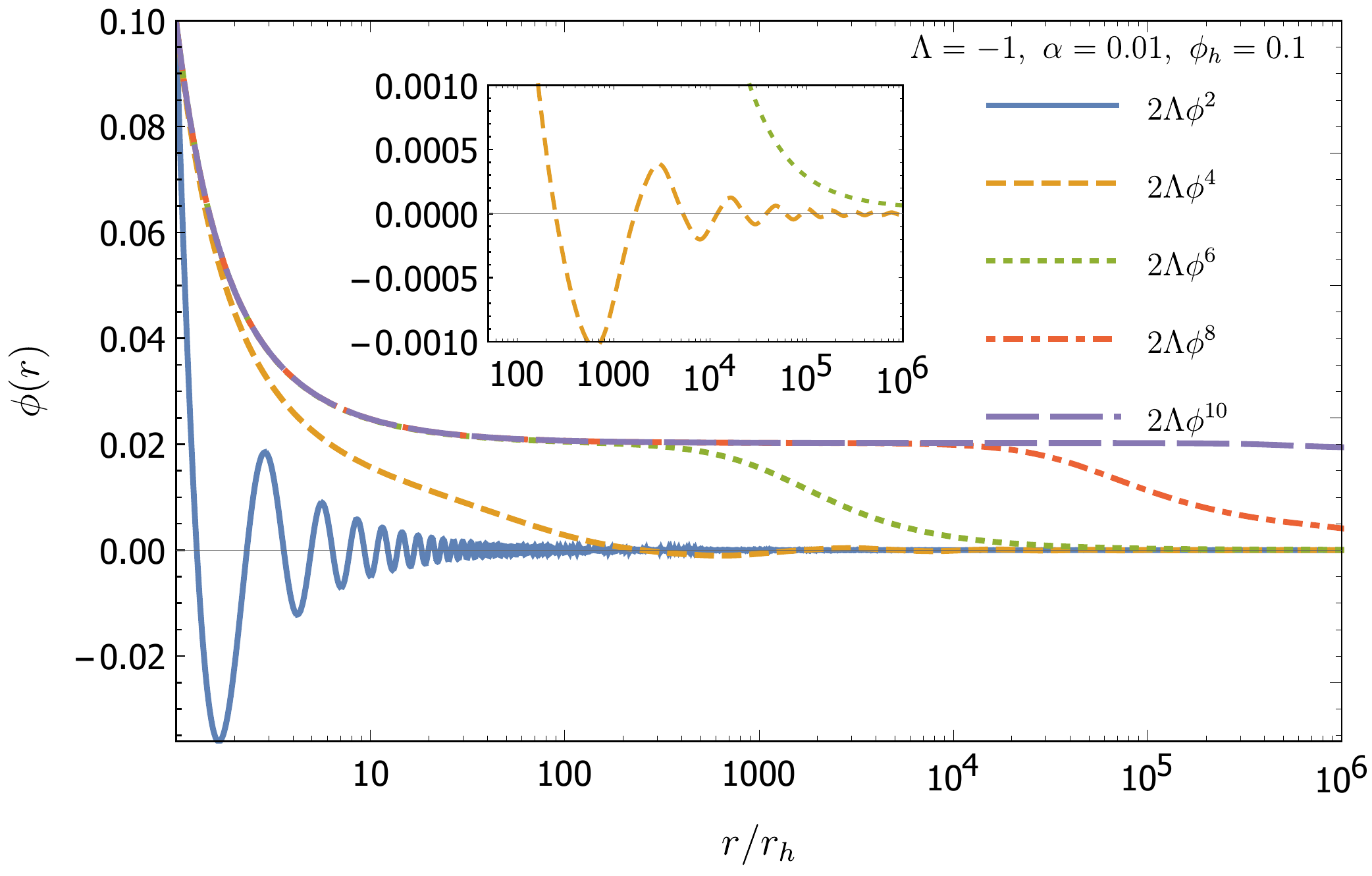}}
\hspace*{-0.2cm} {\includegraphics[width = 0.49 \textwidth]
{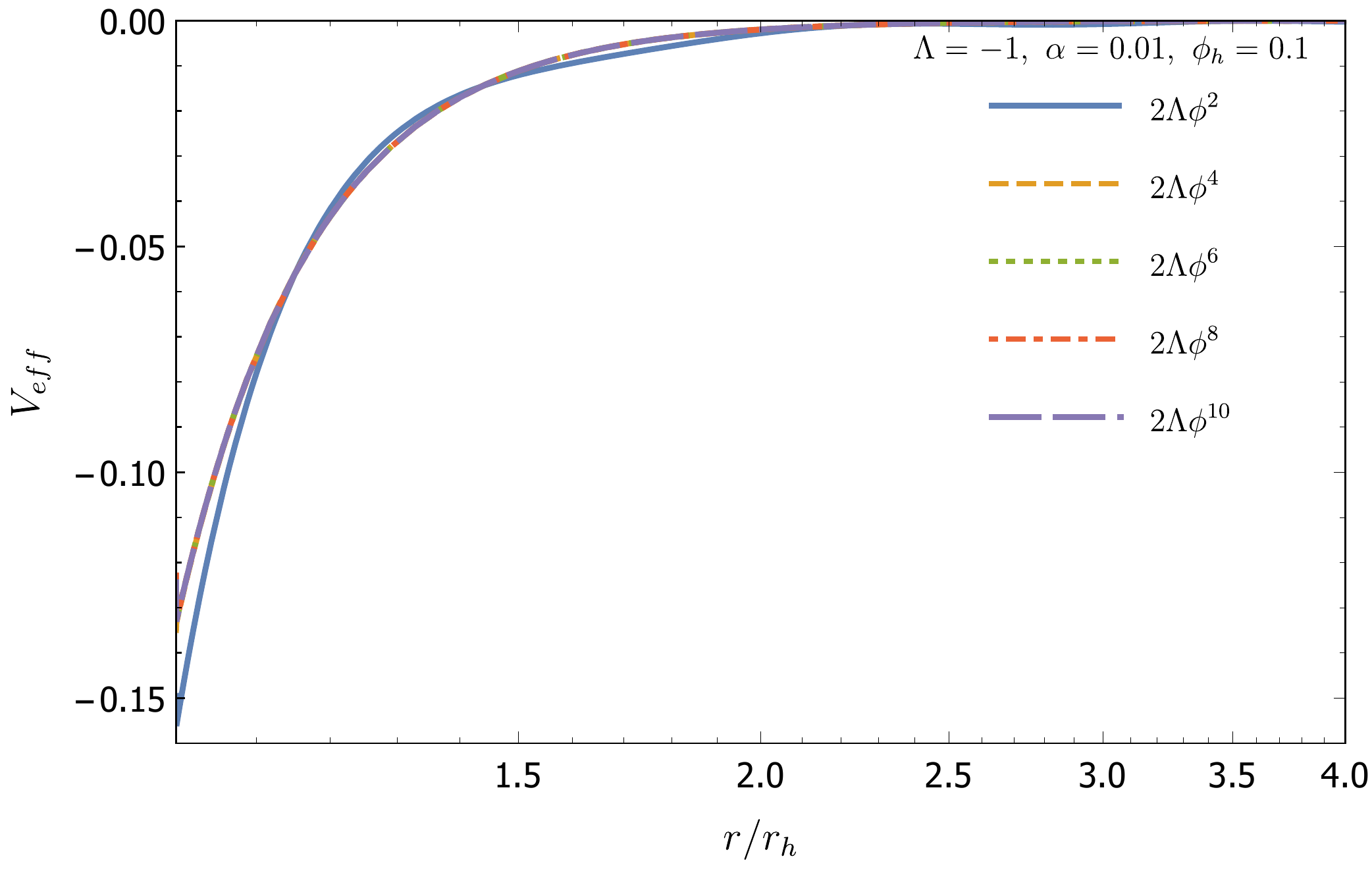}}
\caption{The scalar field (left plot) and its effective potential $V_{eff}$ (right plot) in terms
of the radial coordinate $r$, for $f(\phi)=\alpha e^{\phi}$ and various forms of $V(\phi)$
(with $\Lambda=-1$).}
\label{phi_exp_pot}
  \end{center}
\end{figure}
%%%%%%%%%%%%%%%%%

Let us now turn to the profile of the scalar field which supports the aforementioned behaviour
of the energy-momentum tensor. This is depicted in the left plot of Fig. \ref{phi_exp_pot},
for a variety of forms of its potential $V(\phi)$. As demonstrated in Section 3, the scalar
field is regular near the black-hole horizon independently of the form of its potential.
However, the subsequent evolution does strongly depend on the particular form of
$V(\phi)$. We observe that, despite the negative sign of $\Lambda$, the profile of the
scalar field remains finite in the entire radial regime. For all of the forms of $V(\phi)$
employed, the scalar field decreases at the near-horizon regime and reduces to a 
vanishing value at asymptotic infinity. Once again, no boundary condition is imposed
at large distances on the scalar field, which naturally approaches a zero value. This is
supported also by the behaviour of the effective potential of the scalar field defined as
%%%%%%%%%%%%%
\beq
V_{eff}= -f(\phi)\,R^2_{GB} + 2\Lambda\,V(\phi)\,,
\label{Veff}
\eeq
%%%%%%%%%%%%%
the value of which is presented in the right plot of Fig. \ref{phi_exp_pot}. For all
forms of $V(\phi)$, the effective potential adopts its maximum value near the black hole
horizon, where it supports a non-trivial scalar field, while it decays to zero at asymptotic
infinity leading to a trivial scalar field there. The asymptotic behaviour of the scalar
field is difficult to obtain in an analytic way; employing the asymptotic forms
(\ref{metric_asym}) of the metric functions, the scalar-field equation (\ref{phi-eq})
takes the form
%%%%%%%%%%%%%%%
\beq 
r \phi'' + 2 \phi'-2 r \Lambda \frac{dV}{d\phi}=0\,.
\label{phi-eq-asym}
\eeq
%%%%%%%%%%%%%%%%%%
For $V(\phi)=0$, the above equation leads to the solution (\ref{phi-flat}), which is indeed the
asymptotic behaviour of a scalar field possessing only a coupling to the GB term and no
potential \cite{ABK}. For $V(\phi)=\phi^2$ and $\Lambda=m^2/2$, the solution is given by
Eq. (\ref{phi-massive}) describing the asymptotic form of a massive scalar field
\cite{Doneva-massive}. For the case with $V(\phi)=\phi^2$ and $\Lambda=-m^2/2$ that
we consider here, the solution of Eq. (\ref{phi-eq-asym}) takes the form 
%%%%%%%%%%%%%%%%
\beq
\phi(r) \simeq \frac{1}{r}\,[C_1 \cos (m r) + C_2\,\sin (mr)]\,.
\label{phi-neg-mass}
\eeq
%%%%%%%%%%%%%
The above solution justifies the oscillatory behaviour clearly observed in the lower curve of the
left plot of Fig. \ref{phi_exp_pot}, which corresponds to this case. Unfortunately, Eq. (\ref{phi-eq-asym})
cannot be analytically solved for any of the other forms of the scalar potential $V(\phi)$ that
we study in this work. The similarity in the profile of the effective potential but also the observed
behaviour of the scalar-field curves, depicted in Fig. \ref{phi_exp_pot} for $V(\phi)=\phi^{2n}$,
imply that the scalar field decreases asymptotically to zero in an oscillatory way also
for the cases with $n>1$; however, as $n$ increases, the frequency of the oscillations is strongly
damped and the asymptotic vanishing value is reached at an increasingly larger distance. 

%%%%%%%%%%%%%%%%%%%%%
\begin{figure}[t]
\begin{center}
\mbox{\hspace*{-0.0cm} \includegraphics[width = 0.48 \textwidth] {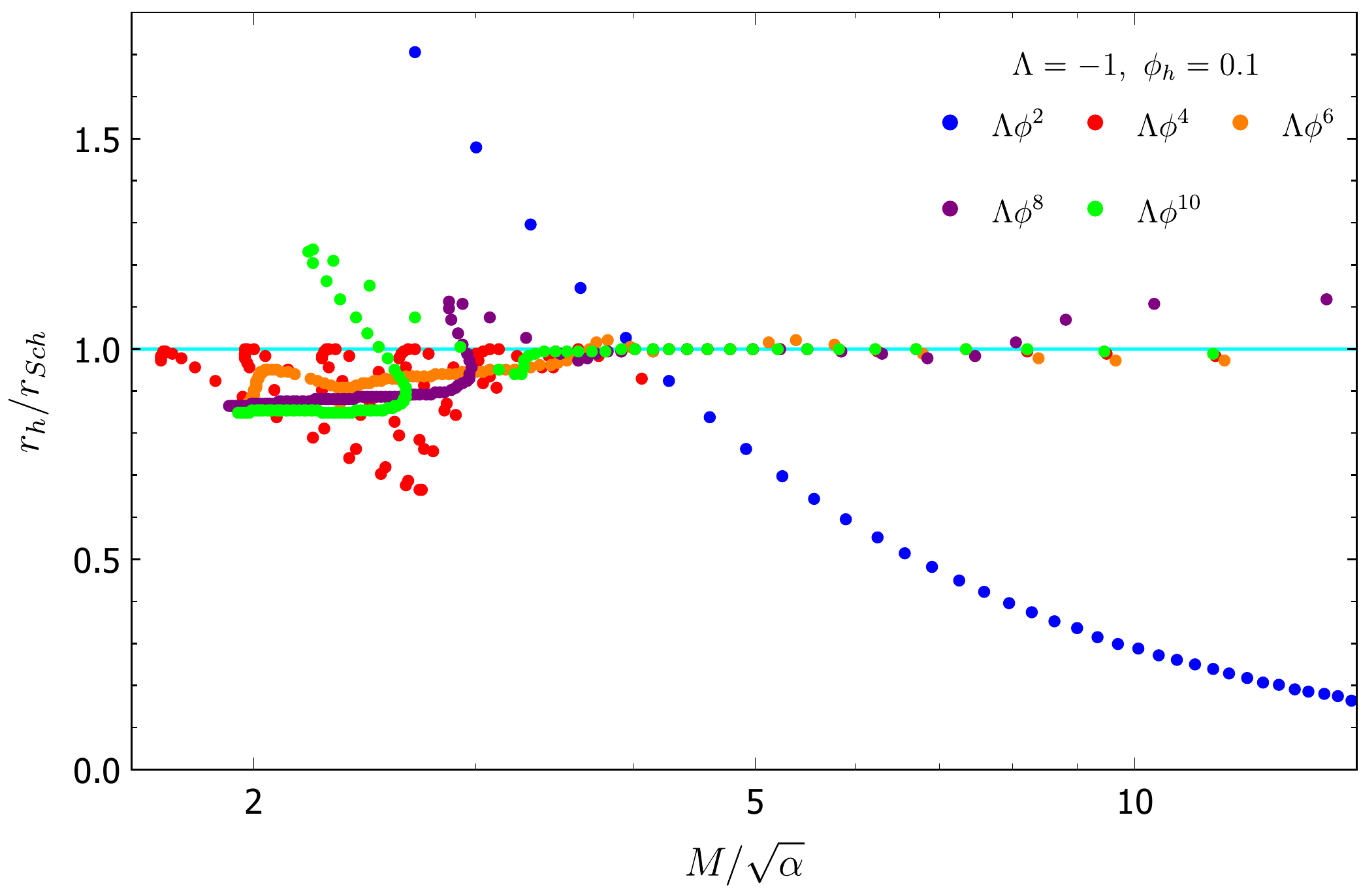}}
\hspace*{-0.0cm} {\includegraphics[width = 0.48 \textwidth]
{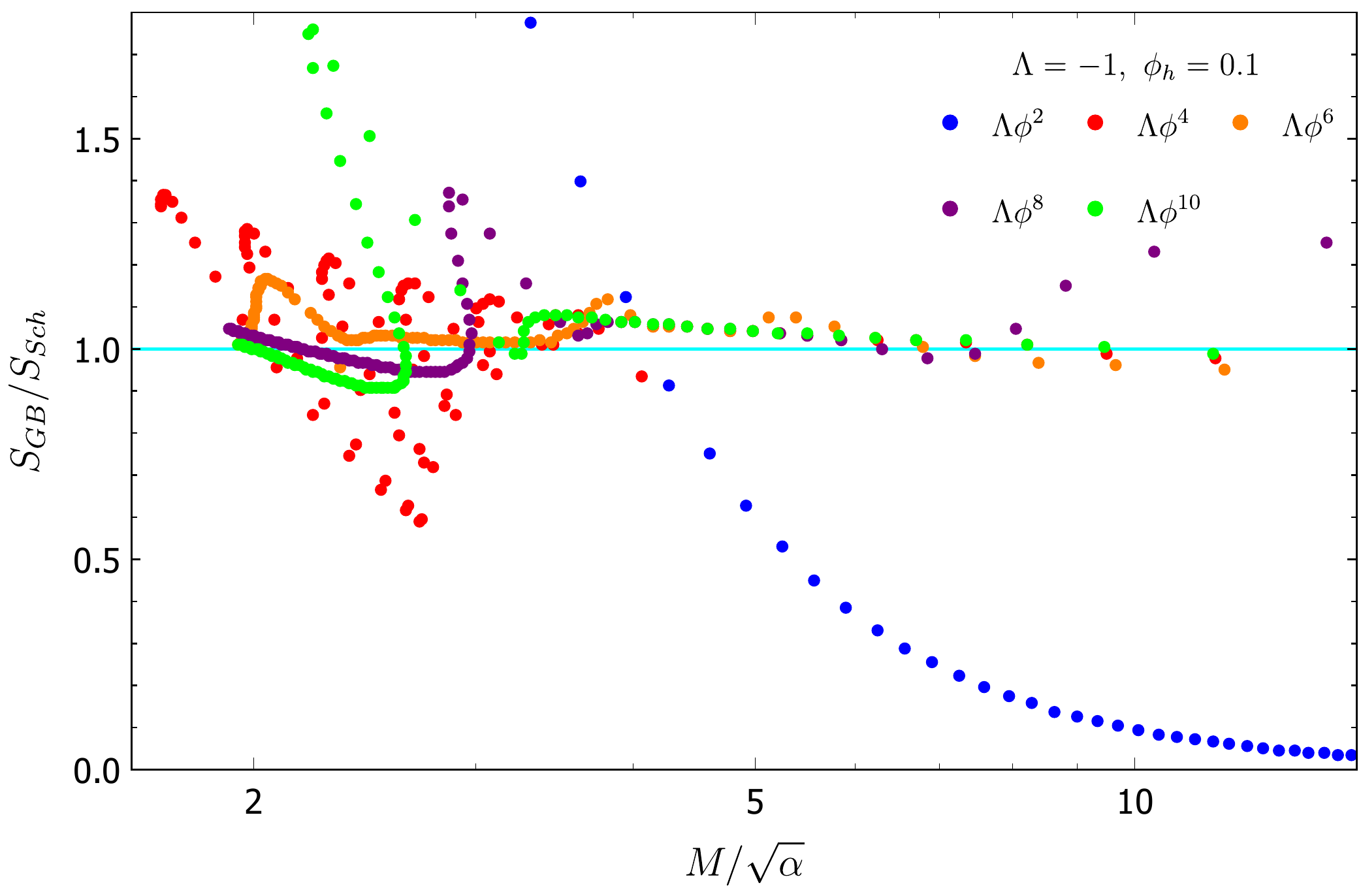}}
    \caption{The horizon radius $r_h$ (left plot) and entropy $S_{GB}$ (right plot) normalised
with respect to the corresponding Schwarzschild values, for $f(\phi)=\alpha e^{\phi}$ and
different forms of $V(\phi)$ (with $\Lambda<0$).}
   \label{r-S}
  \end{center}
\end{figure}
%%%%%%%%%%%%%%%%%

Although this may be true, there seem to be some fundamental differences between the
GB black-hole solutions where the scalar field possesses a quadratic potential and the solutions
where the scalar field has a higher-power potential.
To see this, in Fig. \ref{r-S}, we plot the horizon radius and the entropy of the corresponding
black-hole solutions in terms of their mass $M$. These quantities are normalised to the
corresponding Schwarzschild values, $r_{Sch}=2M$ and $S_{Sch}=4\pi M^2$, for easy
comparison. All solutions for $V(\phi)=\phi^{2n}$ with $n>1$ exhibit the existence of
multiple branches with their horizon radius and entropy
being smaller or larger than the corresponding Schwarzschild values depending on the 
mass range and value of the power $n$. On the other hand, the quadratic case allows only
one branch, which crosses the horizontal line, marking the equality of $r_h$ and $S_{GB}$ with
the Schwarzschild values, at only one point: black holes with masses below that point have
a horizon radius larger than the corresponding Schwarzschild solution and are also
more thermodynamically stable. For all forms of the scalar potential, our
numerical analysis has revealed the existence of an upper bound on the coupling parameter
$\alpha$, which then translates to a lower bound for the black-hole mass $M$, a feature
common for all GB black holes \cite{DBH, ABK, BAK}.

Another distinctive feature of the case with a negative quadratic potential is the behaviour 
of the black-hole solutions in the limit of large mass, a feature that distinguishes this case
both from the ordinary massive case and from the case with a higher-power potential. 
From Fig. \ref{r-S}, we observe that, as the mass of the black-hole solution increases,
the horizon radius and entropy for the case with $V(\phi)=\phi^2$ decrease reaching
eventually a very small value. Thus, in the limit of large mass, a branch of massive, 
ultra-compact black holes seems to emerge. In contrast, for all other forms of the scalar-field
potential, this branch of solutions is absent since the horizon radius and entropy of the
corresponding solutions approach, in the limit of large mass, values which are very close
to the Schwarzschild ones. In the ordinary massive case \cite{Doneva-massive}, the 
GB black holes also approach the Schwarzschild solution in the large-mass limit.
We consider the emergence of this ultra-compact family of black holes as a rather
interesting feature of the theory, therefore, in the next subsection we study in greater
detail the case with a negative quadratic potential.

%%%%%%%%%%%%%%%%%%%%%%%%%%%%%%%%%%%%%%%%%%%%%%%%%%%%%%%%%%%%%

\subsection{Exponential Coupling Function with a Quadratic Potential}

%%%%%%%%%%%%%%%%%%%%%
\begin{figure}[t]
\begin{center}
\mbox{\hspace*{-0.0cm} \includegraphics[width = 0.48 \textwidth] {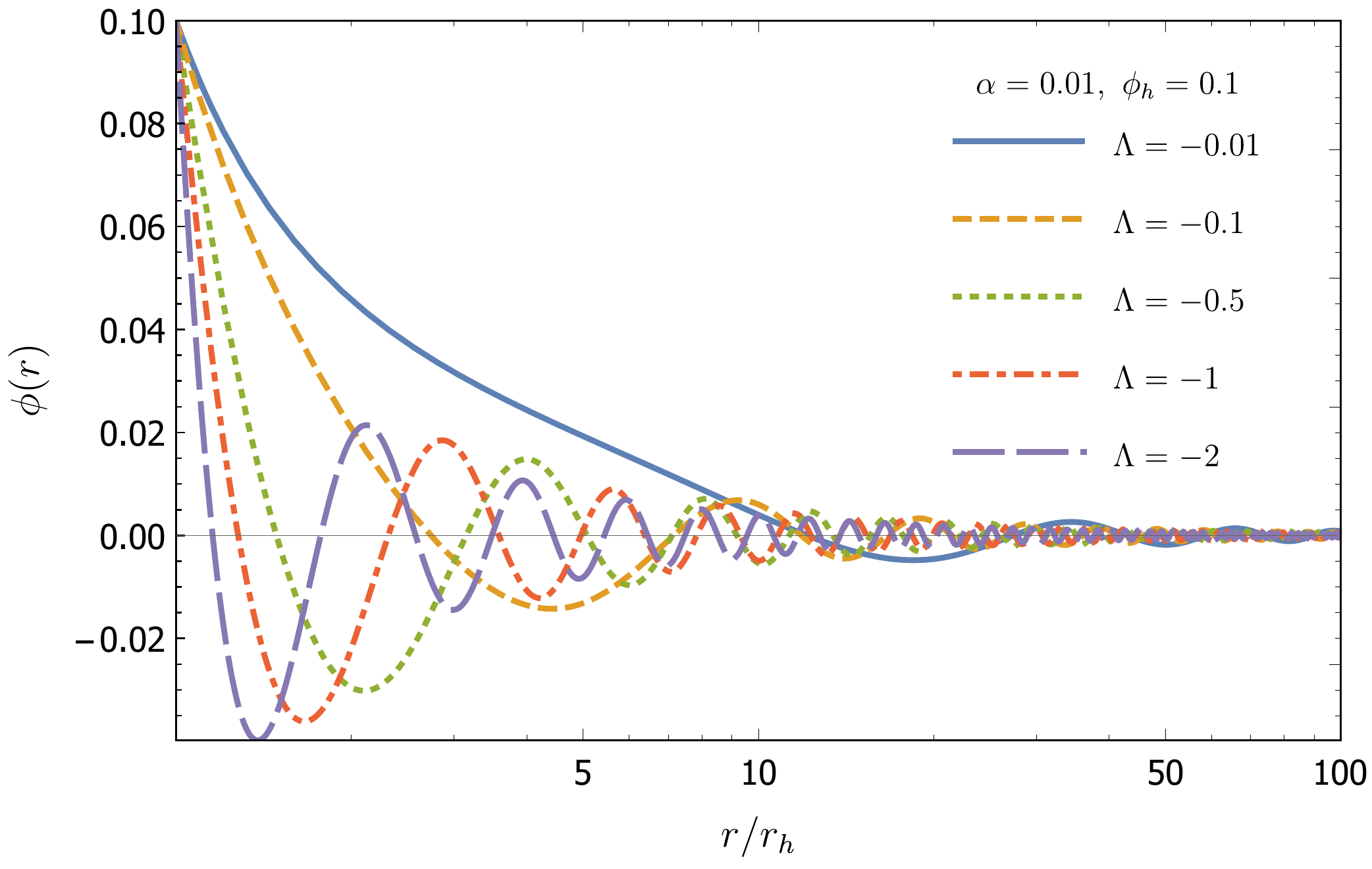}}
\hspace*{-0.0cm} {\includegraphics[width = 0.48 \textwidth]
{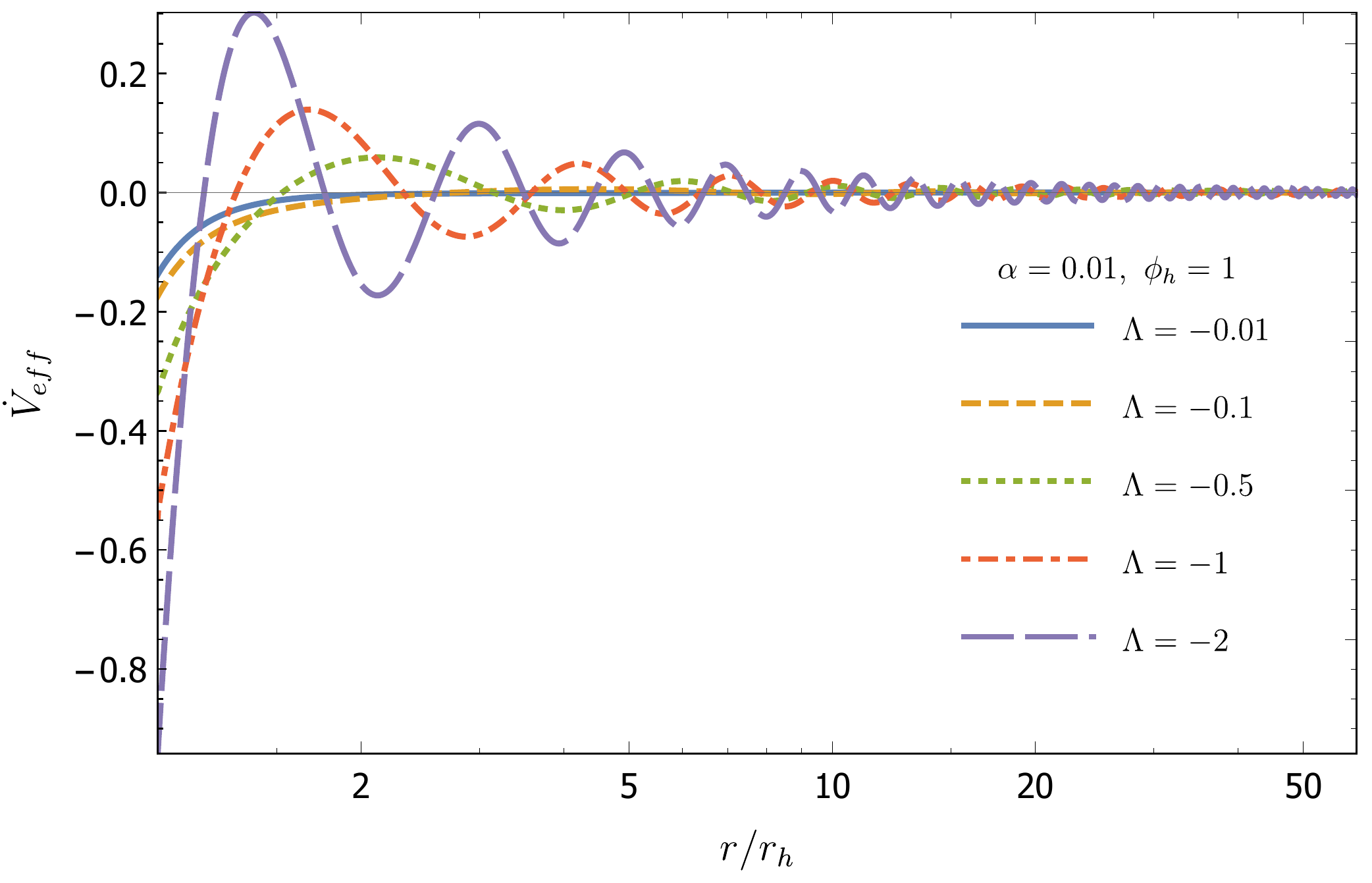}}
    \caption{The scalar field $\phi$ (left plot) and its source term $\dot V_{eff}$ (right plot)
in terms of the radial coordinate $r$, for $f(\phi)=\alpha e^{\phi}$ and $V(\phi)=\phi^2$
(with $\Lambda<0$).}
   \label{phi_quad}
  \end{center}
\end{figure}
%%%%%%%%%%%%%%%%%

If we focus on the EsGB theory with an exponential coupling function and a negative
quadratic potential, the decisive parameter of the theory is the coupling parameter
$\Lambda$. The form of the metric functions $A(r)$ and $B(r)$ are not significantly
affected as $\Lambda$ varies, and they are still accurately described by the profiles
depicted in Fig. \ref{Metric_exp}. However, $\Lambda$ strongly affects the solution
for the scalar field: as may be clearly seen from the left plot of Fig. \ref{phi_quad},
the parameter $\Lambda$ affects both the rate of decrease of the scalar field near
the black-hole horizon and the frequency of the oscillatory phase as the asymptotic
solution at infinity is approached. These dependences are also supported by Eq. (\ref{solphi'})
and Eq. (\ref{phi-neg-mass}), where $m=\sqrt{2|\Lambda|}$, respectively. We note
that both of these quantities increase with $\Lambda$. The same holds for the 
source term $\dot V_{eff}$, which appears in the scalar-field equation (\ref{phi-eq_0})
and determines the profile of the scalar field; this is depicted in the right plot of
Fig. \ref{phi_quad} for various values of $\Lambda$.  

\begin{figure}[b!]
\begin{center}
\mbox{\hspace*{-0.0cm} \includegraphics[width = 0.48 \textwidth] {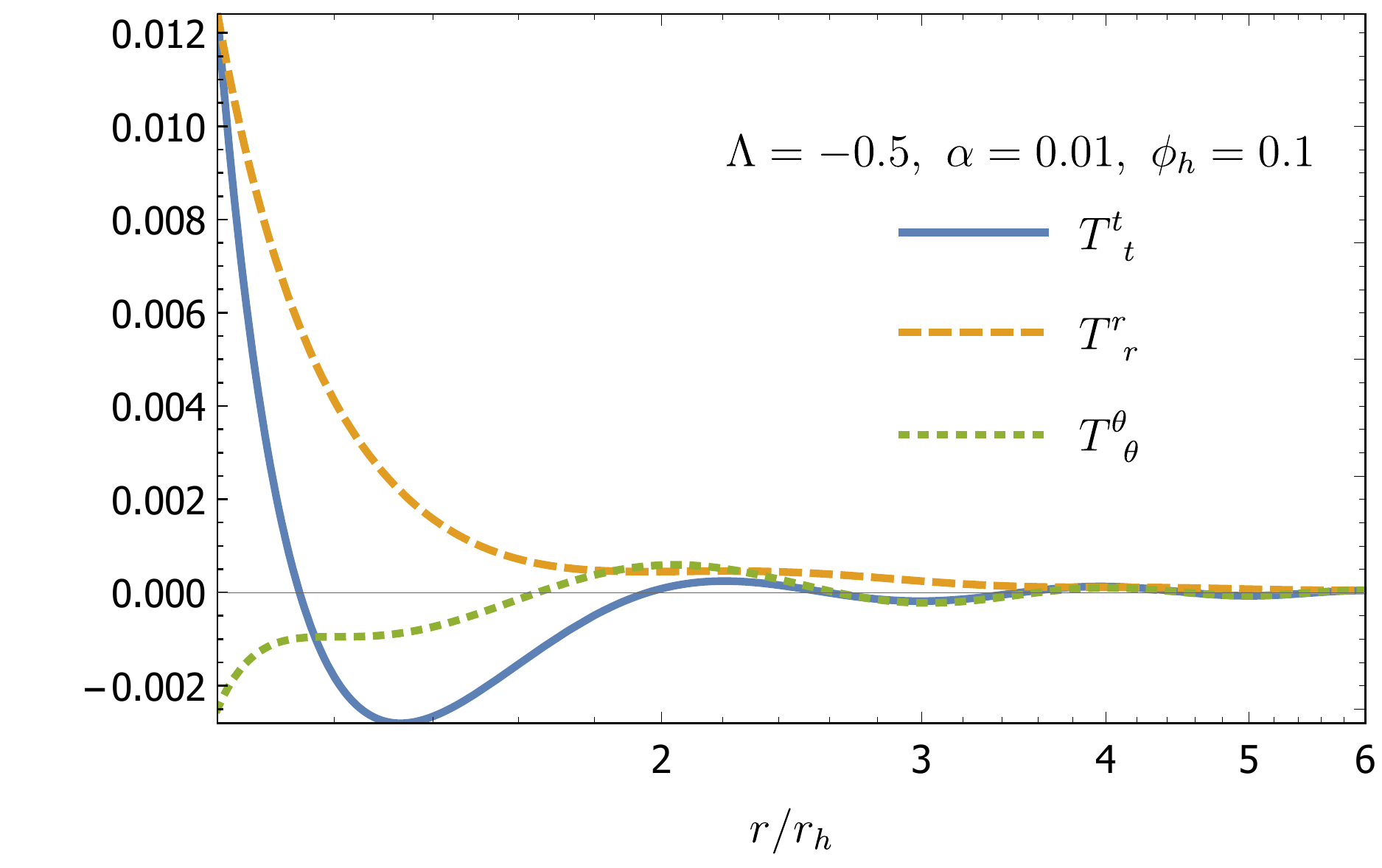}}
\hspace*{-0.0cm} {\includegraphics[width = 0.48 \textwidth]
{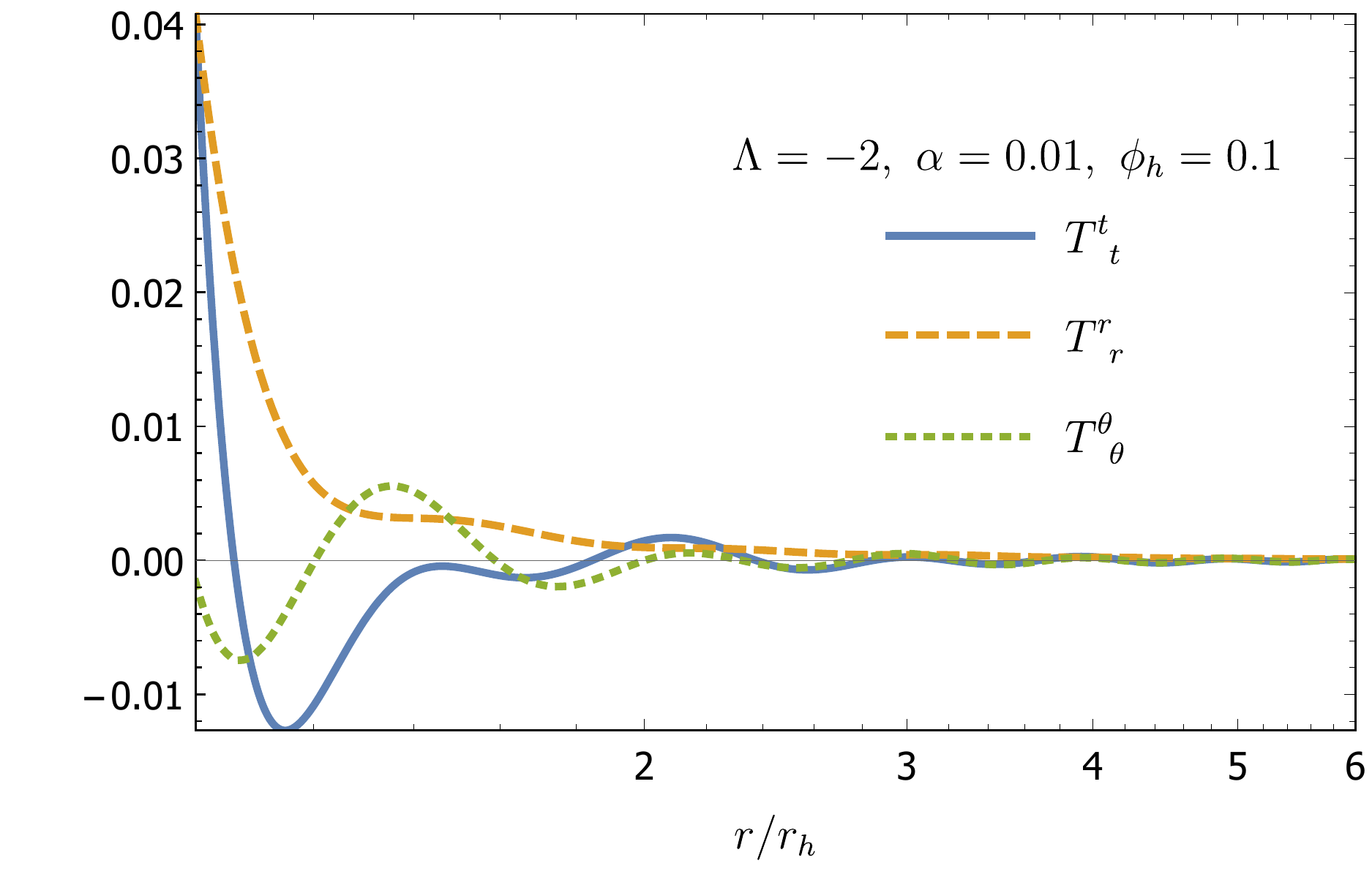}}
    \caption{The energy-momentum tensor $T^\mu_{\;\,\nu}$ components for
$f(\phi)=\alpha e^{\phi}$ and $V(\phi)=\phi^2$, and two different values of $\Lambda$,
namely $\Lambda=-0.5$ (left plot) and $\Lambda=-2$ (right plot).}
   \label{Tmn_quad}
  \end{center}
\end{figure}

Finally, as expected, the same oscillatory behaviour is observed also in the components
of the energy-momentum tensor, and becomes more prominent as $\Lambda$ increases.
In Fig. \ref{Tmn_quad}, we display all three $T^\mu_{\;\,\nu}$ components for
$\Lambda=-0.5$ (left plot) and $\Lambda=-2$ (right plot) with all the other
parameters fixed. The overall behaviour noted in Fig. \ref{GB_Tmn} is observed
also here, with all components remaining finite and approaching zero values at infinity.
In addition, we note that, as $\Lambda$ increases, an oscillatory phase appears in
all components and is preserved until the asymptotic regime at infinity is reached.

%%%%%%%%%%%%%%%%%%%%%
%%%%%%%%%%%%%%%%%

\begin{figure}[t]
\begin{center}
\mbox{\hspace*{-0.0cm} \includegraphics[width = 0.48 \textwidth] {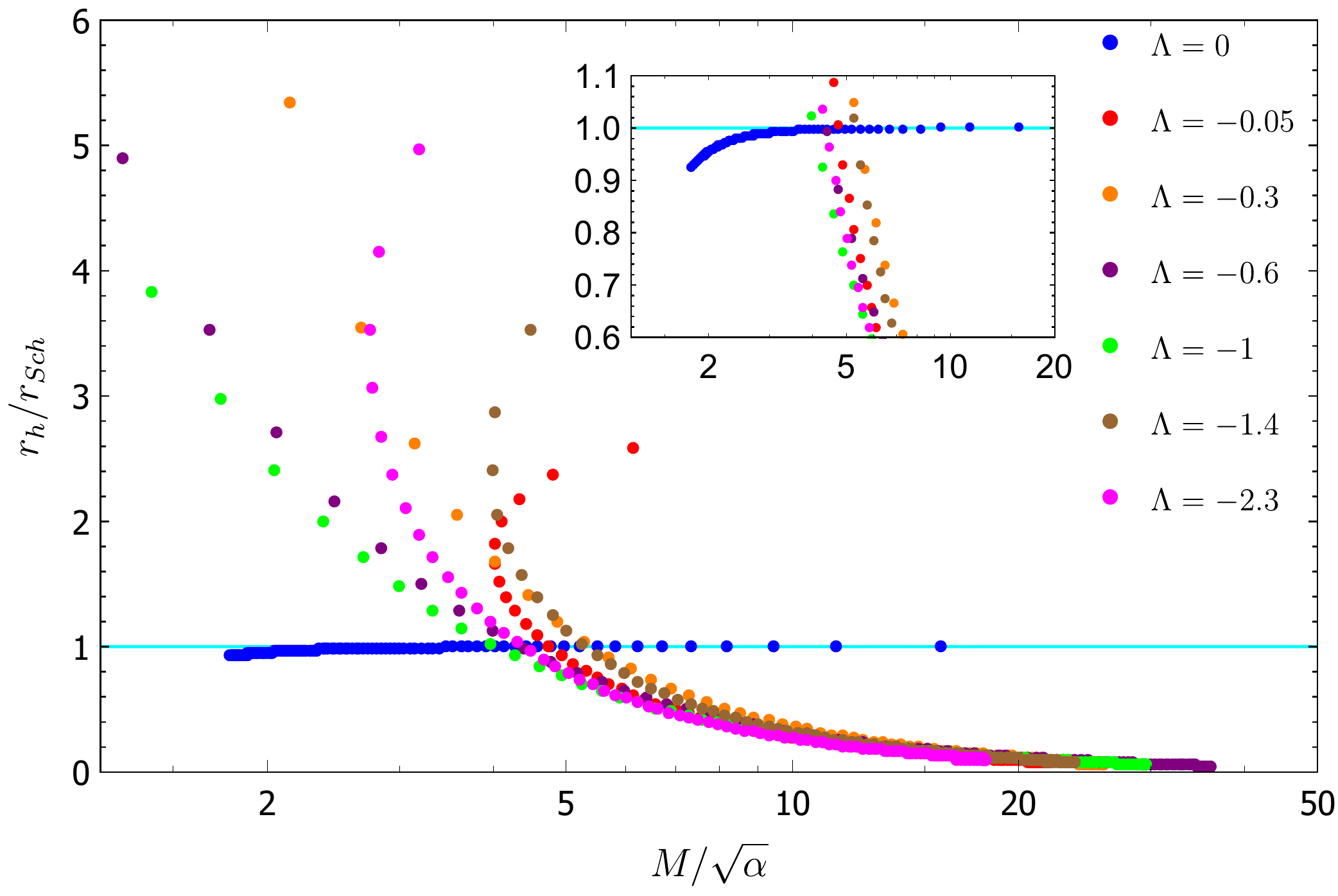}}
\hspace*{-0.0cm} {\includegraphics[width = 0.48 \textwidth]
{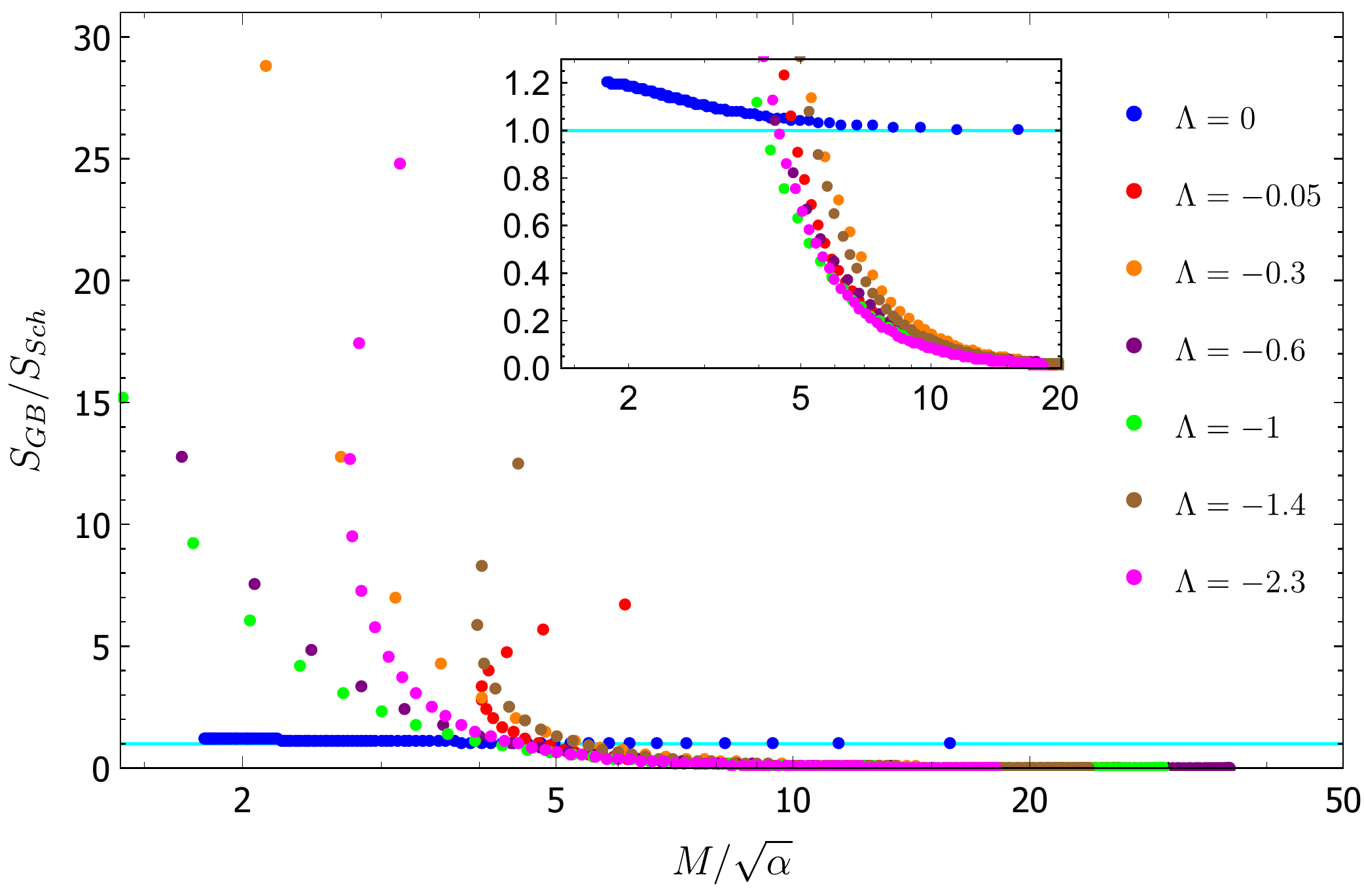}}
    \caption{The horizon radius $r_h$ (left plot) and entropy $S_{GB}$ (right plot) normalised
with respect to the corresponding Schwarzschild values, for $f(\phi)=\alpha e^{\phi}$, 
$V(\phi)=\phi^2$ and various values of $\Lambda<0$. The blue dots correspond to the
dilatonic black holes with no potential ($\Lambda=0$).}
   \label{rh-S-quad}
  \end{center}
\end{figure}

In Fig. \ref{rh-S-quad}, we plot again the horizon radius $r_h$ (left plot) and entropy $S_{GB}$
(right plot) normalised with respect to the corresponding Schwarzschild values, for solutions
obtained for $f(\phi)=\alpha e^{\phi}$, $V(\phi)=\phi^2$ and various negative values of the parameter
$\Lambda$. The case with $\Lambda=0$ corresponding to the asymptotically-flat dilatonic
GB black holes with no potential \cite{DBH, ABK} is also shown for comparison. From the left
plot of Fig. \ref{rh-S-quad}, we observe that whereas the dilatonic black holes are always
smaller than the corresponding
Schwarzschild black holes, the black-hole solutions with a negative quadratic potential may
be either smaller, equal or larger than the Schwarzschild solution with respect to their
horizon values. The same holds for the entropy of these solutions, shown in the right
plot, with the ones being larger in size than the Schwarzschild radius being also more
thermodynamically stable. The new
feature emerging from these plots is that, also for the negative quadratic potential, branches
of solutions with the same mass $M$ but with different horizon radii and entropies appear
for certain regimes of the $\Lambda$ parameter.

%%%%%%%%%%%%%%%%%%%%%
%%%%%%%%%%%%%%%%%

\begin{figure}[t]
\begin{center}
\includegraphics[width = 0.49 \textwidth] {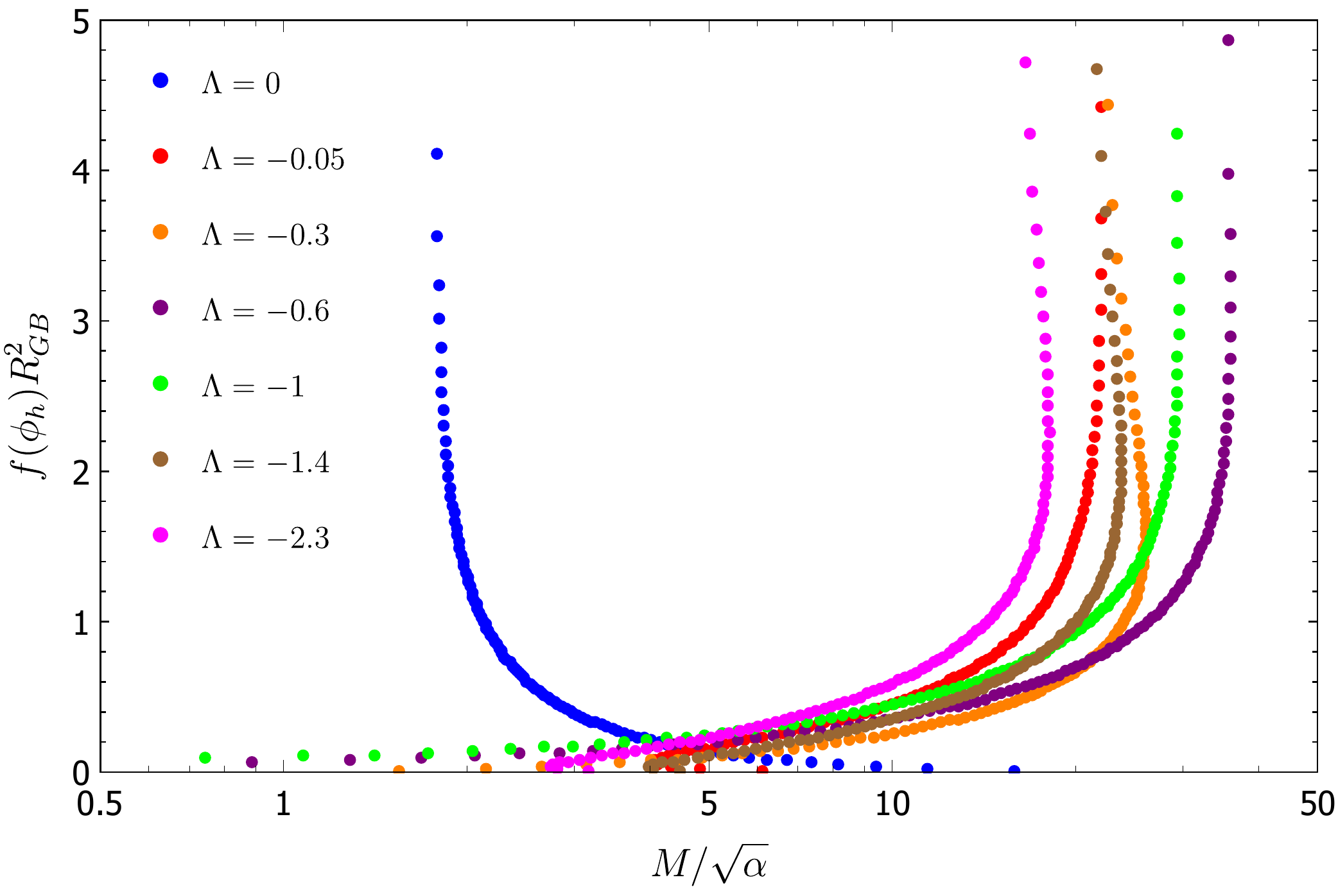}
\hspace*{-0.0cm} {\includegraphics[width = 0.49 \textwidth]
{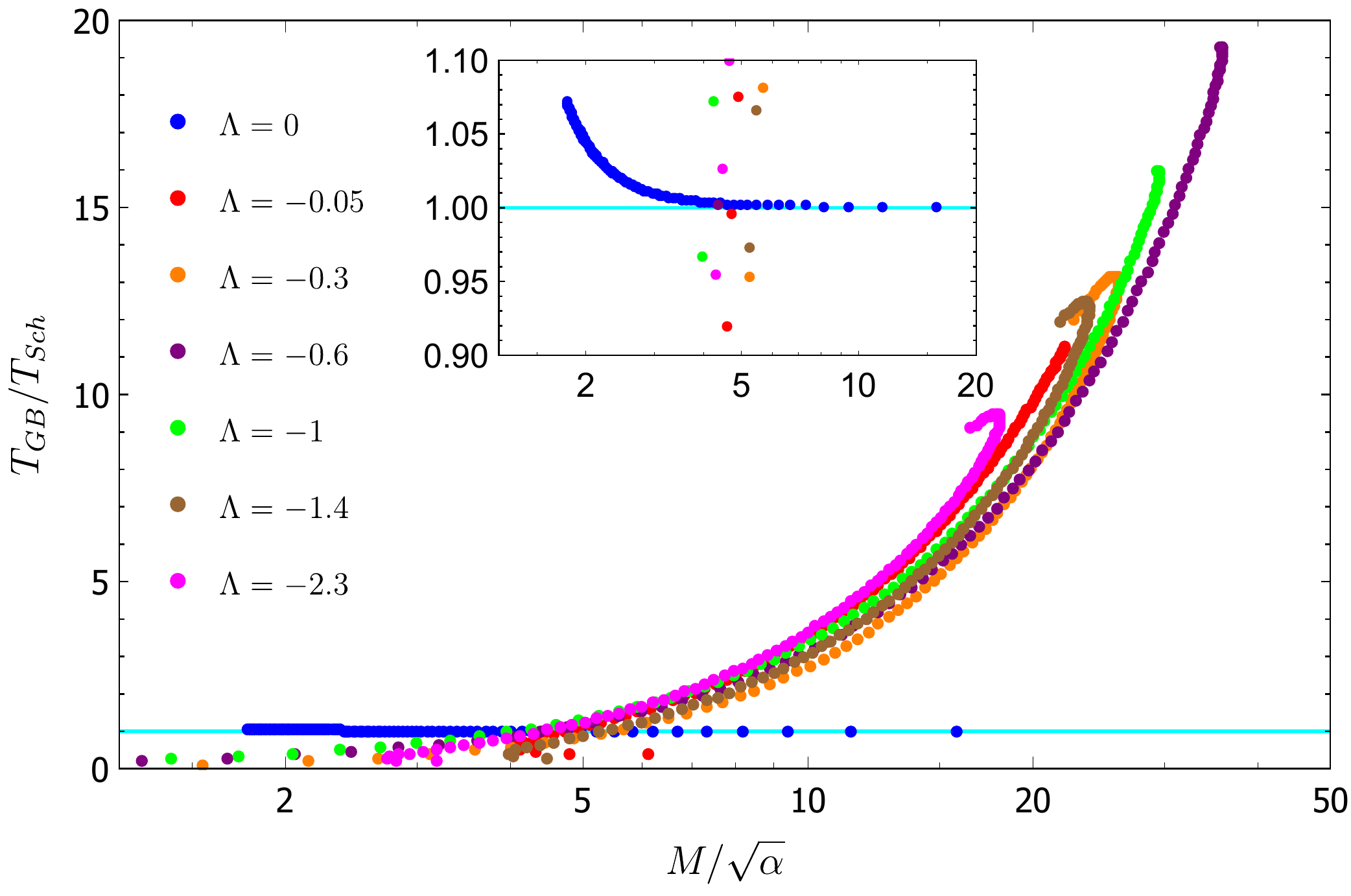}}
    \caption{The combination $f(\phi)\,R^2_{GB}$ at the horizon radius (left plot) and 
the temperature (right plot) in terms of the mass $M$, for $f(\phi)=\alpha e^{\phi}$, 
$V(\phi)=\phi^2$ and various values of $\Lambda<0$. The blue dots correspond again
to the dilatonic black holes with no potential ($\Lambda=0$).}
   \label{GB_mass_quad}
  \end{center}
\end{figure}

The branch of the massive ultra-compact GB black holes is also present for every
non-vanishing, negative value of 
the coupling parameter $\Lambda$. In fact, the exact value of $\Lambda$ determines the
value of $M$ where this branch terminates. As the mass $M$ increases, the horizon radius
gets smaller and its entropy decreases to a very small value. The question emerges
of whether the end point is indeed a black hole with a small but non-vanishing horizon
radius or perhaps a naked singularity. To investigate this, in Fig. \ref{GB_mass_quad} we
plot the values of the GB term at the location of the horizon (multiplied by $f(\phi_h)$
for scaling purposes) in terms of the black-hole mass $M$ of the solutions. The case
of the dilatonic black holes with $\Lambda=0$ is again shown for comparison. 
We observe that, in the presence of a negative potential, a reversal takes place in the
behaviour of the GB term: for $\Lambda=0$, it is the low-mass GB black-hole solutions
that are the most compact, and therefore create the most curved background around them;
for $\Lambda <0$, on the other hand, it is the black-hole solutions near the end point
of the branch of ultra-compact objects, i.e. the black holes with the smallest horizon radius
and the largest mass, that naturally create the most curved background. Although the GB
term reaches a large value, it never becomes infinite -- the solutions at the end of the
branch have a GB term at their horizon which is comparable to the one for a dilatonic
black hole with no potential. This signifies the fact that the horizon radius, although 
very small, remains in fact non-vanishing and therefore these massive, ultra-compact
objects are extremely small-sized black holes. The existence of a minimum horizon radius
is also supported by the constraint $r_h > -2 \dot f_h \phi_h$ discussed in Section 3.
As follows from the left plots of Figs. \ref{phi_exp_pot} and \ref{phi_quad}, the quantity
$\phi'_h$ is always negative while, for all of our choices, $\dot f >0$; as a result, the
aforementioned inequality does impose a lower bound on the horizon value of our
black-hole solutions. For a quadratic potential, this lower bound corresponds to the
end point of the branch of massive, ultra-compact black holes.

%%%%%%%%%%%%%%%%%%%%%
%%%%%%%%%%%%%%%%%

In the right plot of Fig. \ref{GB_mass_quad}, we finally present the temperature
of the black-hole solutions, found in the case of an exponential coupling functions and a
negative quadratic potential, in terms of the mass and for different values of $\Lambda$.
We observe that the temperature of all the solutions belonging to the group of light but
large black holes have a temperature smaller than the Schwarzschild one. In contrast,
all solutions belonging to the group of massive, compact black holes have a 
temperature larger than the Schwarzschild value, which increases as we approach
the end point of the branch of ultra-compact solutions. Therefore, the thermodynamically
more stable solutions, from the point-of-view of the entropy, have a small temperature
while the less stable have a large temperature. In contrast, the behaviour of the 
temperature depends only mildly on the coupling parameter $\Lambda$.

%%%%%%%%%%%%%%%%%%%%%%%%%%%%%%%%%%%%%%%%%%%%%%%%%%%%%%%%%%%%%%%%%%%

\subsection{Different Coupling Functions with a Quadratic Potential}

We would now like to investigate whether the emergence of the aforementioned black-hole
solutions is restricted only in the case of the exponential coupling function between the 
scalar field and the GB term or is a generic feature of the class of EsGB theories. We
have therefore considered a variety of coupling functions, mainly odd and even polynomial
functions of the form $f(\phi)=\alpha \phi^\ell$, with $\ell$ a positive integer. We have also
kept the scalar quadratic potential in order to investigate whether both large and small-sized
black holes continue to emerge. 

%%%%%%%%%%%%%%%%%%%%%
\begin{figure}[t]
\begin{center}
\mbox{\hspace*{-0.3cm} \includegraphics[width = 0.485 \textwidth] {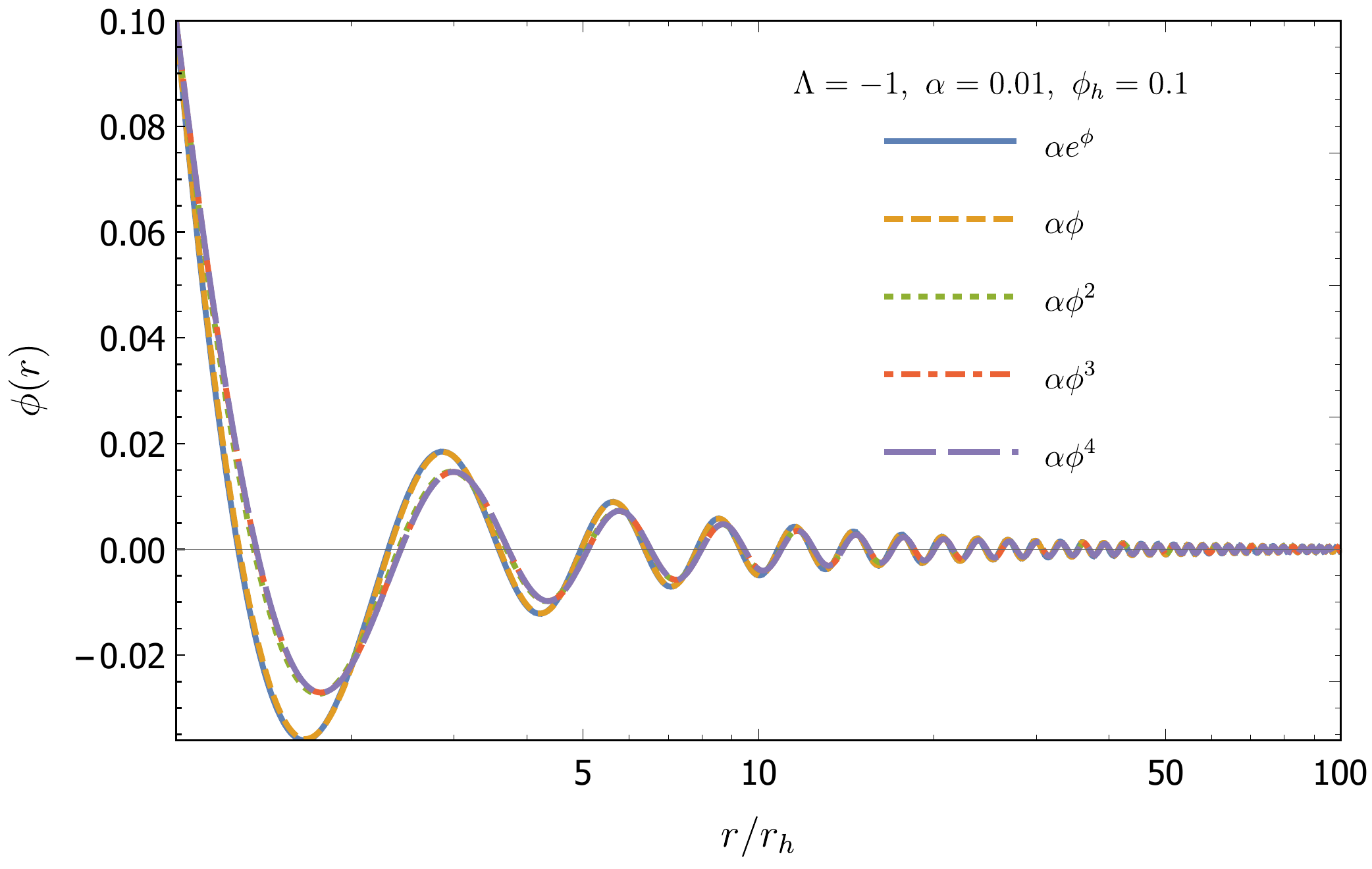}}
\hspace*{-0.2cm} {\includegraphics[width = 0.48 \textwidth]
{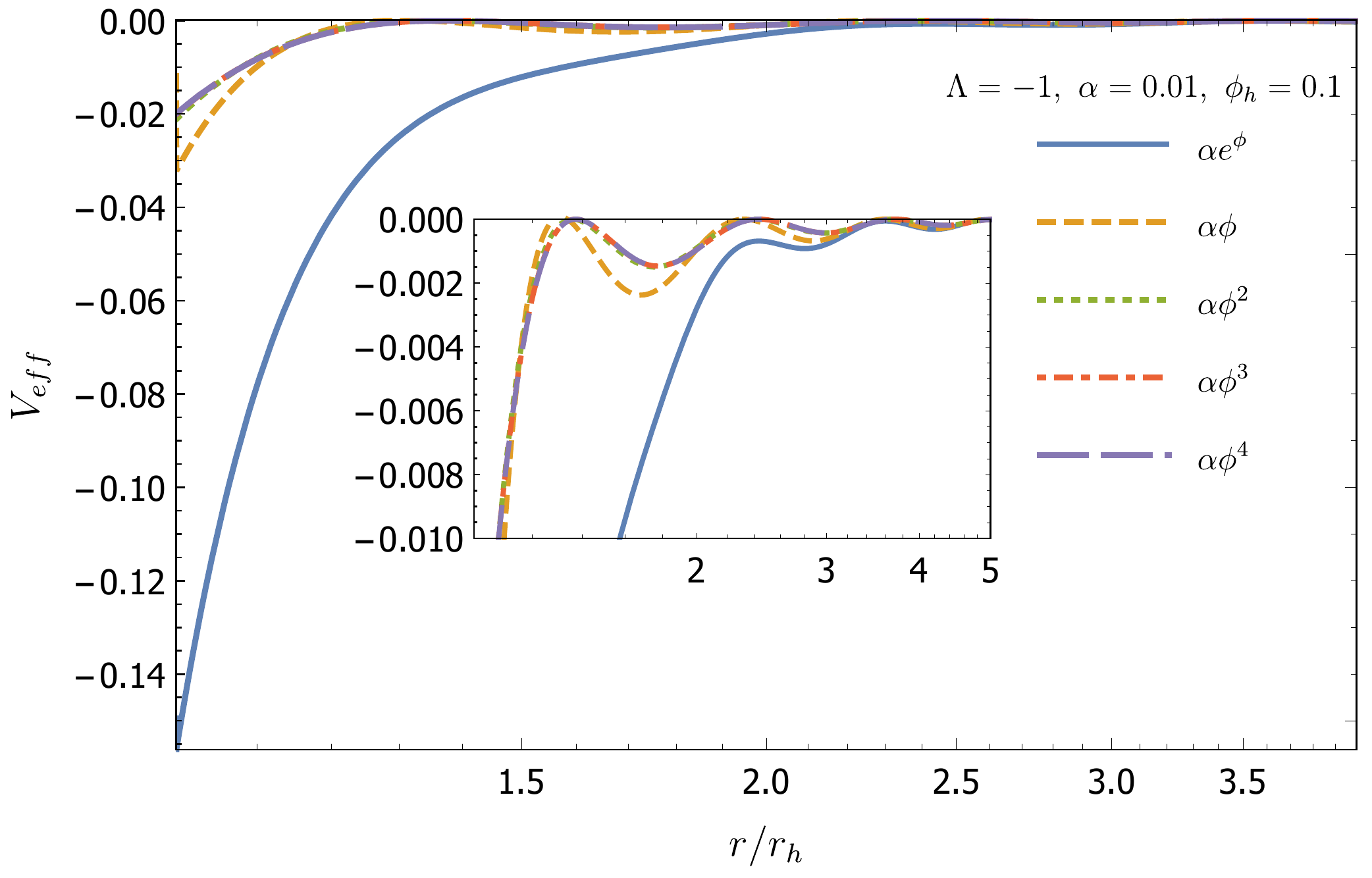}}
    \caption{The scalar field (left plot) and its effective potential $V_{eff}$ (right plot)
in terms of the radial coordinate $r$, for $V(\phi)=\phi^2$ (with $\Lambda=-1$) and a 
variety of coupling functions $f(\phi)$.}
   \label{phi_diff}
  \end{center}
\end{figure}
%%%%%%%%%%%%%%%%%

According to our analysis, classes of black holes similar to the ones presented
in section 5.2 emerge in every single case. The solutions for the metric functions are given
by curves identical to the ones in Fig. \ref{Metric_exp} for every coupling function - as in
the case with no scalar potential \cite{ABK, BAK}, the exact form of $f(\phi)$ is of minor
importance for the solution of the gravitational background. The solution for the scalar
field is more strongly affected by the form of $f(\phi)$, as can be seen in the left plot
of Fig. \ref{phi_diff}. In fact, there seems to be a common behaviour of the solutions
for the scalar field for the coupling
functions $f(\phi)=\alpha e^\phi$ and $f(\phi)=\alpha \phi$ and another one for all
higher polynomials $f(\phi)=\alpha \phi^\ell$, with $\ell>1$. In all cases, the scalar field
is finite and oscillates towards a vanishing value at asymptotic infinity. As in the case
of the exponential coupling function, also here, the value of the coupling parameter
$\Lambda$ affects the near-horizon slope and the frequency of oscillations appearing
in the curve of the scalar field. It also affects in a similar way the effective potential
of the scalar field (whose form for a fixed value of $\Lambda$ is shown in the right
plot of Fig. \ref{phi_diff}) and the energy-momentum tensor components, therefore,
we refrain from presenting additional plots here.

%%%%%%%%%%%%%%%%%%%%%
\begin{figure}[t]
\begin{center}
\mbox{\hspace*{-0.0cm} \includegraphics[width = 0.48 \textwidth] {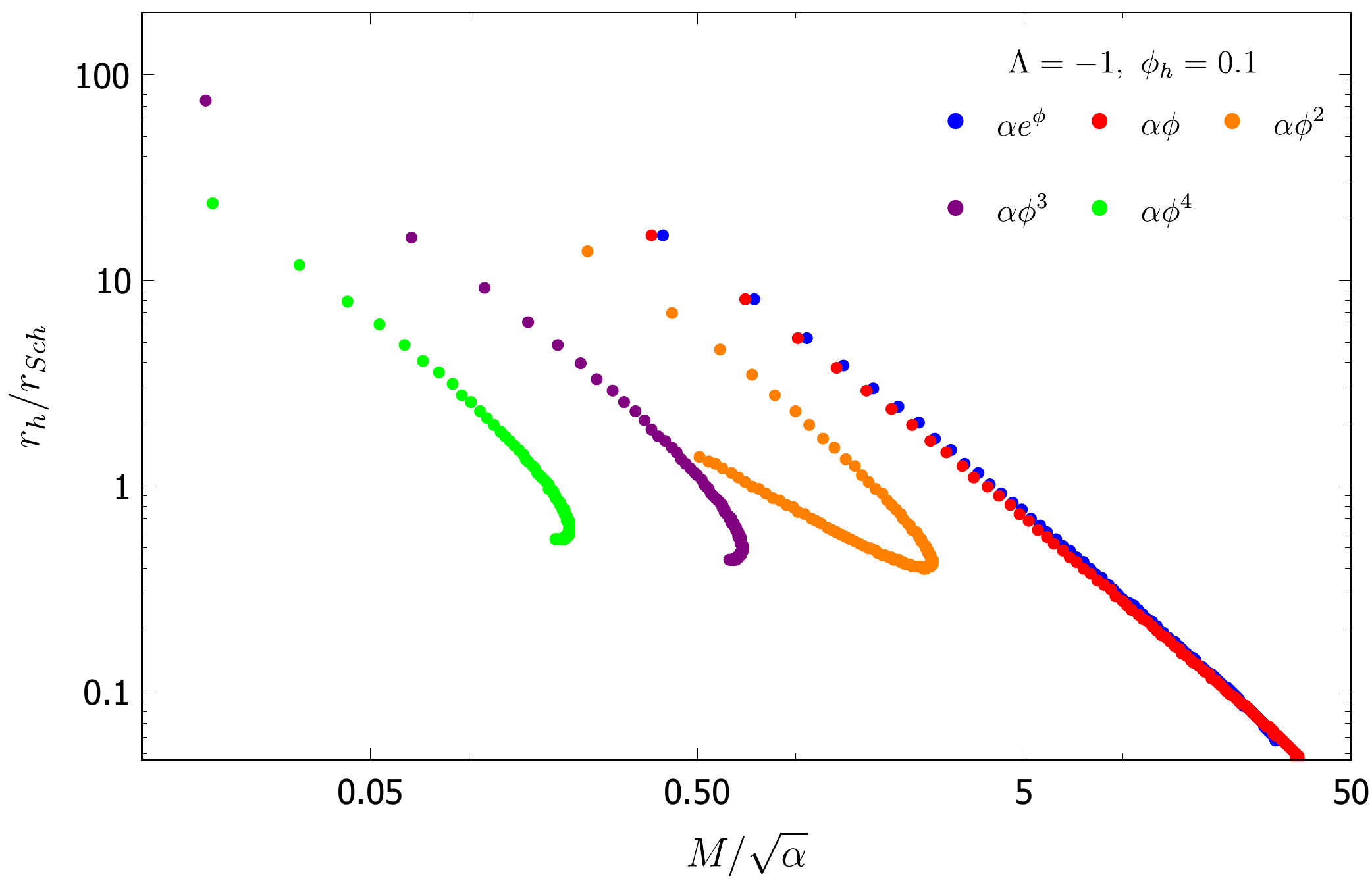}}
\hspace*{-0.0cm} {\includegraphics[width = 0.48 \textwidth]
{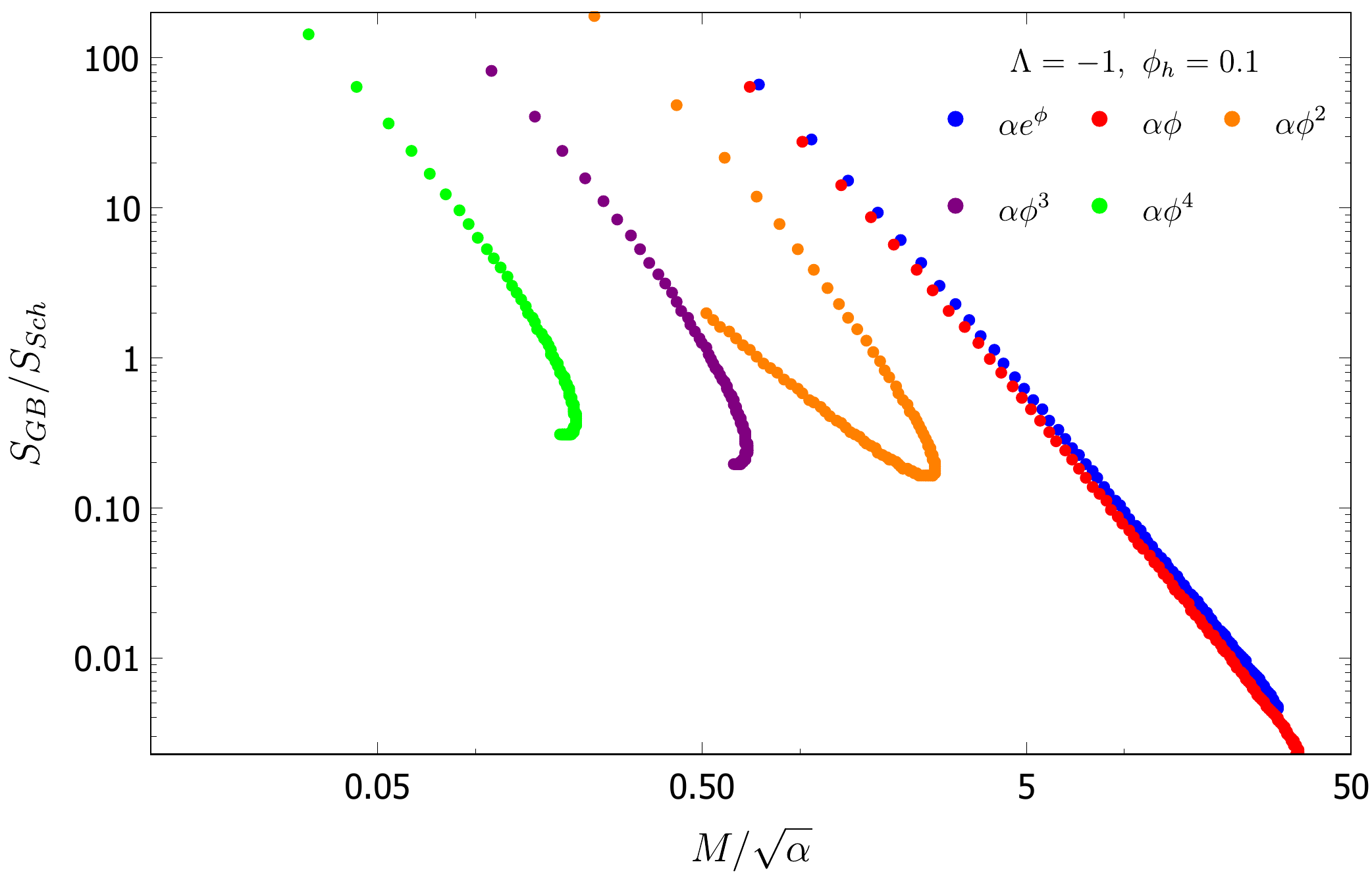}}
    \caption{The horizon radius $r_h$ (left plot) and entropy $S_{GB}$ (right plot) normalised
with respect to the corresponding Schwarzschild values, for $V(\phi)=\phi^2$ (with $\Lambda=-1$)
and a variety of coupling functions $f(\phi)$.}
   \label{rh-S_diff}
  \end{center}
\end{figure}
%%%%%%%%%%%%%%%%%

The horizon radius and entropy of the black-hole solutions obtained for different forms
of the coupling function $f(\phi)$, normalised again to the corresponding Schwarzschild
values, are presented in Fig. \ref{rh-S_diff}. For a quadratic scalar potential, we observe
that all classes of solutions extend between a maximum and a minimum horizon radius,
which in turn corresponds to a minimum and a maximum mass for the black holes. The
subgroup of black holes with a horizon radius larger than the Schwarzschild solution
have also a larger entropy, and thus they are more thermodynamically stable. 
Independently of the form of the coupling function, the smallest in mass GB black holes
with a negative quadratic potential are typically 10 times larger than the corresponding
Schwarzschild solution, and have at least two orders of magnitude larger entropy.

The subgroup of the more massive black holes, on the other hand, terminates for each
coupling function at a different maximum value of the mass, or at a different minimum
value of the horizon radius, as the bound $r_h>-2 \dot f_h \phi'_h$ dictates. The solutions
for the exponential and linear coupling functions have the largest upper value for the mass
and the smallest horizon radius of all, with the latter being approximately the
1/20 of the horizon radius of the Schwarzschild solution with the same mass.  In contrast,
as the integer $m$ in the form of the coupling
function increases, the maximum mass decreases and the minimum horizon radius increases.
Since the small-sized black holes are smaller in entropy compared to the Schwarzschild
solution, they are also less thermodynamically stable. According to the right plot of Fig. \ref{rh-S_diff},
the dilatonic and `linear' GB ultra-compact black holes have two orders of magnitude smaller
entropy than the Schwarzschild solution. This is the same difference in entropy as the one
between the small-mass GB black holes and the Schwarzschild solution. 
In other words, the ultra-compact GB black holes are less thermodynamically stable compared
to the Schwarzschild solution as much as the Schwarzschild solution is less stable compared
to the small-mass GB black holes. On the other hand, the `quadratic' and `higher' GB black holes
have approximately the 1/4 of the entropy of the Schwarzschild solution. This makes them
much more thermodynamically stable than their dilatonic and `linear' analogues but their
compactness is limited being typically only half in size compared to the Schwarzschild
solution.

%%%%%%%%%%%%%%%%%%%%%
\begin{figure}[t]
\begin{center}
\includegraphics[width = 0.49 \textwidth] {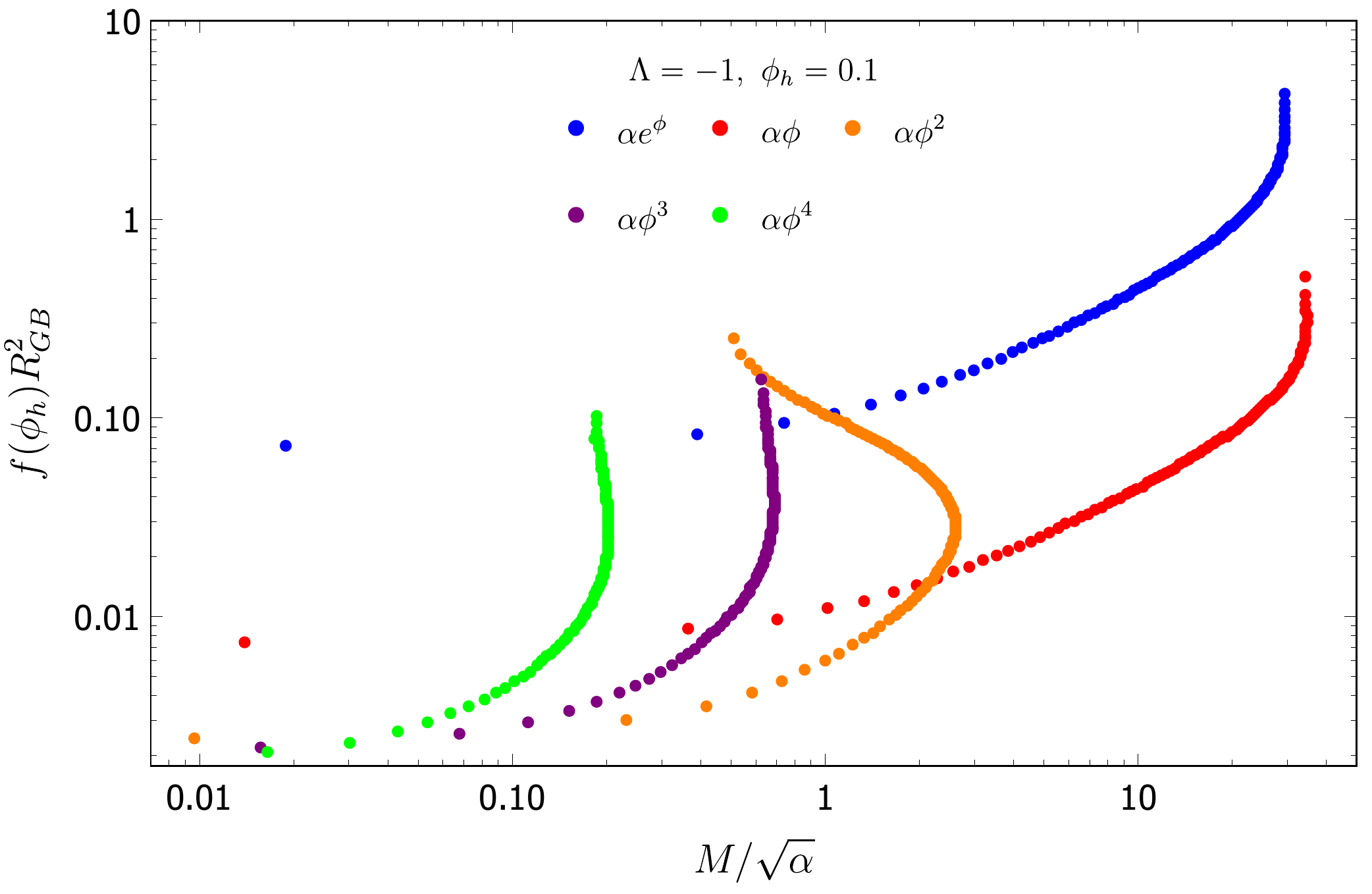}
\hspace*{-0.0cm} {\includegraphics[width = 0.49 \textwidth]
{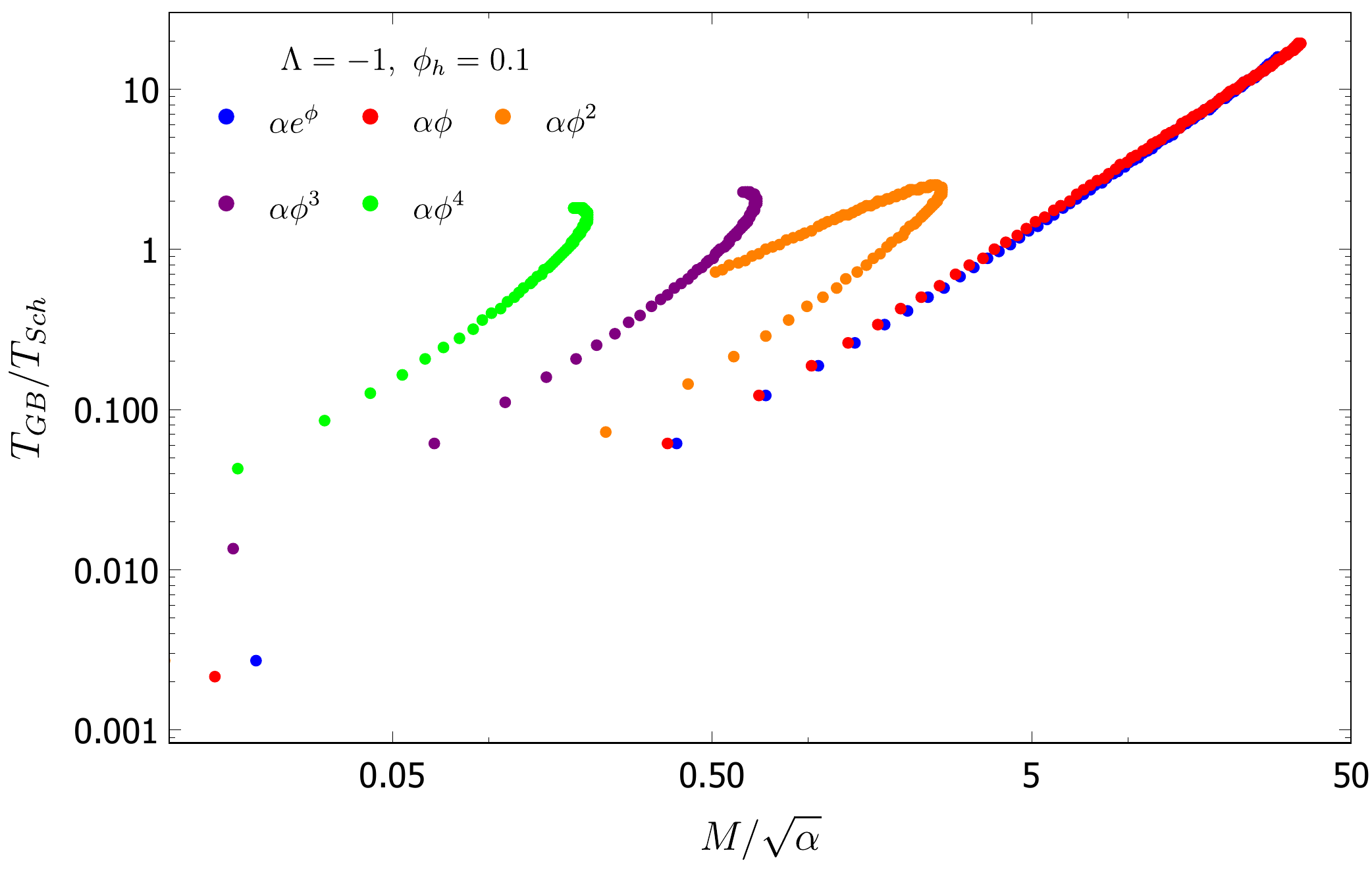}}
    \caption{The combination $f(\phi)\,R^2_{GB}$ at the horizon radius (left plot) and the
temperature (right plot) in terms of the mass $M$, for $V(\phi)=\phi^2$ (and $\Lambda=-1$)
and various forms of the coupling function $f(\phi)$.}
   \label{GB_diff_quad}
  \end{center}
\end{figure}
%%%%%%%%%%%%%%%%%

In Fig. \ref{GB_diff_quad} (left plot), we depict again the combination
$f(\phi)\,R^2_{GB}$ at the horizon of the black-hole solutions, in terms of their mass $M$,
for the quadratic negative potential and for various forms of the coupling function. We observe
again that the GB term remains finite even close to the end point of the branch of the
ultra-compact solutions. In accordance with the previous findings, the curvature of
spacetime is stronger around the small-sized, massive black holes obtained for the 
exponential and linear coupling functions, which constitute the most compact objects
found in this theory. Finally, in the right plot of Fig. \ref{GB_diff_quad}, we present the
temperature of the black-hole solutions, found for a negative quadratic potential and
various coupling functions. The same behaviour found in the case of the exponential
coupling function is also observed here: the light but large black holes have a
temperature smaller than the Schwarzschild one while the massive, compact black
holes have a temperature larger than the Schwarzschild value. Again, it is the exponential
and linear coupling functions which produce the solutions with the largest values of the
temperature. In the opposite limit of small mass, it is also the same coupling functions
that produce the smallest values of the temperature and support the most thermodynamically
stable solutions.

%%%%%%%%%%%%%%%%%%%%%%%%%%%%%%%%%%%%%%%%%%%%%%%%%%%%%%%%%%%%%%%%%%%%%

\section{Conclusions} 

In this work, we have focused on the study of the EsGB theory where the scalar field
possesses also a self-interacting potential. This potential was considered to be negative
and thus to have the opposite sign in the action compared to a traditional self-interacting
term. This choice was motivated by our desire to investigate whether this theory, which 
yielded a plethora of novel black-hole solutions with a non-trivial scalar field in the
presence of a negative cosmological constant \cite{BAK}, could still support similar types
of solutions in the more realistic case where the constant negative distribution of energy
is replaced by a negative field potential. To this end, we have considered a variety of
forms for both the coupling function between the scalar field and the GB term and for the
scalar-field potential, and looked for static, spherically-symmetric, regular black holes
with a non-trivial scalar field. 

We first performed an analytic study of the field equations and demonstrated that,
independently of the forms of the scalar-GB coupling function and of the scalar potential,
an asymptotic solution describing a black-hole horizon with a regular scalar field
always emerges. The particular choices for these two functions do not significantly
affect the form of the gravitational background either close to or far away from
the black-hole horizon. The field equations lead naturally to asymptotically-flat
families of solutions which are everywhere regular, as the profile of the scale-invariant
GB term reveals. The solution for the scalar field does depend on the assumed form
of its potential but it always describes a regular field approaching a vanishing value
at asymptotic infinity. All components of the energy-momentum tensor also remain
finite for every choice of the scalar potential and coupling function, and tend asymptotically
to zero values in agreement with the asymptotic flatness of the solutions. The effective
potential of the scalar field receives contributions from both the GB coupling and the
negative potential, however, both contributions remain bounded and both vanish at large
distances.

Using the case of the exponential scalar-GB coupling function as a prototype, we varied
the form of the scalar potential and studied the properties of the black-hole solutions
obtained, namely the horizon radius, the entropy and the temperature.  We have found
a distinctly different
behaviour of the parameters of the black holes obtained for either $V(\phi)=\phi^2$ or for
$V(\phi)=\phi^{2n}$, with $n>1$. In the latter case, branches of solutions, either smaller
or larger than the Schwarzschild solution depending on their mass and value of $n$,
were found; these solutions were characterised by a minimum mass and approached
the properties of the Schwarzschild solution in the limit of large mass, as all previously
found GB black holes \cite{DBH, ABK, BAK}. In the former case, however, of a negative
quadratic potential, the solutions are divided in two subgroups: the first subgroup 
comprises the small-mass GB black holes which all have a larger horizon radius,
a larger entropy and a smaller temperature compared to the Schwarzschild solution;
the second subgroup includes the more massive black holes the horizon radius of
which gets increasingly smaller as their mass increases -- these solutions have also
a smaller entropy and a larger temperature compared to the Schwarzschild solution.
Whereas the GB black holes obtained in the case of
a positive quadratic potential approach the Schwarzschild solution in the limit of large
mass \cite{Doneva-massive}, the GB solutions supported by a negative quadratic potential
exhibit a regime of very massive but ultra-compact subgroup of black holes. 

The same behaviour is observed when the coupling function between the scalar field
and the GB term assumes a polynomial function, either even or odd, of the scalar 
field. The only differences appear in the values of the minimum and maximum values
of the horizon radius, or correspondingly the mass, of the black-hole solutions obtained
in each case. In all cases, however, the spacetime remains regular even around the
massive, ultra-compact solutions. According to our results, the most compact black
holes emerge for either the exponential (dilatonic) or linear form of the coupling
function with the smallest horizon radius being approximately the 1/20 of the horizon
radius of a Schwarzschild black hole with exactly the same mass. These objects have
only the 1/100 of the entropy of the Schwarzschild solution, while the Schwarzschild
solution has again only the 1/100 of the entropy of the largest (and less massive)
GB black holes in the theory. 

It is worth stressing again the large differences found between the results derived in
this work and those found in the case of a negative cosmological constant \cite{BAK},
as this was our main motivation. In the latter case, the constant, negative $\Lambda$
term worked together with the GB term in order to create a branch of black holes
which started from the Schwarzschild solution, in the limit of large mass, and gave increasingly
smaller black holes (both in terms of the mass and the horizon radius) until the point
of minimum mass, and minimum horizon radius, was reached. When the negative
cosmological constant is replaced by a negative field potential -- different from quadratic
-- the existence of a branch of solutions extending from the Schwarzschild limit down
to a minimum-mass GB black-hole is still observed; however, the horizon radius may be
larger or smaller than the Schwarzschild value depending on the particular value of the
mass of the black hole. When specifically the negative quadratic potential is considered, we have
the appearance of the two distinct subgroups of {\it light} but {\it large} black holes (as in the
case of a positive quadratic potential \cite{Doneva-massive}) and the {\it massive} but
{\it ultra-compact} ones. The reversal in the behaviour of the GB term in this case is also
interesting:
whereas in the asymptotically-flat \cite{ABK} and asymptotically AdS \cite{BAK} cases,
the GB takes its largest value around the black holes with the minimum mass, now it
is the most massive black holes which create the most curved gravitational background
around them; the reason for that is clearly the fact that these are now the most
compact objects in the theory. 

The above solutions owe their existence to the synergy of the GB term
with the negative scalar potential. Black holes emerging in the theory of a minimally-coupled
scalar field with a negative potential would be most likely considered as unnatural objects
-- indeed, to our knowledge, no such black-hole solutions have been derived or studied.
In our analysis, we have demonstrated that, in the presence of the GB term, the negative
potential does not comprise a source of any irregular or destabilising effects in our
solutions. Both the distribution of energy and the spacetime remain finite outside the
black-hole horizon, and support solutions for the scalar field which are always bounded
and naturally die out at large distances. The negativity of the potential may have 
implications on the stability of the solutions under perturbations. Although the form
of the entropy hints towards the fact that a large number of them may in fact be
stable from the thermodynamical point of view, this remains to be investigated. 
Even if a subclass of our solutions turns out to be unstable, their emergence for
a finite amount of time may be in fact associated with interesting phenomena
especially in the strong gravitational-field limit where quadratic terms, such as the
GB, are expected to be important. 

%%%%%%%%%%%%%%%%%%%%%%%%%%%%%%%%%%%%%%%

{\bf Acknowledgments}
We would like to thank Georgios Antoniou for valuable discussions during the early stages
of this work. This research is implemented through the Operational Program
``Human Resources Development, Education and Lifelong Learning" and is co-financed by
the European Union (European Social Fund) and Greek national funds (MIS code: 5006022).

\appendix

\numberwithin{equation}{section}

\section{Set of Differential Equations}
\label{Equations}

Here, we display the explicit expressions of the coefficients $P$, $Q$ and $S$ which
appear in the system of differential equations (\ref{A-sys})-(\ref{phi-sys}).
For notational simplicity, in these expressions we have eliminated, via Eq. (\ref{B'}),
$B'$, which involves $A''$ and $\phi''$, but retained $e^B$. They are:
%%%%%%%%%%%%%%%%%%%%%%%
\begin{align}
P&=-128 e^{4 B} \Lambda ^2 r^3 V^2 \dot f \left(r A'+2 e^B-2\right)+16
   A'^3 \dot f \bigl[     
   -2 e^B \left(-14 e^B+3 e^{2 B}+19\right)
   r \dot f \phi ' \nn \\[1mm]
   &+8\left(-8 e^B+3 e^{2 B}+9\right) \dot f^2
   \phi '^2-e^{2 B} \left(3 e^B-5\right) r^2  \bigr] 
   +4 e^B  A'^2 \Big\{
   e^B r \dot f \Bigl[\left(5 e^B-19\right) r^2
    \phi'^2 \nn\\[0mm]
    &+12 \left(e^B-1\right)^2\Bigr]
  -4 \dot f^2 \phi' \Big[\left(9 e^B-17\right) r^2
   \phi'^2+8 \left(e^B-1\right)^2\Big] +e^{2 B} r^4
   \phi '\Big\}\nn\\[0mm]  
  &+ 4 e^{2B} 2 \,V \L \Bigl\{-e^{2B} r^3 (-2+rA') \f' -16 A' \dot f^2 \f' \left[ 6(3-4e^B +e^{2B}) +(-5 +e^B)r A' \right] \nn\\[0mm]
  &+ 4 e^B \dot f \Bigl[ -3 r^2 A'^2(1+e^B) +4 \Bigl(4(-1+e^B)^2-r^2\f'^2\Bigr)+2rA'(3-3e^B+r^2\f'^2)\Bigr]\Bigr\} \nn\\[0mm]
  &-2e^{2B} r \f' \Bigl\{8\dot f \Big[ 4 r \L \dot V e^B (-1+e^B) -\f'\Big( 4 e^B (-1+e^B) + r^2 \f'^2 (-2+e^B)\Bigr)\Big]\nn\\[0mm]
  &-4re^B(-1+e^B)-r\f'^2\Bigl[ r^2e^B-16\ddot f(-1+e^B) \Bigr]\Bigr\}
  - A' e^B \Bigl\{ -32 r \dot f^2 \f^2 \Big[ 8 r \L \dot V e^{2B} \nn\\[0mm]
  &- \f'(9-4e^B+3e^{2B}) \Bigr]-r^3\f' e^B\Big[4e^B(1+e^B)-\f'^2\Bigl(r^2 e^B+16\ddot f (1+e^B)\Big) \Big]\nn\\[0mm]
  &+8 e^B \dot f \Big[ 4(-1+e^B)^2 + 4 r^3 \L \dot V \f' e^B(1+e^B) +r^2 \f'^2(-7+3e^B)-2\f'^4(r^4 +8r^2\ddot f)\Big]\Big\},
  \end{align}
%%%%%%%%%%%%%%%%%%%%%%%%%%%%%%%%%%%%
\begin{align}
Q&=2304 A'  \dot f^2 \ddot f  \phi'^4-1152
   A'^2 \dot f^3 \phi'^3 + e^B \Big(-144 r^2 A' \dot f^2 \phi'^4+672 r
   A'^2 \dot f^2 \phi '^2+768 A'^2 \dot f^3 \phi '^3\nn\\[0mm]
   &-384 A' \dot f^2 \phi '^2-1024 r A' \dot f \ddot f
   \phi '^3-3584 A' \dot f^2 \ddot f \phi '^4 +480 r \dot f^2 \phi '^4+64 r^2
   \dot f \ddot f \phi '^5-640 \dot f \ddot f \phi '^3 \Big) \nn\\[0mm]
   & + e^{2B} \Big( 128 r^2 A' \ddot f \phi '^2+52 r^3 A' \dot f \phi '^3+
   80 r^2 A' \dot f^2 \phi '^4-128 r^2 A'^2 \dot f \phi '-576 \Lambda  r^2 V A'
   \dot f^2 \phi '^2\nn\\[0mm]
   &-320 r A'^2 \dot f^2 \phi '^2+176 r A' \dot f \phi '-128 A'^2 \dot f^3 \phi '^3+640 A'
   \dot f^2 \phi '^2+1280 r A' \dot f \ddot f \phi '^3+1280 A' \dot f^2 \ddot f \phi '^4\nn\\[1mm]
   &-16 r^3 \ddot f \phi '^4+128 r \ddot f \phi '^2-4 r^4 \dot f \phi '^5-152 r^2
   \dot f \phi '^3-384 \Lambda  r^2 \dot f^2 \phi '^3 \dot V-256 r \dot f^2 \phi'^4
  +384 \Lambda  r V \dot f^2 \phi'^2 \nn\\[1mm]
   &+160 \dot f \phi '-64 r^2 \dot f \ddot f \phi'^5-512 \Lambda  r^2 V \dot f \ddot f \phi '^3
  +1280 \dot f \ddot f \phi '^3\Big] + e^{3B} \Big[ -128 r^2 A' \ddot f \phi '^2
   -12 r^3 A' \dot f \phi '^3\nn\\[1mm]
   &+208 \Lambda  r^3 V A' \dot f \phi '+32 r^2
    A'^2 \dot f \phi '+320 \Lambda  r^2 V A'
   \dot f^2 \phi '^2+32 r A'^2
   \dot f'^2 \phi '^2-224 r A' \dot f \phi '\nn\\[1mm]
   &-256 A' \dot f^2 \phi '^2-256 r A' \dot f \ddot f
   \phi '^3-6 r^4 A' \phi '^2+8 r^3
   A'^2-24 r^2 A'+16 r^3 \ddot f \phi '^4+128
   \Lambda  r^3 V \ddot f \phi '^2\nn\\[1mm]
   &-256 r \ddot f \phi
   '^2+16 \Lambda  r^4 V \dot f \phi '^3+160 \Lambda
    r^3 \dot f \phi '^2 \dot V+24 r^2 \dot f \phi'^3+128 \Lambda  r^2 \dot f^2 \phi
   '^3 \dot V+224 \Lambda  r^2 V \dot f \phi '\nn\\[1mm]
   &+32 r \dot f^2
   \phi '^4-512 \Lambda  r V \dot f^2
   \phi '^2-320 \dot f \phi '\nn+512 \Lambda  r^2 V \dot f \ddot f
   \phi '^3-640 \dot f \ddot f \phi '^3+r^5
   \phi '^4+12 r^3 \phi '^2-32 r \Big]\nn\\[0mm]
   & +e^{4B} \Big[ -48 \Lambda  r^3 V A' \dot f \phi '\nn
    +48 r A' \dot f \phi '-24 \Lambda  r^4
   V A'+24 r^2 A'-128 \Lambda  r^3 V \ddot f \phi '^2+128 r \ddot f \phi '^2\nn\\[0mm]
   &+128 \Lambda ^2 r^4 V^2 \dot f \phi '\nn
  -32\Lambda  r^3 \dot f \phi '^2 \dot V-224 \Lambda  r^2 V \dot f
   \phi '+128 \Lambda  r V \dot f^2 \phi
   '^2+160 \dot f \phi '-4 \Lambda  r^5 V \phi'^2\nn\\[0mm]
   &-16 \Lambda  r^4 \varphi ' V'+4 r^3 \phi
   '^2-64 \Lambda  r^3 V+64 r \Big] +e^{5B} \Big[  -32 \Lambda ^2 r^5 V^2+
64 \Lambda  r^3 V-32 r \Big],
\end{align}
and
%%%%%%%%%%%%%%%%
\begin{align}
S&=2304 A'  \dot f^3 \phi'^2 + 8 e^B \big(  -128 r A'  \dot f^2 \phi '-448 A' \dot f^3
   \phi'^2+32 r^2 \dot f^2 \phi'^3 -80 \dot f^2 \phi' \big) \nn\\[1mm]
   &+8e^{2B} \big( 16 r^2 A' f'+160 r A' \dot f^2 \phi '+160 A'
   \dot f^3 \phi '^2-12 r^3 \dot f \phi '^2-16 r^2 \dot f^2 \phi '^3-
   64 \Lambda  r^2 V \dot f^2 \phi '\nn\\[1mm]
   &+16 r \dot f+160 \dot f^2 \phi ' \big) + 8 e^{3B} \big( -16 r^2 A' \dot f
  -32 r A' \dot f^2 \phi '+4 r^3 \dot f
   \phi '^2+16 \Lambda  r^3 V \dot f \nn \\[1mm]
   &+64 \Lambda  r^2 V \dot f^2 \phi '-32 r \dot f-80 \dot f^2 \phi'
   +r^4 \phi ' \big) +8 e^{4B} \big(  16 r \dot f-16 \Lambda  r^3 V \dot f \big).
\end{align}
%%%%%%%%%%%%%%%%

%%%%%%%%%%%%%%%%%%%%%%%%%%%%%%%%%%%%%%%%%%%%%%%%%%%%%%%%%%%

\section{Scalar Quantities}
\label{scalar}

By employing the metric components of the line-element (\ref{metric}), one may
compute the following scalar-invariant gravitational quantities:
%%%%%%%%%%%%%%%%%%%%%%
\bea
R&=&+\frac{e^{-B}}{2r^2}\left(4e^B-4-r^2A'^2+4rB'-4rA'+r^2A'B'-2r^2A''\right),\label{A1}\\\nonumber\\
R_{\mu\nu}R^{\mu\nu}&=&+\frac{e^{-2B}}{16 r^4}\left[8(2-2e^B+rA'-rB')^2+r^2(rA'^2-4B'-rA'B'+2rA'')^2\right.\nonumber\\
&&\left.+r^2(rA'^2+A'(4-rB')+2rA'')^2\right],\\\nonumber\\
R_{\mu\nu\rho\sigma}R^{\mu\nu\rho\sigma}&=&+\frac{e^{-2B}}{4r^4}\left[r^4A'^4-2r^4A'^3B'-4r^4A'B'A''+r^2A'^2(8+r^2B'^2+4r^2A'')\right.\nonumber\\
&&\left.+16(e^B-1)^2+8r^2B'^2+4r^4A''^2\right],\\\nonumber\\
R_{GB}^2&=&+\frac{2e^{-2B}}{r^2}\left[(e^B-3)A'B'-(e^B-1)A'^2-2(e^B-1)A''\right].\label{A4}
   \eea
%%%%%%%%%%%%%%%%%%%%%%%%%%%%%%%%%%


\begin{thebibliography}{99}

\bibitem{Stelle} K.~S.~Stelle,
  %``Renormalization of Higher Derivative Quantum Gravity,''
  Phys.\ Rev.\ D {\bf 16} (1977) 953.
  %%CITATION = doi:10.1103/PhysRevD.16.953;%%

\bibitem{General} T.~P.~Sotiriou,
  %``Gravity and Scalar Fields,''
  Lect.\ Notes Phys.\  {\bf 892} (2015) 3;\\
  %%CITATION = doi:10.1007/978-3-319-10070-8_1;%%
%%%%%%%%%%%%%%%%%
E.~Berti {\it et al.},
  %``Testing General Relativity with Present and Future Astrophysical Observations,''
  Class.\ Quant.\ Grav.\  {\bf 32} (2015) 243001.

%\cite{LIGO}
\bibitem{LIGO}
https://www.ligo.org/

%\cite{VIRGO}
\bibitem{VIRGO}
http://www.virgo-gw.eu/


\bibitem{Lovelock} 
  D.~Lovelock,
  %``The Einstein tensor and its generalizations,''
  J.\ Math.\ Phys.\  {\bf 12}, 498 (1971).
  doi:10.1063/1.1665613
  %%CITATION = doi:10.1063/1.1665613;%%

\bibitem{NH-scalar} J.~D.~Bekenstein,
  %``Transcendence of the law of baryon-number conservation in black hole physics,''
  Phys.\ Rev.\ Lett.\  {\bf 28} (1972) 452; 
  %%CITATION = doi:10.1103/PhysRevLett.28.452;%%
C.~Teitelboim,
  %``Nonmeasurability of the lepton number of a black hole,''
  Lett.\ Nuovo Cim.\  {\bf 3S2} (1972) 397.
  %%CITATION = doi:10.1007/BF02826050;%%

\bibitem{YM} M.~S.~Volkov and D.~V.~Galtsov,
 %``NonAbelian Einstein Yang-Mills black holes,''
 JETP Lett.\  {\bf 50} (1989) 346;
  %%CITATION = JTPLA,50,346;%% 
P.~Bizon,
%``Colored black holes,''
  Phys.\ Rev.\ Lett.\  {\bf 64} (1990) 2844;
  %%CITATION = doi:10.1103/PhysRevLett.64.2844;%%
B.~R.~Greene, S.~D.~Mathur and C.~M.~O'Neill,
  %``Eluding the no hair conjecture: Black holes in spontaneously broken gauge theories,''
  Phys.\ Rev.\ D {\bf 47} (1993) 2242;
  %%CITATION = doi:10.1103/PhysRevD.47.2242;%%
K.~i.~Maeda, T.~Tachizawa, T.~Torii and T.~Maki,
%``Stability of nonAbelian black holes and catastrophe theory,''
Phys.\ Rev.\ Lett.\  {\bf 72} (1994) 450.
  %%CITATION = doi:10.1103/PhysRevLett.72.450;%%

\bibitem{Skyrmions} H.~Luckock and I.~Moss,
  %``Black Holes Have Skyrmion Hair,''
  Phys.\ Lett.\ B {\bf 176} (1986) 341;
  %%CITATION = doi:10.1016/0370-2693(86)90175-9;%%
S.~Droz, M.~Heusler and N.~Straumann,
  %``New black hole solutions with hair,''
  Phys.\ Lett.\ B {\bf 268} (1991) 371.
  %%CITATION = doi:10.1016/0370-2693(91)91592-J;%%

\bibitem{Conformal} J.~D.~Bekenstein,
  %``Exact solutions of Einstein conformal scalar equations,''
  Annals Phys.\  {\bf 82} (1974) 535; 
  %%CITATION = doi:10.1016/0003-4916(74)90124-9;%%
Annals Phys.\  {\bf 91} (1975) 75.
  %%CITATION = doi:10.1016/0003-4916(75)90279-1;%%

%\cite{Zwiebach:1985uq}
\bibitem{Zwiebach}
  B.~Zwiebach,
  %``Curvature Squared Terms and String Theories,''
  Phys.\ Lett.\  {\bf 156B}, 315 (1985).
  %doi:10.1016/0370-2693(85)91616-8

%\cite{Gross:1986mw}
\bibitem{Gross}
  D.~J.~Gross and J.~H.~Sloan,
  %``The Quartic Effective Action for the Heterotic String,''
  Nucl.\ Phys.\ B {\bf 291}, 41 (1987).
  %doi:10.1016/0550-3213(87)90465-2

%\cite{Metsaev:1987zx}
\bibitem{Metsaev}
  R.~R.~Metsaev, A.~A.~Tseytlin,
  %``Order alpha-prime (Two Loop) Equivalence of the String Equations of Motion and the Sigma Model Weyl Invariance Conditions: Dependence on the Dilaton and the Antisymmetric Tensor,''
  Nucl.\ Phys.\  {\bf B293 }, 385 (1987).

\bibitem{DBH} P.~Kanti, N.~E.~Mavromatos, J.~Rizos, K.~Tamvakis and E.~Winstanley,
  %``Dilatonic black holes in higher curvature string gravity,''
  Phys.\ Rev.\ D {\bf 54} (1996) 5049; 
  %%CITATION = doi:10.1103/PhysRevD.54.5049;%%
Phys.\ Rev.\ D {\bf 57} (1998) 6255.
  %%CITATION = doi:10.1103/PhysRevD.57.6255;%%
  
\bibitem{Gibbons}
  G.~W.~Gibbons and K.~i.~Maeda,
  %``Black Holes and Membranes in Higher Dimensional Theories with Dilaton Fields,''
  Nucl.\ Phys.\ B {\bf 298} (1988) 741.
  %%CITATION = doi:10.1016/0550-3213(88)90006-5;%%
  
\bibitem{Callan}
  C.~G.~Callan, Jr., R.~C.~Myers and M.~J.~Perry,
  %``Black Holes in String Theory,''
  Nucl.\ Phys.\ B {\bf 311} (1989) 673.
  %%CITATION = doi:10.1016/0550-3213(89)90172-7;%%
  
\bibitem{Campbell}
  B.~A.~Campbell, M.~J.~Duncan, N.~Kaloper and K.~A.~Olive,
  %``Axion hair for Kerr black holes,''
  Phys.\ Lett.\ B {\bf 251} (1990) 34;
  %%CITATION = doi:10.1016/0370-2693(90)90227-W;%%
 B.~A.~Campbell, N.~Kaloper and K.~A.~Olive,
  %``Axion hair for dyon black holes,''
  Phys.\ Lett.\ B {\bf 263} (1991) 364.
  %%CITATION = doi:10.1016/0370-2693(91)90474-5;%%

\bibitem{Mignemi}
S.~Mignemi and N.~R.~Stewart,
  %``Charged black holes in effective string theory,''
  Phys.\ Rev.\ D {\bf 47} (1993) 5259.
  %%CITATION = doi:10.1103/PhysRevD.47.5259;%%
  
\bibitem{Kanti1995}
  P.~Kanti and K.~Tamvakis,
  %``Classical moduli O (alpha-prime) hair,''
  Phys.\ Rev.\ D {\bf 52} (1995) 3506.
  %%CITATION = doi:10.1103/PhysRevD.52.3506;%%

\bibitem{Torii}
  T.~Torii, H.~Yajima and K.~i.~Maeda,
  %``Dilatonic black holes with Gauss-Bonnet term,''
  Phys.\ Rev.\ D {\bf 55} (1997) 739.
  %%CITATION = doi:10.1103/PhysRevD.55.739;%%
  
\bibitem{KT}
  P.~Kanti and K.~Tamvakis,
  %``Colored black holes in higher curvature string gravity,''
  Phys.\ Lett.\ B {\bf 392} (1997) 30; 
  %%CITATION = doi:10.1016/S0370-2693(96)01521-3;%%
P.~Kanti and E.~Winstanley,
  %``Do stringy corrections stabilize colored black holes?,''
  Phys.\ Rev.\ D {\bf 61} (2000) 084032.
  %%CITATION = doi:10.1103/PhysRevD.61.084032;%%

\bibitem{Guo} Z.~K.~Guo, N.~Ohta and T.~Torii,
  %``Black Holes in the Dilatonic Einstein-Gauss-Bonnet Theory in Various Dimensions. I. Asymptotically Flat Black Holes,''
 Prog.\ Theor.\ Phys.\  {\bf 120} (2008) 581;
  %%CITATION = doi:10.1143/PTP.120.581;%%
 K.~i.~Maeda, N.~Ohta and Y.~Sasagawa,
  %``Black Hole Solutions in String Theory with Gauss-Bonnet Curvature Correction,''
  Phys.\ Rev.\ D {\bf 80} (2009) 104032;
  %%CITATION = doi:10.1103/PhysRevD.80.104032;%%
N.~Ohta and T.~Torii,
  %``Global Structure of Black Holes in String Theory with Gauss-Bonnet Correction in Various Dimensions,''
  Prog.\ Theor.\ Phys.\  {\bf 124} (2010) 207.
  %%CITATION = doi:10.1143/PTP.124.207;%%
  
 \bibitem{Kleihaus}
  B.~Kleihaus, J.~Kunz and E.~Radu,
  %``Rotating Black Holes in Dilatonic Einstein-Gauss-Bonnet Theory,''
  Phys.\ Rev.\ Lett.\  {\bf 106} (2011) 151104;
  %%CITATION = doi:10.1103/PhysRevLett.106.151104;%%
  B.~Kleihaus, J.~Kunz, S.~Mojica and E.~Radu,
  %``Spinning black holes in Einstein–Gauss-Bonnet–dilaton theory: Nonperturbative solutions,''
  Phys.\ Rev.\ D {\bf 93} (2016) no.4,  044047.
  %%CITATION = doi:10.1103/PhysRevD.93.044047;%%\bibitem{Mignemi:1992nt}
  
   \bibitem{Pani}
  P.~Pani, C.~F.~B.~Macedo, L.~C.~B.~Crispino and V.~Cardoso,
  %``Slowly rotating black holes in alternative theories of gravity,''
  Phys.\ Rev.\ D {\bf 84} (2011) 087501;
  %%CITATION = doi:10.1103/PhysRevD.84.087501;%%
  %%%%%%
  P.~Pani, E.~Berti, V.~Cardoso and J.~Read,
  %``Compact stars in alternative theories of gravity. Einstein-Dilaton-Gauss-Bonnet gravity,''
  Phys.\ Rev.\ D {\bf 84} (2011) 104035.
  %%CITATION = doi:10.1103/PhysRevD.84.104035;%%
  
\bibitem{Herdeiro}
  C.~A.~R.~Herdeiro and E.~Radu,
  %``Kerr black holes with scalar hair,''
  Phys.\ Rev.\ Lett.\  {\bf 112} (2014) 221101.
  %%CITATION = doi:10.1103/PhysRevLett.112.221101;%%
  
  \bibitem{Ayzenberg}
  D.~Ayzenberg and N.~Yunes,
  %``Slowly-Rotating Black Holes in Einstein-Dilaton-Gauss-Bonnet Gravity: Quadratic Order in Spin Solutions,''
  Phys.\ Rev.\ D {\bf 90} (2014) 044066
   Erratum: [Phys.\ Rev.\ D {\bf 91} (2015) no.6,  069905].
  %%CITATION = doi:10.1103/PhysRevD.91.069905, 10.1103/PhysRevD.90.044066;%%

\bibitem{Win-review} E.~Winstanley,
  %``Classical Yang-Mills black hole hair in anti-de Sitter space,''
  Lect.\ Notes Phys.\  {\bf 769} (2009) 49.
  %%CITATION = doi:10.1007/978-3-540-88460-6_2;%%

  \bibitem{Charmousis-rev}
  C.~Charmousis,
  %``Higher order gravity theories and their black hole solutions,''
  Lect.\ Notes Phys.\  {\bf 769} (2009) 299.
  %%CITATION = doi:10.1007/978-3-540-88460-6_8;%%
  
 \bibitem{Herdeiro-review}
  C.~A.~R.~Herdeiro and E.~Radu,
  %``Asymptotically flat black holes with scalar hair: a review,''
  Int.\ J.\ Mod.\ Phys.\ D {\bf 24} (2015) no.09,  1542014.
  %%CITATION = doi:10.1142/S0218271815420146;%%
  
   \bibitem{Blazquez}
  J.~L.~Blazquez-Salcedo {\it et al.},
  %``Black holes in Einstein-Gauß-Bonnet-dilaton theory,''
  IAU Symp.\  {\bf 324} (2017) 265.
  %%CITATION = doi:10.1017/S1743921316012965;%%

\bibitem{Bekenstein} J.~D.~Bekenstein,
  %``Novel ¡¡no-scalar-hair¢¢ theorem for black holes,''
  Phys.\ Rev.\ D {\bf 51} (1995) no.12,  R6608.
  %%CITATION = doi:10.1103/PhysRevD.51.R6608;%%

\bibitem{Horndeski} G.~W.~Horndeski,
  %``Second-order scalar-tensor field equations in a four-dimensional space,''
  Int.\ J.\ Theor.\ Phys.\  {\bf 10} (1974) 363.
  %%CITATION = doi:10.1007/BF01807638;%%

\bibitem{Galileon}
  A.~Nicolis, R.~Rattazzi and E.~Trincherini,
  %``The Galileon as a local modification of gravity,''
  Phys.\ Rev.\ D {\bf 79} (2009) 064036.
  %%CITATION = doi:10.1103/PhysRevD.79.064036;%%
  
\bibitem{SF} T.~P.~Sotiriou and V.~Faraoni,
  %``Black holes in scalar-tensor gravity,''
  Phys.\ Rev.\ Lett.\  {\bf 108} (2012) 081103.
  %%CITATION = doi:10.1103/PhysRevLett.108.081103;%%

\bibitem{HN}  L.~Hui and A.~Nicolis,
  %``No-Hair Theorem for the Galileon,''
  Phys.\ Rev.\ Lett.\  {\bf 110} (2013) 241104.
  %%CITATION = doi:10.1103/PhysRevLett.110.241104;%%
  
\bibitem{SZ}  T.~P.~Sotiriou and S.~Y.~Zhou,
  %``Black hole hair in generalized scalar-tensor gravity,''
  Phys.\ Rev.\ Lett.\  {\bf 112} (2014) 251102.
  %%CITATION = doi:10.1103/PhysRevLett.112.251102;%%
  
\bibitem{Babichev}
  E.~Babichev and C.~Charmousis,
  %``Dressing a black hole with a time-dependent Galileon,''
  JHEP {\bf 1408} (2014) 106.
  %%CITATION = doi:10.1007/JHEP08(2014)106;%%
   
\bibitem{Benkel}  T.~P.~Sotiriou and S.~Y.~Zhou,
  %``Black hole hair in generalized scalar-tensor gravity: An explicit example,''
  Phys.\ Rev.\ D {\bf 90} (2014) 124063;
  %%CITATION = doi:10.1103/PhysRevD.90.124063;%%
R.~Benkel, T.~P.~Sotiriou and H.~Witek,
  %``Black hole hair formation in shift-symmetric generalised scalar-tensor gravity,''
  Class.\ Quant.\ Grav.\  {\bf 34} (2017) no.6,  064001;
  %%CITATION = doi:10.1088/1361-6382/aa5ce7;%%
Phys.\ Rev.\ D {\bf 94} (2016) no.12,  121503.
  %%CITATION = doi:10.1103/PhysRevD.94.121503;%%

\bibitem{Yunes2011}
  N.~Yunes and L.~C.~Stein,
  %``Non-Spinning Black Holes in Alternative Theories of Gravity,''
  Phys.\ Rev.\ D {\bf 83} (2011) 104002.
  %%CITATION = doi:10.1103/PhysRevD.83.104002;%%
    
\bibitem{ABK} G. Antoniou, A. Bakopoulos and P. Kanti, 
  %``Evasion of No-Hair Theorems and Novel Black-Hole Solutions in Gauss-Bonnet Theories,''
  Phys.\ Rev.\ Lett.\  {\bf 120} (2018) no.13,  131102;
  %%CITATION = doi:10.1103/PhysRevLett.120.131102;%%
%``Black-Hole Solutions with Scalar Hair in Einstein-Scalar-Gauss-Bonnet Theories,''
  Phys.\ Rev.\ D {\bf 97} (2018) no.8,  084037.
  %%CITATION = doi:10.1103/PhysRevD.97.084037;%%

\bibitem{Doneva} D.~D.~Doneva and S.~S.~Yazadjiev,
  %``New Gauss-Bonnet Black Holes with Curvature-Induced Scalarization in Extended Scalar-Tensor Theories,''
  Phys.\ Rev.\ Lett.\  {\bf 120} (2018) no.13,  131103.
  %%CITATION = doi:10.1103/PhysRevLett.120.131103;%%

\bibitem{Silva} H.~O.~Silva, J.~Sakstein, L.~Gualtieri, T.~P.~Sotiriou and E.~Berti,
  %``Spontaneous scalarization of black holes and compact stars from a Gauss-Bonnet coupling,'
  Phys.\ Rev.\ Lett.\  {\bf 120} (2018) no.13,  131104.
  %%CITATION = doi:10.1103/PhysRevLett.120.131104;%%

\bibitem{Bardoux}
  Y.~Bardoux, M.~M.~Caldarelli and C.~Charmousis,
  %``Shaping black holes with free fields,''
  JHEP {\bf 1205} (2012) 054.
  %%CITATION = doi:10.1007/JHEP05(2012)054;%%

\bibitem{Ayzen} K.~Yagi, L.~C.~Stein, N.~Yunes and T.~Tanaka,
  %``Post-Newtonian, Quasi-Circular Binary Inspirals in Quadratic Modified Gravity,''
 Phys.\ Rev.\ D {\bf 85} (2012) 064022
 Erratum: [Phys.\ Rev.\ D {\bf 93} (2016) no.2,  029902];
 %%CITATION = doi:10.1103/PhysRevD.93.029902, 10.1103/PhysRevD.85.064022;%%
  D.~Ayzenberg, K.~Yagi and N.~Yunes,
  %``Linear Stability Analysis of Dynamical Quadratic Gravity,''
  Phys.\ Rev.\ D {\bf 89} (2014) no.4,  044023.
  %%CITATION = doi:10.1103/PhysRevD.89.044023;%%

\bibitem{Charmousis}
  C.~Charmousis, T.~Kolyvaris, E.~Papantonopoulos and M.~Tsoukalas,
  %``Black Holes in Bi-scalar Extensions of Horndeski Theories,''
  JHEP {\bf 1407} (2014) 085.
  %%CITATION = doi:10.1007/JHEP07(2014)085;%%

\bibitem{Correa}
  F.~Correa, M.~Hassaine and J.~Oliva,
  %``Black holes in New Massive Gravity dressed by a (non)minimally coupled scalar field,''
  Phys.\ Rev.\ D {\bf 89} (2014) no.12,  124005.
  %%CITATION = doi:10.1103/PhysRevD.89.124005;%%

\bibitem{Dolan}
  S.~R.~Dolan, S.~Ponglertsakul and E.~Winstanley,
  %``Stability of black holes in Einstein-charged scalar field theory in a cavity,''
  Phys.\ Rev.\ D {\bf 92} (2015) no.12,  124047.
  %%CITATION = doi:10.1103/PhysRevD.92.124047;%%

\bibitem{Kunz} 
J.~L.~Blazquez-Salcedo, C.~F.~B.~Macedo, V.~Cardoso, V.~Ferrari, L.~Gualtieri, F.~S.~Khoo, J.~Kunz and P.~Pani,
  %``Perturbed black holes in Einstein-dilaton-Gauss-Bonnet gravity: Stability, ringdown,
  %and gravitational-wave emission,''
Phys.\ Rev.\ D {\bf 94} (2016) no.10,  104024.
   %%CITATION = doi:10.1103/PhysRevD.94.104024;%%

\bibitem{Bhatta} S.~Bhattacharya and S.~Chakraborty,
  %``Constraining some Horndeski gravity theories,''
  Phys.\ Rev.\ D {\bf 95} (2017) no.4,  044037;
   %%CITATION = doi:10.1103/PhysRevD.95.044037;%%
  I.~Banerjee, S.~Chakraborty and S.~SenGupta,
  %``Excavating black hole continuum spectrum: Possible signatures of scalar hairs and of higher dimensions,''
  Phys.\ Rev.\ D {\bf 96} (2017) no.8,  084035.
%%CITATION = doi:10.1103/PhysRevD.96.084035;%%

\bibitem{Doneva-NS} D.~D.~Doneva and S.~S.~Yazadjiev,
  %``Neutron star solutions with curvature induced scalarization in the extended Gauss-Bonnet
%scalar-tensor theories,''
  JCAP {\bf 1804} (2018) no.04,  011.
  %%CITATION = doi:10.1088/1475-7516/2018/04/011;%%

\bibitem{Motohashi} H.~Motohashi and M.~Minamitsuji,
  %``General Relativity solutions in modified gravity,''
  Phys.\ Lett.\ B {\bf 781} (2018) 728;
  %%CITATION = doi:10.1016/j.physletb.2018.04.041;%%
  %``Stealth Schwarzschild solution in shift symmetry breaking theories,''
  Phys.\ Rev.\ D {\bf 98} (2018) no.8,  084027.
  %%CITATION = doi:10.1103/PhysRevD.98.084027;%%

\bibitem{Radu} C.~A.~R.~Herdeiro, E.~Radu, N.~Sanchis-Gual and J.~A.~Font,
  %``Spontaneous Scalarization of Charged Black Holes,''
  Phys.\ Rev.\ Lett.\  {\bf 121} (2018) no.10,  101102;
  %%CITATION = doi:10.1103/PhysRevLett.121.101102;%%
T.~Delsate, C.~Herdeiro and E.~Radu,
  %``Non-perturbative spinning black holes in dynamical Chern-Simons gravity,''
  Phys.\ Lett.\ B {\bf 787} (2018) 8;
  %%CITATION = doi:10.1016/j.physletb.2018.09.060;%%
  Y.~Brihaye, C.~Herdeiro and E.~Radu,
  %``The scalarised Schwarzschild-NUT spacetime,''
  Phys.\ Lett.\ B {\bf 788} (2019) 295.
%%CITATION = doi:10.1016/j.physletb.2018.11.022;%%

\bibitem{Doneva-Papa}
  D.~D.~Doneva, S.~Kiorpelidi, P.~G.~Nedkova, E.~Papantonopoulos and S.~S.~Yazadjiev,
  %``Charged Gauss-Bonnet black holes with curvature induced scalarization in the extended scalar-tensor theories,''
  Phys.\ Rev.\ D {\bf 98} (2018) no.10,  104056.
  %%CITATION = doi:10.1103/PhysRevD.98.104056;%%

\bibitem{Butler} M.~Butler, A.~M.~Ghezelbash, E.~Massaeli and M.~Motaharfar,
  %``Atiyah-Hitchin in Five Dimensional Einstein-Gauss-Bonnet Gravity,''
  arXiv:1808.03217 [hep-th].
  %%CITATION = ARXIV:1808.03217;%%

\bibitem{Danila}
  B.~Danila, T.~Harko, F.~S.~N.~Lobo and M.~K.~Mak,
  %``Spherically symmetric static vacuum solutions in hybrid metric-Palatini gravity,''
  arXiv:1811.02742 [gr-qc].
  %%CITATION = ARXIV:1811.02742;%%

\bibitem{Stetsko}
  M.~M.~Stetsko,
  %``Slowly rotating black hole solution in the scalar-tensor theory with nonminimal derivative coupling and its thermodynamics,''
  arXiv:1811.05030 [hep-th].
  %%CITATION = ARXIV:1811.05030;%%

\bibitem{Tatter}
  O.~J.~Tattersall, P.~G.~Ferreira and M.~Lagos,
  %``Speed of gravitational waves and black hole hair,''
  Phys.\ Rev.\ D {\bf 97} (2018) no.8,  084005.
  %%CITATION = doi:10.1103/PhysRevD.97.084005;%%

\bibitem{Mukherjee}
S.~Mukherjee and S.~Chakraborty,
  %``Horndeski theories confront the Gravity Probe B experiment,''
  Phys.\ Rev.\ D {\bf 97} (2018) no.12,  124007.
  %%CITATION = doi:10.1103/PhysRevD.97.124007;%% 

\bibitem{Chakra}
S.~Chakrabarti,
  %``Collapsing spherical star in Scalar-Einstein-Gauss-Bonnet gravity with a quadratic coupling,''
  Eur.\ Phys.\ J.\ C {\bf 78} (2018) no.4,  296.
  %%CITATION = doi:10.1140/epjc/s10052-018-5798-9;%%

\bibitem{Berti} E.~Berti, K.~Yagi and N.~Yunes,
  %``Extreme Gravity Tests with Gravitational Waves from Compact Binary Coalescences: (I) Inspiral-Merger,''
  Gen.\ Rel.\ Grav.\  {\bf 50} (2018) no.4,  46.
  %%CITATION = doi:10.1007/s10714-018-2362-8;%%

\bibitem{Brihaye}
  Y.~Brihaye and B.~Hartmann,
  %``Critical phenomena of charged Einstein-Gauss-Bonnet black holes with charged scalar hair,''
  Class.\ Quant.\ Grav.\  {\bf 35} (2018) no.17,  175008.
  %%CITATION = doi:10.1088/1361-6382/aad389;%%

\bibitem{Prabhu} K.~Prabhu and L.~C.~Stein,
  %``Black hole scalar charge from a topological horizon integral in Einstein-dilaton-Gauss-Bonnet gravity,''
  Phys.\ Rev.\ D {\bf 98} (2018) no.2,  021503.
  %%CITATION = doi:10.1103/PhysRevD.98.021503;%%

\bibitem{Myung} Y.~S.~Myung and D.~C.~Zou,
  %``Gregory-Laflamme instability of black hole in Einstein-scalar-Gauss-Bonnet theories,''
  Phys.\ Rev.\ D {\bf 98} (2018) no.2,  024030;
  %%CITATION = doi:10.1103/PhysRevD.98.024030;%%
  %``Quasinormal modes of scalarized black holes in the Einstein-Maxwell-Scalar theory,''
  arXiv:1812.03604 [gr-qc].
  %%CITATION = ARXIV:1812.03604;%%

\bibitem{Don-Kunz}
  J.~L.~Blazquez-Salcedo, D.~D.~Doneva, J.~Kunz and S.~S.~Yazadjiev,
  %``Radial perturbations of the scalarized Einstein-Gauss-Bonnet black holes,''
  Phys.\ Rev.\ D {\bf 98} (2018) no.8,  084011;
  %%CITATION = doi:10.1103/PhysRevD.98.084011;%%
J.~L.~Blazquez-Salcedo, Z.~Altaha Motahar, D.~D.~Doneva, F.~S.~Khoo, J.~Kunz, S.~Mojica, K.~V.~Staykov and S.~S.~Yazadjiev,
  %``Quasinormal modes of compact objects in alternative theories of gravity,''
  arXiv:1810.09432 [gr-qc].
  %%CITATION = ARXIV:1810.09432;%%

\bibitem{Benkel2018}
  R.~Benkel, N.~Franchini, M.~Saravani and T.~P.~Sotiriou,
  %``Causal structure of black holes in shift-symmetric Horndeski theories,''
  Phys.\ Rev.\ D {\bf 98} (2018) no.6,  064006.
%%CITATION = doi:10.1103/PhysRevD.98.064006;%%

\bibitem{Iorio} L.~Iorio and M.~L.~Ruggiero,
  %``Constraining some $r^{-n}$ extra-potentials in modified gravity models with LAGEOS-type laser-ranged geodetic satellites,''
  JCAP {\bf 1810} (2018) no.10,  021.
%%CITATION = doi:10.1088/1475-7516/2018/10/021;%%

\bibitem{Ovalle}
  J.~Ovalle, R.~Casadio, R.~da Rocha, A.~Sotomayor and Z.~Stuchlik,
  %``Einstein-Klein-Gordon system by gravitational decoupling,''
  EPL {\bf 124} (2018) no.2,  20004;
  %%CITATION = doi:10.1209/0295-5075/124/20004;%%}
  J.~Ovalle,
  %``Decoupling gravitational sources in general relativity: The extended case,''
  Phys.\ Lett.\ B {\bf 788} (2019) 213.
  %%CITATION = doi:10.1016/j.physletb.2018.11.029;%%

\bibitem{Barack}
  L.~Barack {\it et al.},
  %``Black holes, gravitational waves and fundamental physics: a roadmap,''
   Class.\ Quant.\ Grav.\  {\bf 36} (2019) no.14,  143001.
  %%CITATION = ARXIV:1806.05195;%%

\bibitem{Gao} Y.~X.~Gao, Y.~Huang and D.~J.~Liu,
  %``Scalar perturbations on the background of Kerr black holes in the quadratic dynamical Chern-Simons gravity,''
  Phys.\ Rev.\ D {\bf 99} (2019) no.4,  044020.
  %%CITATION = ARXIV:1808.01433;%%

\bibitem{Lee} B.~H.~Lee, W.~Lee and D.~Ro,
  %``Expanded evasion of the black hole no-hair theorem in dilatonic Einstein-Gauss-Bonnet theory,''
  Phys.\ Rev.\ D {\bf 99} (2019) no.2,  024002.
  %%CITATION = ARXIV:1809.05653;%%

\bibitem{Witek2018}
  H.~Witek, L.~Gualtieri, P.~Pani and T.~P.~Sotiriou,
  %``Black holes and binary mergers in scalar Gauss-Bonnet gravity: scalar field dynamics,''
  Phys.\ Rev.\ D {\bf 99} (2019) no.6,  064035.
  %%CITATION = ARXIV:1810.05177;%%

\bibitem{Moto} H.~Motohashi and S.~Mukohyama,
  %``Shape dependence of spontaneous scalarization,''
  Phys.\ Rev.\ D {\bf 99} (2019) no.4,  044030.
  %%CITATION = ARXIV:1810.12691;%%

\bibitem{Kazanas}
  J.~Sultana and D.~Kazanas,
  %``A no-hair theorem for spherically symmetric black holes in $R^2$ gravity,''
  Gen.\ Rel.\ Grav.\  {\bf 50} (2018) no.11,  137.
 %%CITATION = doi:10.1007/s10714-018-2463-4;%%

\bibitem{Nojiri-Odintsov-Oikonomou}
S.~Nojiri, S.~D.~Odintsov and V.~K.~Oikonomou,
%``Ghost-free Gauss-Bonnet Theories of Gravity, ''
Phys.\ Rev. \ D {\bf 99} (2019) 044050.
%%CITATION = doi:10.1103/PhysRevD.99.044050;%%

\bibitem{Ghodsi:2018vhq}
  S.~Qolibikloo and A.~Ghodsi,
  %``More on phase transition and Renyi entropy,''
  Eur.\ Phys.\ J.\ C {\bf 79} (2019) no.5,  406.
  %%CITATION = doi:10.1140/epjc/s10052-019-6927-9;%%

%\cite{Cunha:2019dwb}
\bibitem{Cunha} 
  P.~V.~P.~Cunha, C.~A.~R.~Herdeiro and E.~Radu,
  %``Spontaneously Scalarized Kerr Black Holes in Extended Scalar-Tensor–Gauss-Bonnet Gravity,''
  Phys.\ Rev.\ Lett.\  {\bf 123}, no. 1, 011101 (2019).
  %doi:10.1103/PhysRevLett.123.011101
  %[arXiv:1904.09997 [gr-qc]].

\bibitem{Minamitsuji-Ikeda_Dec2018}
M.~Minamitsuji and T.~Ikeda,
%``Scalarized black holes in the presence of the coupling to Gauss-Bonnet gravity, ''
Phys.\ Rev. \ D {\bf 99} (2019) 044017;
%%CITATION = doi:10.1103/PhysRevD.99.044017;%%
%\bibitem{Minamitsuji_Apr2019}
%M.~Minamitsuji and T.~Ikeda,
%``Spontaneous scalarization of black holes in the Horndeski theory, ''
Phys.\ Rev. \ D {\bf 99} (2019) 104069.
%%CITATION = doi:10.1103/PhysRevD.99.104069;%%

\bibitem{Stetsko_Dec2018}
M.~M.~Stetsko,
%``Topological black hole in the theory with nonminimal derivative coupling with power-law Maxwell field and its thermodynamics, ''
Phys.\ Rev. \ D {\bf 99} (2019) 044028.
%%CITATION = doi:10.1103/PhysRevD.99.044028;%%

\bibitem{Myung_Dec2018}
Y.~S.~Myung and D.-C.~Zou,
%``Scalarized black holes in the presence of the coupling to Gauss-Bonnet gravity, ''
Phys.\ Lett. \ B {\bf 790} (2019) 400.
%%CITATION = doi:10.1016/j.physletb.2019.01.046;%%

\bibitem{Brihaye-Ducobu}
Y.~Brihaye and L.~Ducobu,
%``Hairy black holes, boson stars and non-minimal coupling to curvature invariants, ''
Phys.\ Lett. \ B {\bf 795} (2019) 135.
%%CITATION = doi:10.1016/j.physletb.2019.06.006;%%

\bibitem{Herdeiro-Radu}
C.~A.~R.~Herdeiro and E.~Radu,
%``Black hole scalarization from the breakdown of scale invariance, ''
Phys.\ Rev. \ D {\bf 99} (2019) 084039.
%%CITATION = doi:10.1103/PhysRevD.99.084039;%%

\bibitem{Kobayashi}
T.~Kobayashi,
%``Horndeski theory and beyond: a review, ''
Rept.\ Prog. \ Phys. {\bf 82} (2019) 086901.
%%CITATION = doi:10.1088/1361-6633/ab2429;%%

\bibitem{Macedo_Dec2018}
H.~O.~Silva, C.~F.~B.~Macedo, T.~P.~Sotiriou, L.~Gualtieri, J.~Sakstein and E.~Berti,
%``Stability of scalarized black hole solutions in scalar-Gauss-Bonnet gravity, ''
Phys.\ Rev. \ D {\bf 99} (2019) 104041.
%%CITATION = doi:10.1103/PhysRevD.99.104041;%%

\bibitem{Dombriz}
A.~de la Cruz-Dombriz and F.~J.M.~Torralba,
%``Birkhoff's theorem for stable torsion theories, ''
JCAP {\bf 1903} (2019) 002.
%%CITATION = doi:10.1088/1475-7516/2019/03/002;%%

\bibitem{Wang-Shen-Xie}
C.-Y.~Wang, Y.-F.~Shen and Y.~Xie,
%``Weak and strong deflection gravitational lensings by a charged Horndeski black hole, ''
JCAP {\bf 1904} (2019) 022.
%%CITATION = doi:10.1088/1475-7516/2019/04/022;%%

\bibitem{Cano-Ruiperez}
P.-A.~Cano and A.~Ruiperez,
%``Leading higher-derivative corrections to Kerr geometry, ''
JHEP {\bf 1905} (2019) 189.
%%CITATION = doi:10.1007/JHEP05(2019)189;%%

\bibitem{Ramazanoglu_Jan2019}
F.~M.~Ramazanoglu,
%``Spontaneous tensorization from curvature coupling and beyond, ''
Phys.\ Rev. \ D {\bf 99} (2019) 084015.
%%CITATION = doi:10.1103/PhysRevD.99.084015;%%

\bibitem{Fernandes}
P.~G.~S.~Fernandes, C.~A.~R.~Herdeiro, A.~M.~Pombo, E.~Radu and N.~Sanchis-Gual,
%``Spontaneous Scalarisation of Charged Black Holes: Coupling Dependence and Dynamical Features, ''
Class.\ Quant. \ Grav. {\bf 36} (2019) 134002.
%%CITATION = doi:10.1088/1361-6382/ab23a1;%%

\bibitem{Brihaye_Feb2019}
Y.~Brihaye and B.~Hartmann,
%``Spontaneous scalarization of charged black holes at the approach to extremality, ''
Phys. \ Lett. \ B {\bf 792} (2019) 244.
%%CITATION = doi:10.1016/j.physletb.2019.03.043;%%

\bibitem{Saravani-Sotiriou}
M.~Saravani and T.~P.~Sotiriou,
%``Classification of shift-symmetric Horndeski theories and hairy black holes, ''
Phys.\ Rev. \ D {\bf 99} (2019) 124004.
%%CITATION = doi:10.1103/PhysRevD.99.124004;%%

\bibitem{Doneva-massive}
D.~D.~Doneva, K.~V.~Staykov and S.~S.~Yazadjiev,
%``Gauss-Bonnet black holes with a massive scalar field, ''
Phys.\ Rev. \ D {\bf 99} (2019) 104045.
%%CITATION = doi:10.1103/PhysRevD.99.104045;%%

\bibitem{Macedo_potential}
C.~F.~B.~Macedo, J.~Sakstein, E.~Berti, L.~Gualtieri, H.~O.~Silva and T.~P.~Sotiriou,
%``Self-interactions and Spontaneous Black Hole Scalarization, ''
Phys.\ Rev. \ D {\bf 99} (2019) 104041.
%%CITATION = doi:10.1103/PhysRevD.99.104041;%%

\bibitem{Saffer}
A.~Saffer, H.~O.~Silva and N.~Yunes,
%``Exterior spacetime of relativistic stars in scalar-Gauss-Bonnet gravity, ''
Phys.\ Rev. \ D {\bf 100} (2019) 044030.
%%CITATION = doi:10.1103/PhysRevD.100.044030;%%

\bibitem{Anson}
T.~Anson, E.~Babichev, C.~Charmousis and S.~Ramazanov,
%``Cosmological instability of scalar-Gauss-Bonnet theories exhibiting scalarization, ''
JCAP {\bf 1906} (2019) 023.
%%CITATION = doi:10.1088/1475-7516/2019/06/023;%%

\bibitem{Myung_Mar2019}
Y.~S.~Myung and D.-C.~Zou,
%``Black holes in Gauss–Bonnet and Chern–Simons-scalar theory, ''
Int.\ J. \ Mod. \ Phys. \ D {\bf 28} (2019) 1950114.
%%CITATION = doi:10.1142/S0218271819501141;%%

\bibitem{Brihaye-Hartmann_Mar2019}
Y.~Brihaye and B.~Hartmann,
%``Spontaneous scalarization of boson stars, ''
JHEP {\bf 1909} (2019) 049.
%%CITATION = doi:10.1007/JHEP09(2019)049;%%

\bibitem{Tattersall}
O.~J.~Tattersall and P.~G.~Ferreira,
%``Spontaneous scalarization in generalized scalar-tensor theory, ''
Phys.\ Rev. \ D {\bf 99} (2019) 104082.
%%CITATION = doi:10.1103/PhysRevD.99.104082;%%

\bibitem{Andreou}
N.~Andreou, N.~Franchini, G.~Ventagli and T.~P.~Sotiriou,
%``Spontaneous scalarization in generalized scalar-tensor theory, ''
Phys.\ Rev. \ D {\bf 99} (2019) 124022.
%%CITATION = doi:10.1103/PhysRevD.99.124022;%%

\bibitem{Liang-Sakstein-Trodden}
Q.~Liang, J.~Sakstein and M.~Trodden,
%``Baryogenesis via gravitational spontaneous symmetry breaking, ''
Phys.\ Rev. \ D {\bf 100} (2019) 063518.
%%CITATION = doi:10.1103/PhysRevD.100.0635187;%%

\bibitem{Charmousis:2019vnf}
  C.~Charmousis, M.~Crisostomi, R.~Gregory and N.~Stergioulas,
  %``Rotating Black Holes in Higher Order Gravity,''
  Phys.\ Rev.\ D {\bf 100} (2019) no.8,  084020.
  %%CITATION = doi:10.1103/PhysRevD.100.084020;%%

\bibitem{Hui-Kabat}
L.~Hui, D.~Kabat, X.~Li, L.~Santoni and S.~S.~C.~Wong,
%``Black Hole Hair from Scalar Dark Matter, ''
JCAP {\bf 1906} (2019) 038.
%%CITATION = doi:10.1088/1475-7516/2019/06/038;%%

\bibitem{Tuan}
D.~Q.~Tuan and S.~H.~Q.~Nguyen,
%``No small hairs in anisotropic power-law Gauss-Bonnet inflation, ''
Commun.\ Phys. {\bf 29} (2019) 173.
%%CITATION = doi:10.15625/0868-3166/29/2/13677;%%

\bibitem{Tuan Do}
Tuan Do \textit{et al},
%``Relativistic redshift of the star S0-2 orbiting the Galactic center supermassive black hole, ''
Science {\bf 365} (2019) 6454.
%%CITATION = doi:10.1126/science.aav8137;%%

\bibitem{Fernandes-Herdeiro-Pombo-Radu-Gual_Jul2019}
P.~G.~S.~Fernandes, C.~A.~R.~Herdeiro, A.~M.~Pombo, E.~Radu and N.~Sanchis-Gual,
%``Charged black holes with axionic-type couplings: classes of solutions and dynamical scalarisation, ''
Phys.\ Rev. \ D {\bf 100} (2019) 084045.
%%CITATION = doi:10.1103/PhysRevD.100.084045;%%

\bibitem{Konoplya-Zhidenko}
R.~A.~Konoplya and A.~Zhidenko,
%``Analytical representation for metrics of scalarized Einstein-Maxwell black holes and their shadows, ''
Phys.\ Rev. \ D {\bf 100} (2019) 044015.
%%CITATION = doi:10.1103/PhysRevD.100.044015;%%


\bibitem{Franchini-Sotiriou}
N.~Franchini and T.~P.~Sotiriou,
%``Cosmology with subdominant Horndeski scalar field, ''
arXiv:1903.05427 [gr-qc].
%%CITATION = ARXIV:1903.05427;%%

\bibitem{Hees}
A.~Hees, O.~Minazzoli, E.~Savalle, Y.~V.~Stadnik, P.~Wolf and B.~Roberts,
%``Violation of the equivalence principle from light scalar fields: from Dark Matter candidates to scalarized black holes, ''
arXiv:1905.08524 [gr-qc].
%%CITATION = ARXIV:1905.08524;%%

\bibitem{Anson-Babichev-Ramazanov}
T.~Anson, E.~Babichev and S.~Ramazanov,
%``Reconciling spontaneous scalarization with cosmology, ''
arXiv:1905.10393 [gr-qc].
%%CITATION = ARXIV:1905.10393;%%

\bibitem{Khalil}
M.~Khalil, N.~Sennett, J.~Steinhoff and A.~Buonanno,
%``Theory-agnostic framework for dynamical scalarization of compact binaries, ''
arXiv:1906.08161 [gr-qc].
%%CITATION = ARXIV:1906.08161;%%

\bibitem{Rham}
C.~de Rham and J.~Zhang,
%``Perturbations of Stealth Black Holes in DHOST Theories, ''
arXiv:1907.006992 [hep-th].
%%CITATION = ARXIV:1907.10112;%%

\bibitem{Perez-Cruz-Lepe-Rivera}
G.~Aguilar-Perez, M.~Cruz, S.~Lepe and I.~Moran-Rivera,
%``Hairy black hole stability under odd parity perturbations in the Einstein-Gauss-Bonnet model, ''
arXiv:1907.06168 [gr-qc].
%%CITATION = ARXIV:1907.06168;%%

\bibitem{Konoplya-Pappas-Zhidenko}
R.~A.~Konoplya, T.~Pappas and A.~Zhidenko,
%``Einstein--scalar--Gauss--Bonnet black holes: Analytical approximation for the metric and applications to calculations of shadows, ''
arXiv:1907.10112 [gr-qc].
%%CITATION = ARXIV:1907.10112;%%

\bibitem{Gao-Liu_Aug2019}
Y.-X.~Gao and D.-J.~Liu,
%``Analytically approximated scalarized black holes and their thermodynamic stability, ''
arXiv:1908.01346 [gr-qc].
%%CITATION = ARXIV:1908.01346;%%

\bibitem{Ikeda-Nakamura-Minamitsuji}
T.~Ikeda, T.~Nakamura and M.~Minamitsuji,
%``Spontaneous scalarization of charged black holes in the Scalar-Vector-Tensor theory, ''
arXiv:1908.09394 [gr-qc].
%%CITATION = ARXIV:1908.09394;%%

\bibitem{Julie-Berti}
F.-L.~Julie and E.~Berti,
%``Post-Newtonian dynamics and black hole thermodynamics in Einstein-scalar-Gauss-Bonnet gravity, ''
arXiv:1909.05258 [gr-qc].
%%CITATION = ARXIV:1909.05258;%%

\bibitem{Ramazanoglu_Oct2019}
F.~M.~Ramazanoglu and K.~I.~Unluturk,
%``Generalized disformal coupling leads to spontaneous tensorization,''
arXiv:1910.02801 [gr-qc].
%%CITATION = ARXIV:1910.02801;%%

\bibitem{Aelst}
K.~V.~Aelst, E.~Gourgoulhon, P.~Grandclement and C. Charmousis,
%``Hairy rotating black holes in cubic Galileon theory,''
arXiv: 1910.08451 [gr-qc].
%%CITATION = ARXIV:1910.08451;%%

\bibitem{Barrientos}
J.~Barrientos, F.~Cordonier-Tello, C.~Corral, F.~Izaurieta, P.~Medina, E.~Rodriguez and O.~Valdivia,
%``Luminical Propagation of Gravitational Waves in Scalar-tensor Theories: The Case for Torsion, ''
arXiv:1910.00148 [gr-qc].
%%CITATION = ARXIV:1910.00148;%%

\bibitem{Ramazanoglu:2019jrr}
  F.~M.~Ramazanoglu and K.~I.~Unluturk,
  %``Generalized disformal coupling leads to spontaneous tensorization,''
  Phys.\ Rev.\ D {\bf 100} (2019) no.8,  084026.
  %%CITATION = doi:10.1103/PhysRevD.100.084026;%%

\bibitem{Grunau:2019bsd}
  S.~Grunau and M.~Kruse,
  %``Motion of charged particles around a scalarized black hole in Kaluza-Klein theory,''
  Phys.\ Rev.\ D {\bf 101} (2020) no.2,  024051.
  %%CITATION = doi:10.1103/PhysRevD.101.024051;%%

\bibitem{Peng:2019qrl}
  Y.~Peng,
  %``Scalarization of compact stars in the scalar-Gauss-Bonnet gravity,''
  JHEP {\bf 1912} (2019) 064.
  %%CITATION = doi:10.1007/JHEP12(2019)064;%%

\bibitem{Bakopoulos:2019tvc}
  A.~Bakopoulos, P.~Kanti and N.~Pappas,
  %``Existence of solutions with a horizon in pure scalar-Gauss-Bonnet theories,''
  Phys.\ Rev.\ D {\bf 101} (2020) no.4,  044026.
  %%CITATION = doi:10.1103/PhysRevD.101.044026;%%

\bibitem{Bernardo:2019yxp}
  R.~C.~Bernardo, J.~Celestial and I.~Vega,
  %``Stealth black holes in shift symmetric kinetic gravity braiding,''
  Phys.\ Rev.\ D {\bf 101} (2020) no.2,  024036.
  %%CITATION = doi:10.1103/PhysRevD.101.024036;%%

\bibitem{Blazquez-Salcedo:2019nwd}
  J.~L.~Blazquez-Salcedo, S.~Kahlen and J.~Kunz,
  %``Quasinormal modes of dilatonic Reissner-Nordstrom black holes,''
  Eur.\ Phys.\ J.\ C {\bf 79} (2019) no.12,  1021.
  %%CITATION = doi:10.1140/epjc/s10052-019-7535-4;%%

\bibitem{Tattersall:2019nmh}
  O.~J.~Tattersall,
  %``Quasi-Normal Modes of Hairy Scalar Tensor Black Holes: Odd Parity,''
  arXiv:1911.07593 [gr-qc].
  %%CITATION = ARXIV:1911.07593;%%

\bibitem{Zou:2019ays}
  D.~C.~Zou and Y.~S.~Myung,
  %``Scalar hairy black holes in Einstein-Maxwell-conformally coupled scalar theory,''
  arXiv:1911.08062 [gr-qc].
  %%CITATION = ARXIV:1911.08062;%%

\bibitem{Noller:2019chl}
  J.~Noller, L.~Santoni, E.~Trincherini and L.~G.~Trombetta,
  %``Black Hole Ringdown as a Probe for Dark Energy,''
  arXiv:1911.11671 [gr-qc].
  %%CITATION = ARXIV:1911.11671;%%

\bibitem{Singh:2019wpu}
  D.~V.~Singh, S.~G.~Ghosh and S.~D.~Maharaj,
  %``Bardeen-like regular black holes in $5D$ Einstein-Gauss-Bonnet gravity,''
  Annals Phys.\  {\bf 412} (2020) 168025.
  %%CITATION = doi:10.1016/j.aop.2019.168025;%%

\bibitem{Collodel:2019kkx}
  L.~G.~Collodel, B.~Kleihaus, J.~Kunz and E.~Berti,
  %``Spinning and excited black holes in Einstein-scalar-Gauss-Bonnet theory,''
  arXiv:1912.05382 [gr-qc].
  %%CITATION = ARXIV:1912.05382;%%

\bibitem{Kim:2019hfp}
  H.~C.~Kim, B.~H.~Lee, W.~Lee and Y.~Lee,
  %``Rotating black holes with an anisotropic matter field,''
  arXiv:1912.09709 [gr-qc].
  %%CITATION = ARXIV:1912.09709;%%

\bibitem{Peng:2019cmm}
  Y.~Peng,
  %``Scalarization of horizonless reflecting stars: neutral scalar fields non-minimally coupled to Maxwell fields,''
  arXiv:1912.11989 [gr-qc]; arXiv:2002.01892 [gr-qc].
  %%CITATION = ARXIV:1912.11989;%%
 %``Analytical investigations on formations of hairy neutral reflecting shells in the scalar-Gauss-Bonnet gravity,''
  %%CITATION = ARXIV:2002.01892;%%

\bibitem{Cuyubamba:2019qtz}
  M.~A.~Cuyubamba Espinoza,
  %``Black Holes and Wormholes in Higher-Curvature Corrected Theories of Gravity,''
  arXiv:1912.08382 [hep-th].
  %%CITATION = ARXIV:1912.08382;%%

\bibitem{Zou:2020rlv}
  D.~C.~Zou and Y.~S.~Myung,
  %``Black hole with primary scalar hair in Einstein-Weyl-Maxwell-conformal scalar theory,''
  arXiv:2001.01351 [gr-qc].
  %%CITATION = ARXIV:2001.01351;%%

\bibitem{Stetsko:2020ntv}
  M.~M.~Stetsko,
  %``Static topological black hole with nonminimal derivative coupling and nonlinear electromagnetic field of Born-Infeld type,''
  arXiv:2001.03574 [hep-th].
  %%CITATION = ARXIV:2001.03574;%%

\bibitem{Konoplya:2020hyk}
  R.~A.~Konoplya and A.~Zhidenko,
  %``General parametrization of black holes: the only parameters that matter,''
  arXiv:2001.06100 [gr-qc].
  %%CITATION = ARXIV:2001.06100;%%

\bibitem{Barausse:2020rsu}
  E.~Barausse {\it et al.},
  %``Prospects for Fundamental Physics with LISA,''
  arXiv:2001.09793 [gr-qc].
  %%CITATION = ARXIV:2001.09793;%%

\bibitem{Alexeyev:2020cuq}
  S.~Alexeyev and M.~Sendyuk,
  %``Black Holes and Wormholes in Extended Gravity,''
  Universe {\bf 6} (2020) no.2,  25.
  doi:10.3390/universe6020025

\bibitem{Kase:2020yhw}
  R.~Kase, M.~Minamitsuji and S.~Tsujikawa,
  %``Neutron stars with a generalized Proca hair and spontaneous vectorization,''
  arXiv:2001.10701 [gr-qc].
  %%CITATION = ARXIV:2001.10701;%%

\bibitem{Delgado:2020rev}
  J.~F.~M.~Delgado, C.~A.~R.~Herdeiro and E.~Radu,
  %``Spinning black holes in shift-symmetric Horndeski theory,''
  arXiv:2002.05012 [gr-qc].
  %%CITATION = ARXIV:2002.05012;%%

\bibitem{Hees:2020gda}
  A.~Hees {\it et al.},
  %``Search for a Variation of the Fine Structure around the Supermassive Black Hole in Our Galactic Center,''
  Phys.\ Rev.\ Lett.\  {\bf 124} (2020) no.8,  081101.
  %%CITATION = doi:10.1103/PhysRevLett.124.081101;%%

\bibitem{Macedo:2020tbm}
  C.~F.~B.~Macedo,
  %``Scalar modes, spontaneous scalarization and circular null-geodesics of black holes in scalar-Gauss-Bonnet gravity,''
  arXiv:2002.12719 [gr-qc].
  %%CITATION = ARXIV:2002.12719;%%

\bibitem{Carson}
  Z.~Carson and K.~Yagi,
  %``Probing Einstein-dilaton Gauss-Bonnet Gravity with the inspiral and ringdown of gravitational waves,''
  arXiv:2003.00286 [gr-qc].
  %%CITATION = ARXIV:2003.00286;%%

%%%%%%%%%%%%%%%%%%%%%%%%%%%%%%%%%%%%%%%%%%%%%%%%%%%%%%%%%%%%%%

\bibitem{Martinez-deSitter}
  C.~Martinez, R.~Troncoso and J.~Zanelli,
  %``De Sitter black hole with a conformally coupled scalar field in four-dimensions,''
  Phys.\ Rev.\ D {\bf 67} (2003) 024008.
  %%CITATION = doi:10.1103/PhysRevD.67.024008;%%

\bibitem{Harper}
  T.~J.~T.~Harper, P.~A.~Thomas, E.~Winstanley and P.~M.~Young,
  %``Instability of a four-dimensional de Sitter black hole with a conformally coupled scalar field,''
  Phys.\ Rev.\ D {\bf 70} (2004) 064023.
  %%CITATION = doi:10.1103/PhysRevD.70.064023;%%

\bibitem{Narita}
  T.~Torii, K.~Maeda and M.~Narita,
  %``No scalar hair conjecture in asymptotic de Sitter space-time,''
  Phys.\ Rev.\ D {\bf 59} (1999) 064027.
  %%CITATION = doi:10.1103/PhysRevD.59.064027;%%

\bibitem{Winstanley-Nohair} E.~Winstanley,
  %``On the existence of conformally coupled scalar field hair for black holes in (anti-)de Sitter space,''
  Found.\ Phys.\  {\bf 33} (2003) 111;
  %%CITATION = doi:10.1023/A:1022871809835;%%
  %``Dressing a black hole with non-minimally coupled scalar field hair,''
  Class.\ Quant.\ Grav.\  {\bf 22} (2005) 2233.
  %%CITATION = doi:10.1088/0264-9381/22/11/020;%%

\bibitem{Bhatta2007}
  S.~Bhattacharya and A.~Lahiri,
  %``Black-hole no-hair theorems for a positive cosmological constant,''
  Phys.\ Rev.\ Lett.\  {\bf 99} (2007) 201101.
  %%CITATION = doi:10.1103/PhysRevLett.99.201101;%%

\bibitem{Martinez} M.~Henneaux, C.~Martinez, R.~Troncoso and J.~Zanelli,
  %``Asymptotically anti-de Sitter spacetimes and scalar fields with a logarithmic branch,''
  Phys.\ Rev.\ D {\bf 70} (2004) 044034;
  %%CITATION = doi:10.1103/PhysRevD.70.044034;%%
  C.~Martinez, R.~Troncoso and J.~Zanelli,
  %``Exact black hole solution with a minimally coupled scalar field,''
  Phys.\ Rev.\ D {\bf 70} (2004) 084035;
  %%CITATION = doi:10.1103/PhysRevD.70.084035;%%
  C.~Erices and C.~Martinez,
  %``Rotating hairy black holes in arbitrary dimensions,''
  Phys.\ Rev.\ D {\bf 97} (2018) no.2,  024034.
  %%CITATION = doi:10.1103/PhysRevD.97.024034;%%

\bibitem{Radu-Win}
  E.~Radu and E.~Winstanley,
  %``Conformally coupled scalar solitons and black holes with negative cosmological constant,''
  Phys.\ Rev.\ D {\bf 72} (2005) 024017.
  %%CITATION = doi:10.1103/PhysRevD.72.024017;%%

\bibitem{Anabalon}
  A.~Anabalon and H.~Maeda,
  %``New Charged Black Holes with Conformal Scalar Hair,''
  Phys.\ Rev.\ D {\bf 81} (2010) 041501.
  %%CITATION = doi:10.1103/PhysRevD.81.041501;%%

\bibitem{Hosler}
  D.~Hosler and E.~Winstanley,
  %``Higher-dimensional solitons and black holes with a non-minimally coupled scalar field,''
  Phys.\ Rev.\ D {\bf 80} (2009) 104010.
  %%CITATION = doi:10.1103/PhysRevD.80.104010;%%

\bibitem{Kolyvaris} C.~Charmousis, T.~Kolyvaris and E.~Papantonopoulos,
  %``Charged C-metric with conformally coupled scalar field,''
  Class.\ Quant.\ Grav.\  {\bf 26} (2009) 175012;
  %%CITATION = doi:10.1088/0264-9381/26/17/175012;%%
  T.~Kolyvaris, G.~Koutsoumbas, E.~Papantonopoulos and G.~Siopsis,
  %``A New Class of Exact Hairy Black Hole Solutions,''
  Gen.\ Rel.\ Grav.\  {\bf 43} (2011) 163.
  %%CITATION = doi:10.1007/s10714-010-1079-0;%%

\bibitem{Ohta}
  K.~i.~Maeda, N.~Ohta and Y.~Sasagawa,
  %``AdS Black Hole Solution in Dilatonic Einstein-Gauss-Bonnet Gravity,''
  Phys.\ Rev.\ D {\bf 83} (2011) 044051;
  %%CITATION = doi:10.1103/PhysRevD.83.044051;%%
Z.~K.~Guo, N.~Ohta and T.~Torii,
  %``Black Holes in the Dilatonic Einstein-Gauss-Bonnet Theory in Various Dimensions II. Asymptotically AdS Topological Black Holes,''
  Prog.\ Theor.\ Phys.\  {\bf 121} (2009) 253;
  %%CITATION = doi:10.1143/PTP.121.253;%%
N.~Ohta and T.~Torii,
  %``Black Holes in the Dilatonic Einstein-Gauss-Bonnet Theory in Various Dimensions. III. Asymptotically AdS Black Holes with k = +-1,''
  Prog.\ Theor.\ Phys.\  {\bf 121} (2009) 959; 
  %%CITATION = doi:10.1143/PTP.121.959;%%
 N.~Ohta and T.~Torii,
  %``Black Holes in the Dilatonic Einstein-Gauss-Bonnet Theory in Various Dimensions IV: Topological Black Holes with and without Cosmological Term,''
  Prog.\ Theor.\ Phys.\  {\bf 122} (2009) 1477.
%%CITATION = doi:10.1143/PTP.122.1477;%%

\bibitem{Saenz}
  S.~G.~Saenz and C.~Martinez,
  %``Anti-de Sitter massless scalar field spacetimes in arbitrary dimensions,''
  Phys.\ Rev.\ D {\bf 85} (2012) 104047.
  %%CITATION = doi:10.1103/PhysRevD.85.104047;%%

\bibitem{Caldarelli}
  M.~M.~Caldarelli, C.~Charmousis and M.~Hassaine,
  %``AdS black holes with arbitrary scalar coupling,''
  JHEP {\bf 1310} (2013) 015.
  %%CITATION = doi:10.1007/JHEP10(2013)015;%%

\bibitem{Gonzalez}
  P.~A.~Gonzalez, E.~Papantonopoulos, J.~Saavedra and Y.~Vasquez,
  %``Four-Dimensional Asymptotically AdS Black Holes with Scalar Hair,''
  JHEP {\bf 1312} (2013) 021.
  %%CITATION = doi:10.1007/JHEP12(2013)021;%%

\bibitem{Gaete}
M.~Bravo Gaete and M.~Hassaine,
  %``Topological black holes for Einstein-Gauss-Bonnet gravity with a nonminimal scalar field,''
  Phys.\ Rev.\ D {\bf 88} (2013) 104011;
  %%CITATION = doi:10.1103/PhysRevD.88.104011;%%
  M.~Bravo Gaete and M.~Hassaine,
  %``Planar AdS black holes in Lovelock gravity with a nonminimal scalar field,''
  JHEP {\bf 1311} (2013) 177.
  %%CITATION = doi:10.1007/JHEP11(2013)177;%%

\bibitem{Giribet}
  G.~Giribet, M.~Leoni, J.~Oliva and S.~Ray,
  %``Hairy black holes sourced by a conformally coupled scalar field in D dimensions,''
  Phys.\ Rev.\ D {\bf 89} (2014) no.8,  085040.
  %%CITATION = doi:10.1103/PhysRevD.89.085040;%%

\bibitem{BenAchour}
  J.~Ben Achour and H.~Liu,
  %``Stealth Schwarzschild-(A)dS black hole in DHOST theories after GW170817: Linear
  %time-dependent scalar dressing,''
  arXiv:1811.05369 [gr-qc].
  %%CITATION = ARXIV:1811.05369;%%

\bibitem{Hartmann} Y.~Brihaye, B.~Hartmann and J.~Urrestilla,
  %``Solitons and black hole in shift symmetric scalar-tensor gravity with cosmological constant,''
  JHEP {\bf 1806} (2018) 074;
  %%CITATION = doi:10.1007/JHEP06(2018)074;%%
  Y.~Brihaye and B.~Hartmann,
  %``Charged scalar-tensor solitons and black holes with (approximate) Anti-de Sitter asymptotics,''
  arXiv:1810.05108 [gr-qc].
  %%CITATION = ARXIV:1810.05108;%%

\bibitem{BAK} A.~Bakopoulos, G.~Antoniou and P.~Kanti,
  %``Novel Black-Hole Solutions in Einstein-Scalar-Gauss-Bonnet Theories with a Cosmological Constant,''
  Phys.\ Rev.\ D {\bf 99} (2019) no.6,  064003.
%  doi:10.1103/PhysRevD.99.064003
%  [arXiv:1812.06941 [hep-th]].
  %%CITATION = doi:10.1103/PhysRevD.99.064003;%%} 

\bibitem{Achour-Liu}
J.~B.~Achour and H.~Liu,
%``Hairy Schwarzschild-(A)dS black hole solutions in degenerate higher order scalar-tensor theories beyond shift symmetry, ''
Phys.\ Rev. \ D {\bf 99} (2019) 064042.
%%CITATION = doi:10.1103/PhysRevD.99.064042;%%

\bibitem{KBP_Corfu}
P.~Kanti, A.~Bakopoulos and N.~Pappas,
%``Scalar-Gauss-Bonnet Theories: Evasion of No-Hair Theorems and novel black-hole solutions, ''
PoS \ CORFU2018 (2019) 091.
%%CITATION = POSCI,CORFU2018,091;%%

\bibitem{BHR}
  Y.~Brihaye, C.~Herdeiro and E.~Radu,
  %``Black Hole Spontaneous Scalarisation with a Positive Cosmological Constant,''
  Phys.\ Lett.\ B {\bf 802} (2020) 135269.
  %%CITATION = doi:10.1016/j.physletb.2020.135269;%%

\bibitem{Brihaye_holo}
  Y.~Brihaye, B.~Hartmann, N.~P.~Aprile and J.~Urrestilla,
  %``Scalarization of asymptotically Anti-de Sitter black holes with applications to holographic phase transitions,''
  arXiv:1911.01950 [gr-qc].
  %%CITATION = ARXIV:1911.01950;%%

\bibitem{Tang}
  Z.~Y.~Tang, B.~Wang and E.~Papantonopoulos,
  %``Exact charged black hole solutions in D-dimensions in f(R) gravity,''
  arXiv:1911.06988 [gr-qc].
  %%CITATION = ARXIV:1911.06988;%%



%%%%%%%%%%%%%%%%%%%%%%%%%%%%%%%%%%%%%%%%%

%\cite{Kanti:2011jz}
\bibitem{KKK1}
%\bibitem{KKK}
  P.~Kanti, B.~Kleihaus and J.~Kunz,
  %``Wormholes in Dilatonic Einstein-Gauss-Bonnet Theory,''
  Phys.\ Rev.\ Lett.\  {\bf 107} (2011) 271101;
%  doi:10.1103/PhysRevLett.107.271101
%  [arXiv:1108.3003 [gr-qc]].
%
%\cite{Kanti:2011yv}
%  P.~Kanti, B.~Kleihaus and J.~Kunz,
  %``Stable Lorentzian Wormholes in Dilatonic Einstein-Gauss-Bonnet Theory,''
  Phys.\ Rev.\ D {\bf 85} (2012) 044007.
%  doi:10.1103/PhysRevD.85.044007
%  [arXiv:1111.4049 [hep-th]].

%\cite{Antoniou:2019awm}
\bibitem{ABKKK} 
  G.~Antoniou, A.~Bakopoulos, P.~Kanti, B.~Kleihaus and J.~Kunz,
  %``Novel Wormhole Solutions in Einstein-Scalar-Gauss-Bonnet Theories,''
  arXiv:1904.13091 [hep-th].


\bibitem{KKK2}
  B.~Kleihaus, J.~Kunz and P.~Kanti,
  %``Particle-like solutions in Einstein-scalar-Gauss-Bonnet theories,''
  arXiv:1910.02121 [gr-qc].
  %%CITATION = ARXIV:1910.02121;%%

\bibitem{Herdeiro-Oliveira}
C.~A.~R.~Herdeiro and J.~M.~S.~Oliveira,
%``On the inexistence of solitons in Einstein–Maxwell-scalar models, ''
Class.\ Quant. \ Grav. {\bf 36} (2019) 105015.
%%CITATION = doi:10.1088/1361-6382/ab1859;%%

\bibitem{Afonso}
  V.~I.~Afonso, G.~J.~Olmo, E.~Orazi and D.~Rubiera-Garcia,
  %``New scalar compact objects in Ricci-based gravity theories,''
  arXiv:1906.04623 [hep-th].
  %%CITATION = ARXIV:1906.04623;%%
  %3 citations counted in INSPIRE as of 30 Oct 2019

\bibitem{Radu-part}
  C.~A.~R.~Herdeiro, J.~M.~S.~Oliveira and E.~Radu,
  %``A class of solitons in Maxwell-scalar and Einstein-Maxwell-scalar models,''
  Eur.\ Phys.\ J.\ C {\bf 80} (2020) no.1,  23.
  %%CITATION = ARXIV:1910.11021;%%

\bibitem{Canate1}
P.~Canate, J.~Sultana and D.~Kazanas,
%``Ellis wormhole without a phantom scalar field, ''
Phys.\ Rev. \ D {\bf 100} (2019) 064007.
%%CITATION = doi:10.1103/PhysRevD.100.064007;%%

\bibitem{Canate2}
P.~Canate and N.~Breton,
%``New exact traversable wormhole solution to the Einstein-scalar-Gauss-Bonnet equations coupled to power-Maxwell electrodynamics, ''
Phys.\ Rev. \ D {\bf 100} (2019) 064067.
%%CITATION = doi:10.1103/PhysRevD.100.064067;%%

%%%%%%%%%%%%%%%%%%%%%%%%%%%%%%%%%%

\bibitem{York} J.~W.~York, Jr.,
  %``Black Hole in Thermal Equilibrium With a Scalar Field: The Back Reaction,''
  Phys.\ Rev.\ D {\bf 31} (1985) 775.
  %%CITATION = doi:10.1103/PhysRevD.31.775;%%

\bibitem{GK} G.~W.~Gibbons and R.~E.~Kallosh,
  %``Topology, entropy and Witten index of dilaton black holes,''
  Phys.\ Rev.\ D {\bf 51} (1995) 2839.
  %%CITATION = doi:10.1103/PhysRevD.51.2839;%% 

\bibitem{GH} G.~W.~Gibbons and S.~W.~Hawking,
  %``Action Integrals and Partition Functions in Quantum Gravity,''
  Phys.\ Rev.\ D {\bf 15} (1977) 2752.
  %%CITATION = doi:10.1103/PhysRevD.15.2752;%%

\bibitem{HP} S.~W.~Hawking and D.~N.~Page,
  %``Thermodynamics of Black Holes in anti-De Sitter Space,''
  Commun.\ Math.\ Phys.\  {\bf 87} (1983) 577.
  %%CITATION = doi:10.1007/BF01208266;%%

\bibitem{Dutta}
  S.~Dutta and R.~Gopakumar,
  %``On Euclidean and Noetherian entropies in AdS space,''
  Phys.\ Rev.\ D {\bf 74} (2006) 044007.
  %%CITATION = doi:10.1103/PhysRevD.74.044007;%%

\bibitem{Wald}
  R.~M.~Wald,
  %``Black hole entropy is the Noether charge,''
  Phys.\ Rev.\ D {\bf 48} (1993) no.8,  R3427.
  %%CITATION = doi:10.1103/PhysRevD.48.R3427;%%

\bibitem{Iyer}
  V.~Iyer and R.~M.~Wald,
  %``Some properties of Noether charge and a proposal for dynamical black hole entropy,''
  Phys.\ Rev.\ D {\bf 50} (1994) 846.
  %%CITATION = doi:10.1103/PhysRevD.50.846;%%

%%%%%%%%%%%%%%%%%%%%%%%%%%%%%%%%%%%%%%%%%%%%%%%%



\end{thebibliography}
\end{document}